\documentclass[letterpaper, appendixfloats]{emulateapj}

\usepackage{epsfig}
\usepackage{amsmath}
\usepackage{rotating}
\usepackage{natbib}
\usepackage{lscape}
\usepackage{enumerate}
\usepackage{multirow}
\usepackage{array}
\usepackage{appendix}
\usepackage{comment}

\def\plotone#1{\centering \leavevmode
\includegraphics[clip=, width=.85\columnwidth]{#1}}
\def\plotoneShrinkVSmall#1{\centering \leavevmode
\includegraphics[clip=, width=.45\columnwidth]{#1}}

\def\plotoneShrinkBig#1{\centering \leavevmode
\includegraphics[clip=, width=.65\columnwidth]{#1}}
\def\plottwo#1#2{\centering \leavevmode
\includegraphics[width=.45\columnwidth]{#1} \hfil
\includegraphics[width=.45\columnwidth]{#2}}

\def\plottwob#1#2{\centering \leavevmode 
\includegraphics[width=.49\columnwidth]{#1} \hfil
\includegraphics[width=.49\columnwidth]{#2}}

\newcommand{\cN}[1]{\mathcal{N}}

\def\gsim{\;\rlap{\lower 2.5pt
 \hbox{$\sim$}}\raise 1.5pt\hbox{$>$}\;}
\def\lsim{\;\rlap{\lower 2.5pt
   \hbox{$\sim$}}\raise 1.5pt\hbox{$<$}\;}

\begin{document}

\title{Spectral and Photometric Diagnostics of Giant Planet Formation Scenarios}

\author{
David S. Spiegel\altaffilmark{1,2},
Adam Burrows\altaffilmark{2}
}

\affil{$^1$Institute for Advanced Study, Princeton, NJ 08540}

\affil{$^2$Department of Astrophysical Sciences, Peyton Hall,
  Princeton University, Princeton, NJ 08544}

\vspace{0.5\baselineskip}

\email{\\
dave@ias.edu\\
burrows@astro.princeton.edu
}

\begin{abstract}
Gas-giant planets that form via core accretion might have very
different characteristics from those that form via disk-instability.
Disk-instability objects are typically thought to have higher
entropies, larger radii, and (generally) higher effective temperatures
than core-accretion objects.  In this paper, we provide a large set of
models exploring the observational consequences of high-entropy (hot)
and low-entropy (cold) initial conditions, in the hope that this will
ultimately help to distinguish between different physical mechanisms
of planet formation.  However, the exact entropies and radii of
newly-formed planets due to these two modes of formation cannot, at
present, be precisely predicted.  It is possible that the distribution
of properties of core-accretion-formed planets and the distribution of
properties of disk-instability-formed planets overlap.  We, therefore,
introduce a broad range of ``Warm Start'' gas-giant planet models.
Between the hottest and the coldest models that we consider,
differences in radii, temperatures, luminosities, and spectra persist
for only a few million to a few tens of millions of years for planets
that are a few times Jupiter's mass or less.  For planets that are
$\sim$five times Jupiter's mass or more, significant differences
between hottest-start and coldest-start models persist for on the
order of 100~Myrs.  We find that out of the standard infrared bands
($J$, $H$, $K$, $L'$, $M$, $N$) the $K$ and $H$ bands are the most
diagnostic of the initial conditions.  A hottest-start model can be
from $\sim$4.5 magnitudes brighter (at Jupiter's mass) to $\sim$9
magnitudes brighter (at ten times Jupiter's mass) than a coldest-start
model in the first few million years.  In more massive objects, these
large differences in luminosity and spectrum persist for much longer
than in less massive objects.  Finally, we consider the influence of
atmospheric conditions on spectra, and find that the presence or
absence of clouds, and the metallicity of an atmosphere, can affect an
object's apparent brightness in different bands by up to several
magnitudes.
\end{abstract}

\keywords{planetary systems -- radiative transfer -- stars: low-mass,
  brown dwarfs -- stars: evolution}

\section{Introduction}
\label{sec:intro}
Objects in the giant planet and brown dwarf mass ranges probably form
in at least two main ways.  On theoretical grounds, core accretion
seems necessary to explain planets like Jupiter and Saturn
\citep{pollack_et_al1996}, while direct gravitational collapse via
disk instability might be unavoidable for objects that form more than
several tens to hundreds of AU from their stars \citep{rafikov2007,
kratter_et_al2010, rafikov2011a}.  If the different modes of formation
lead to different initial planet properties (such as heat content,
radius, temperature, etc.) then there should be observables that, in
principle, could be used to determine the mode of formation of giant
planets and brown dwarfs \citep{fortney_et_al2005, marley_et_al2007,
fortney_et_al2008b}.

Ambitious observing programs for the next decade hope to distinguish
between core-accretion and disk-instability objects.  Examples of such
campaigns include the Gemini Planet Imager (GPI,
\citealt{macintosh_et_al2006, mcbride_et_al2011}), the Near-Infrared
Coronagraphic Imager (NICI) on Gemini South \citep{liu_et_al2010}, the
Spectro-Polarimetric High-contrast Exoplanet REsearch instrument
(SPHERE) on the Very Large Telescope \citep{beuzit_et_al2008}, the
Strategic Exploration of Exoplanets and Disks (SEEDS) experiment on
Subaru \citep{tamura_et_al2009, janson_et_al2011b}, and Project 1640
on Palomar \citep{hinkley_et_al2008, hinkley_et_al2011}.  When
low-mass stars and brown dwarfs occur in close binaries, dynamical
mass estimates can be derived, which can then be used to inform the
initial conditions employed in theoretical models
\citep{konopacky_et_al2010}.  It might be very difficult to obtain
similar mass estimates for directly imaged (wide-separation)
planetary-mass binary companions to stars.  Nevertheless, spectra or
multiband imaging of young planets, together with estimates of the
ages of the systems in which they are found, might be sufficient to
derive constraints on both masses and initial conditions
simultaneously.

Several authors have explored ways to observationally discriminate
between different formation scenarios.  \citeauthor{marley_et_al2007}
(2007, hereafter M07) examined the photometric signatures of ``hot
start'' and ``cold start'' formation scenarios.  In their framework,
the hot-start objects represent disk-instability planets and the
cold-start ones represent those that form via core accretion.
\citet{fortney_et_al2008b} examined the spectral signatures that
distinguish this particular set of ``hot-start'' and ``cold-start''
models, and the influence of metallicity on emergent
spectra.\footnote{An alternative potential observational discriminant
  between formation scenarios is related to the metallicity of stars
  where directly imaged giant planets are found.  Giant planets in the
  inner few AU around their stars presumably formed via core
  accretion, and these objects seem to occur more frequently around
  more metal-rich stars \citep{fischer+valenti2005,
    johnson_et_al2010b}.  \citet{crepp+johnson2011}, therefore,
  suggest that finding a similar correlation between planet occurance
  and metallicity for more distant (directly imaged) objects might
  provide evidence that they form via core accretion, as well.}

In this paper, we build on previous work by examining the photometric
and spectral signatures of a wide range of initial conditions, and the
influence on spectra of different atmosphere types.  Because it is not
clear what precise initial conditions ought to be expected for either
the core accretion or the disk instability mode of formation, we
introduce a class of ``warm-start'' models.  It is possible that both
the core accretion and the disk instability modes of formation produce
initial conditions that span a range, with disk instability forming
planets generally -- but not always -- larger, hotter, and having
higher entropies.  Our warm-start models, then, might be
representative of possible outcomes of either mode of formation.  In
\S\ref{sec:form}, we describe the dependence of planetary radius on
mass and entropy, and the range of expected initial radii and
entropies as a function of mass.  In \S\ref{sec:evcool}, we describe
the cooling and shrinkage of young objects with a range of masses and
initial entropies.  In \S\ref{sec:hotcold}, we discuss the evolution
of hot- and cold-start planet models.  In \S\ref{sec:cont}, we present
a continuum of warm-start models and discuss how observations could be
used to constrain mass, initial entropy/radius, and atmosphere.
Finally, in \S\ref{sec:conc}, we summarize our conclusions.

\section{Formation Scenarios}
\label{sec:form}

There are two main ideas about how objects form in the mass range of
giant planets and brown dwarfs (from a fraction of Jupiter's mass --
$M_J$ -- to tens of $M_J$).\footnote{Planets are often distinguished
  from brown dwarfs via the deuterium-burning criterion, often said to
  be $\sim$13 Jupiter masses \citep{burrows_et_al1997,
    burrows_et_al2001, chabrier+baraffe2000, chabrier_et_al2005}.
  However, this mass depends on metallicity, helium fraction, the
  presence or absence of clouds etc. \citep{saumon+marley2008,
    spiegel_et_al2011a}.  Mode of formation is arguably a better,
  albeit less readily observable, taxonomical discriminant
  \citep{burrows_et_al2001, chabrier_et_al2005, spiegel_et_al2011a,
    schneider_et_al2011}.}  One is that solid cores, similar to, or
somewhat more massive than, the terrestrial planets, form first, and
then a runaway process of gas accretion follows.  In this picture,
once a solid core of $\sim$10 Earth masses has developed, the rate at
which it accretes ``atmosphere'' increases dramatically.  Surrounding
gas from the disk flows inward and forms an accretion shock near the
boundary of the growing planet.  This shock can be quite luminous
\citep{hubickyj_et_al2005, marley_et_al2007}, and might allow the
infalling gas to radiate away much of its initial heat.  Some recent
observational work (e.g., \citealt{janson_et_al2011}) suggests that
core accretion is indeed the predominant mode of planet formation.
The other idea is that the protoplanetary disk becomes gravitationally
unstable and collapses into fragments, forming planets directly.  The
former idea, called ``core accretion'' or ``nucleation instability,''
was suggested more than three decades ago \citep{harris1978,
  mizuno_et_al1978, mizuno1980}, and has been further developed more
recently (e.g., \citealt{ikoma_et_al2000, ikoma_et_al2001}).  The
second idea, called ``disk instability,'' has existed for at least six
decades \citep{kuiper1951}.  In recent years, this mechanism, which
remains controversial in the inner $\sim$10~AU of a planetary system
\citep{rafikov2005}, has been advocated largely by \citet{boss2000,
  boss2007, boss2010} and \citet{mayer_et_al2002, mayer_et_al2004}.

One obvious difference between the mechanisms is that core accretion
involves accretion onto a core.  This could lead objects that form
through core accretion to have higher bulk metallicity than those that
form through disk instability.  Higher metallicity has two competing
effects.  For a given bulk entropy, it causes an object to have a
smaller radius \citep{burrows_et_al2000, guillot_et_al2006,
  leconte_et_al2009, ibgui_et_al2010}.  However, it also leads to
greater atmospheric opacity, which retards the rate at which an object
cools and loses its initial entropy \citep{burrows_et_al2007}.

A second difference between core accretion and disk instability has to
do with the timescale of formation.  Core accretion involves two
distinct steps -- the formation of a rocky core, and the subsequent
accretion of gas.  Simulations suggest that these processes take on
the order of several to tens of millions of years
\citep{pollack_et_al1996, hubickyj_et_al2005}.  Core accretion
simulations of the formation of Jupiter by \citet{pollack_et_al1996}
and \citet{lissauer_et_al2009} indicate that the formation timescale
depends sensitively on the areal surface density ($\sigma_{\rm init}$)
of the protoplanetary disk. The \citet{pollack_et_al1996} simulations
indicate that the time required to form Jupiter is of order several
million years (although for different values of $\sigma_{\rm init}$
the formation timescale can vary by an order of magnitude).  The
opacities of grains in the protoplanetary nebula can also have an
important influence on the core-accretion formation process
\citep{movshovitz_et_al2010}.  Direct collapse, on the other hand, is
thought to proceed on an orbital timescale or a small multiple thereof
-- orders of magnitude faster than core accretion \citep{boss2000}.

A final difference involves the heat of formation that remains in the
object once it is fully formed.  The specific entropy of gas in the
protoplanetary disk is quite high compared with that of a bound
object, such as a planet or a brown dwarf.  But how much of its
initial entropy does the gas lose in becoming part of such an object?
In the core-accretion scenario, it is thought that an accretion disk
forms with an accretion shock near the planet's boundary
\citep{hubickyj_et_al2005}.  As gas flows through the luminous
accretion shock, it loses much of its initial entropy.  For instance,
in work by \citet{hubickyj_et_al2005} and M07, the gas is assumed to
accrete at the entropy of the atmosphere of the forming protoplanet;
i.e., it is taken to lose all its initial (excess) entropy.  Core
accretion is sometimes, therefore, referred to as a ``cold start''
mechanism.  In contrast, it is thought that disk instability leads gas
to retain much more of its initial entropy, thus leading to the ``hot
start'' appelation.  For a given isolated object, a difference in
initial entropy is observationally similar to a difference in age (the
time required to cool from the high entropy condition to the low
entropy condition).  If the age of an object is known precisely
enough, therefore, it might be possible to distinguish a cold start
from a hot start based on the apparent current entropy (and
observables that it influences, such as radius, effective temperature,
luminosity, and spectrum).  Since the relative precision with which
age must be determined to make such a discrimination increases with
increasing age, younger objects are far more useful than older objects
for testing formation models.

The truth, however, might be somewhat more complicated than the simple
``cold-start''/``hot-start'' idea just presented.  M07 pointed out
that the exact post-accretion luminosity of core-accretion objects
depends on uncertain details of the accretion process; however, they
emphasize that there should be a large separation between the initial
entropies of core-accretion planets and disk-instability planets.
This conclusion might turn out to be true, but is perhaps not yet
definitively established.  The accretion shock of the core-accretion
scenario might not be 100\% efficient in removing the initial entropy
\citep{bromley+kenyon2011a, mordasini_et_al2011}.  Conversely, an
accretion shock might form in the disk-instability scenario as well,
radiating away a significant amount of an object's initial heat.  In
other words, the cold start might not be as cold as previously
thought, nor the hot start as hot.  In principle, this could lead to a
spectrum of ``warm start'' initial conditions.  Nevertheless, there
are generally thought to be some broad differences between the early
properties of objects that form via core accretion and those that form
via disk instability.  The infalling gas initially has a significant
excess specific entropy relative to the atmosphere of the forming
planet onto which it is accreting.  If this infalling gas retains some
of its initial excess entropy, the initial entropy of a newly-formed
planet would depend on this (unknown) fraction.  If, on the other
hand, the gas radiates away all its initial entropy down to the
entropy of the gas that is already present in the protoplanet, the
object's entropy at the conclusion of the accretion process depends on
the (unknown) entropy of the gas that was already present.  In either
case, there is not an unambiguous prediction for the initial entropy
of a ``cold start'' object that forms via core accretion.

In order to explore the range of initial entropies that are reasonable
within the core-accretion context, we have computed a large number of
new evolutionary models.  We employ the boundary conditions of
\citet{burrows_et_al1997}, and we assume with M07 that the infalling
gas radiates until it is at the same entropy as the gas already
present.  Figure~\ref{fig:initSR1} displays our new models and the M07
ones.  The thick yellow bands in the top (entropy) and bottom (radius)
panels represent so-called ``hot-start'' models, whose initial
entropies range between those of M07 (open circles) and the
mustard-colored curve (in the top panel) that represents our
``hottest-start'' model.  The blue bands in each panel represent
``cold-start'' models, bounded by a pair of our new core-accretion
models (``Acc: 3$\times$10$^{-5}$'' and ``Acc: 3$\times$10$^{-6}$, Z:
10$^{-2}$'', respectively).  The various new core-accretion models are
described in \S\ref{ssec:initial}.

\subsection{The Influence of Radiating Efficiency During Formation}
\label{ssec:influence_rad}
Different model assumptions about the process of accretion influence
the properties of newly-formed model planets.  We, therefore, consider
a range of assumed radiating efficiencies of the accretion shock.
\citet{bromley+kenyon2011a} and Kenyon \& Bromley (2011 in prep.)
quantify this efficiency as $1-\eta$, where $\eta$ is the fraction of
accretion energy that a parcel of gas retains when it hits the
planet's photosphere, and Kenyon (2011, private communication) has
provided us with several models of forming planets with various values
of $\eta$.  We estimate the specific entropies of these model planets
by finding the values of specific entropy that best explain (within
the context of the \citealt{burrows_et_al1997} models) their
combinations of mass, radius, and luminosity.  These models used
$\eta$ values of 0.15, 0.35, 0.50, and 0.75.  The three models with a
low value of $\eta=0.15$ (which is, nevertheless, larger than the
implied $\eta$ value of 0 in both the M07-clone models and our newly
calculated core-accretion models presented in Fig.~\ref{fig:initSR1})
span the mass range from 0.43~$M_J$ to 4.4~$M_J$, and for these, we
infer specific entropy values from 5.2 to 6.3.  These and all
subsequent specific (per-baryon) entropy values are in units of
Boltzmann's constant per baryon.  For comparison, Jupiter has a
specific entropy of $\sim$6.  Though it might initially be surprising
that newly formed planets should appear as cold as Jupiter is at
$\sim$4.5 billion years, the differences between the equation of state
(EOS) used in the \citet{bromley+kenyon2011a} models and the one
\citep{saumon_et_al1995} used in both the M07 models and in our own
might explain this apparent mystery.  Since
\citet{bromley+kenyon2011a} assume that infalling gas retains 15\% of
its excess internal energy, we presumably would have inferred a
specific entropy closer to 9 (for reasons discussed below) if they had
used the \citet{saumon_et_al1995} EOS.

It is particularly instructive to examine the effect of larger $\eta$.
Inferring the specific entropy from mass and radius as before, we
estimate that a 0.88~$M_J$ object formed with $\eta=0.35$ has a
specific entropy of 7.5; a 1.2~$M_J$ object formed with $\eta=0.50$
has a specific entropy of 10.1; and a 1.6~$M_J$ object formed with
$\eta=0.75$ has a specific entropy of 10.6.  Given that the
$\eta=0.15$ objects appeared to be roughly 3 units lower in specific
entropy than the cold-start band in Fig.~\ref{fig:initSR1}
(encompassing our own models and those of \citealt{marley_et_al2007}),
these specific entropy values at larger values of $\eta$ might
similarly be reduced below what our models would calculate if embedded
in the same formation scenario.  Specific entropy is, of course, not
linearly additive, and so the 10.6 value cannot be adjusted upwards in
a simple way.  Still, this suggests that if the gas cools fairly
inefficiently while falling onto a forming giant planet, the final
assembly might be significantly hotter than core accretion models
typically assume.

\subsection{Evidence for Hot Starts?}
\label{ssec:ev_hot}
Indeed, there is mounting evidence that suggests that particularly
massive young objects form fairly hot.  Several years ago, the brown
dwarf eclipsing binary 2MASS J05352184-0546085 was discovered.  The
two objects in this system have masses $57~M_J$ and $36~M_J$, and are
thought to be quite young, less than a few million years old
\citep{stassun_et_al2006, stassun_et_al2007}.  Their radii ($6.5 R_J$
and $5.0 R_J$, respectively, where $R_J$ is Jupiter's radius) and
effective temperatures (2650~K and 2790~K, respectively) are
indicative of the high-entropy initial conditions expected in the case
of direct gravitational collapse.  The masses and orbits of the
objects in this system are more reminiscent of very low-mass binary
stars than of a star-planet system, and so the formation process might
be expected to be more similar to star formation than to core
accretion \citep{joergens2008}.

Observations suggest that some lower mass objects also form hot.  For
instance, the planets found in the HR8799 system
\citep{marois_et_al2008, marois_et_al2010} appear to be inconsistent
with classical ``cold start'' models \citep{dodson-robinson_et_al2009,
  bowler_et_al2010, janson_et_al2010, currie_et_al2011,
  madhusudhan_et_al2011}.  The planetary companion to $\beta$-Pictoris
($\beta$-Pic~b) also appears to have formed hotter than cold-start
models would have predicted \citep{lagrange_et_al2010,
  quanz_et_al2010}.  Other young substellar objects, such as GQ Lupi
b, can similarly inform our understanding of the character of their
geneses and early evolution \citep{mcelwain_et_al2007,
  neuhauser_et_al2008}.  Given both the theoretical prejudice against
disk instability as the dominant mode of giant planet formation and
the recent observational evidence that objects do not form as cold as
some core-accretion models suggest, the time is ripe to revisit what
observations can tell us about the properties of very young planets.

\section{Evolutionary Cooling}
\label{sec:evcool}

As a self-gravitating sphere ages, it cools and shrinks.  The cooling
is rapid at first, and slows as the object reaches lower effective
temperatures.  More massive objects retain their heat for longer,
since their gravitational energy scales with $M^2$ (and by Virial
equilibrium, the thermal energy scales with the gravitational energy).

\begin{figure}[t!]
\plotoneShrinkBig
{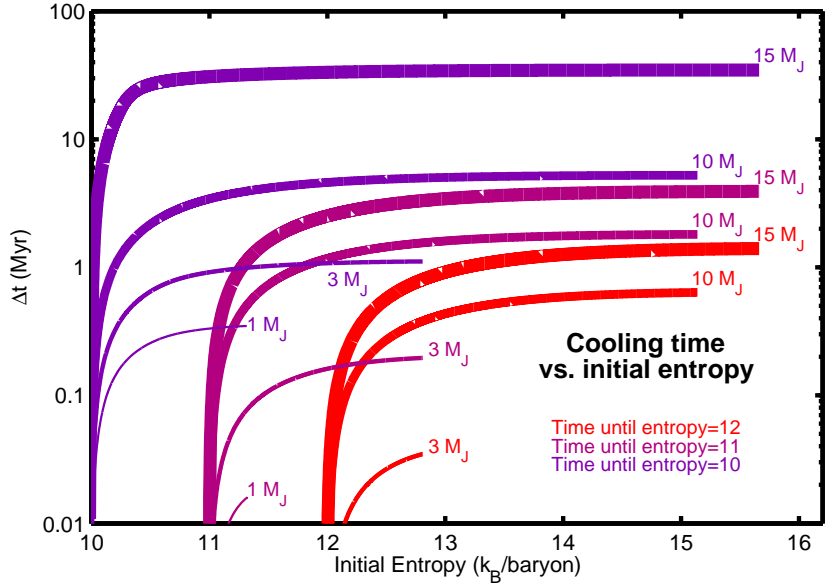}
\caption{Cooling time as a function of initial entropy for
  \citet{burrows_et_al1997} models, for 1, 3, 10, and 15~$M_J$.  Line
  thickness indicates mass, with thicker lines denoting more massive
  objects.  Red, magenta, and blue curves have final specific
  entropies of 12, 11, and 10, respectively.  Objects even as massive
  as 10~$M_J$ cool to an entropy of 12 very quickly (in less than a
  million years after the end of accretion), regardless of their
  initial entropy.  Cooling to an entropy of 11 takes a factor of
  several longer ($\sim$2~Myr for 10-$M_J$ models that start very
  hot).  Cooling to an entropy of 10 requires $\sim$5~Myr for 10-$M_J$
  models.  Cooling times are a factor of $\sim$4 less for 3-$M_J$
  models and another factor of $\sim$3 less for 1-$M_J$ models.}
\label{fig:cooling}
\end{figure}

Figure~\ref{fig:cooling} vividly portrays how the cooling time scales
with mass and with initial entropy, for \citet{burrows_et_al1997}
models.  For objects of 1, 3, 10, and 15 times Jupiter's mass, we plot
the time required to reach various threshold specific entropies (10,
11, and 12), as a function of the object's initial entropy.  15-$M_J$
objects require several tens of Myr to cool to $S=10$ (where $S$ is
specific entropy, in units of Boltzmann constant per baryon), several
Myr to cool to $S=11$, and about 1~Myr to cool to $S=12$.  Less
massive objects cool to $S=10$ in a few Myr or less.  Objects up to
10~$M_J$ cool to an entropy of 12 extremely quickly (in less than a
million years after ``formation'' -- the end of accretion),
irrespective of initial entropy.  Reaching an entropy of 11 takes a
factor of several longer ($\sim$2~Myr for 10-$M_J$ models that start
very hot).  Reaching an entropy of 10 takes 10-$M_J$ models
$\sim$5~Myr.  The cooling times are a factor of $\sim$4 less for
3-$M_J$ models and another factor of $\sim$3 less for 1-$M_J$ models.

\begin{figure}[t!]
\plotone
{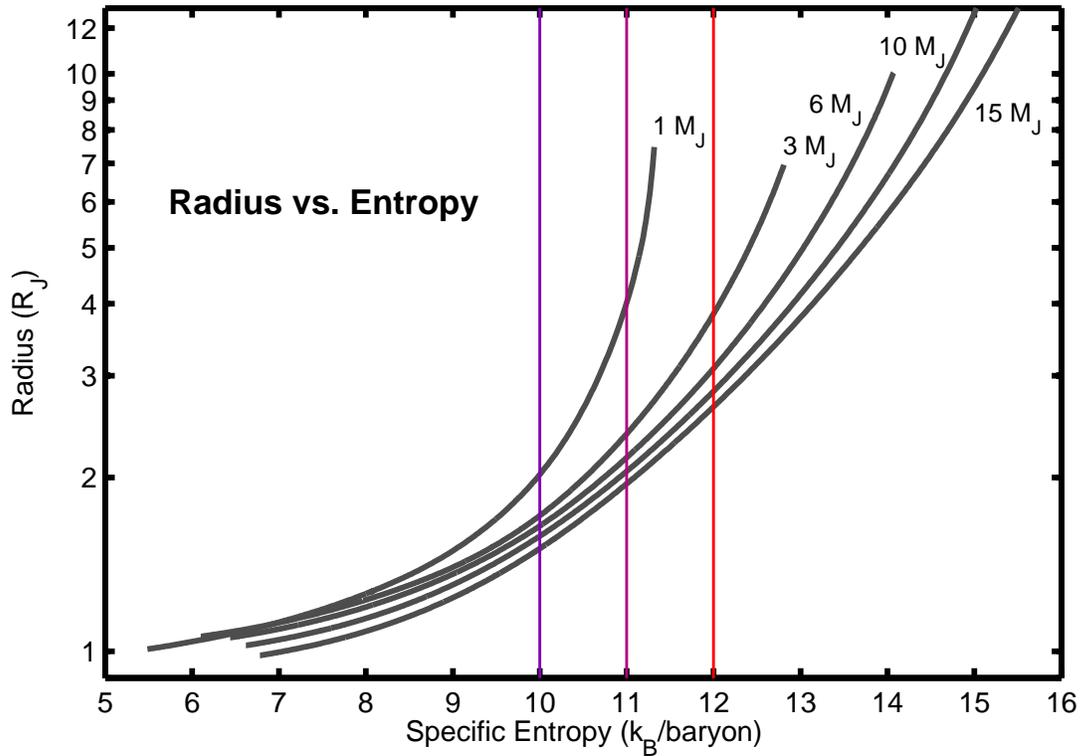}\\
\caption{Radius vs. Entropy for objects of mass 1, 3, 6, 10, and
  15~$M_J$.  Radius increases rapidly with increasing entropy, more so
  for lower mass objects.  Vertical lines (in colors corresponding to
  those in Fig.~\ref{fig:cooling}) mark where the entropy per baryon
  has values of 10, 11, and 12.}
\label{fig:SvsR}
\end{figure}

As an object loses entropy, its radius shrinks.  Figure~\ref{fig:SvsR}
demonstrates the dependence of radius on entropy for hydrogen-helium
mixtures of mass 1, 3, 6, 10, and 15~$M_J$.  The radius increases
rapidly with increasing entropy, and this rapid scaling is more
dramatic for lower mass objects.  The intersections of the
radius/entropy curves with vertical lines at $S=10$, 11, and 12 show
the radii that may be expected for (coreless) objects at the indicated
entropies.  At $S=12$, radii of objects from $3M_J$ to $15M_J$ range
from $\sim$4~$R_J$ to $\sim$3~$R_J$.  At $S=10$, radii range from
$\sim$2~$R_J$ (at $1M_J$) to $\sim$1.5~$R_J$ (at $15M_J$).  In an
Appendix, we present an analytic fit to the radius-entropy-mass
relation.

\section{``Hot Start''/``Cold Start'' Scenarios}
\label{sec:hotcold}

\subsection{Initial Hot and Cold Start Conditions}
\label{ssec:initial}
\citet{marley_et_al2007} published a set of ``Hot Start'' and ``Cold
Start'' models, where the initial entropy had a tuning-fork shape as a
function of mass.  In their framework, the cold-start models were
intended to represent the result of a core-accretion process, and the
hot-start models were intended to represent objects formed via disk
instability.

\begin{figure}[t!]
\plotone
{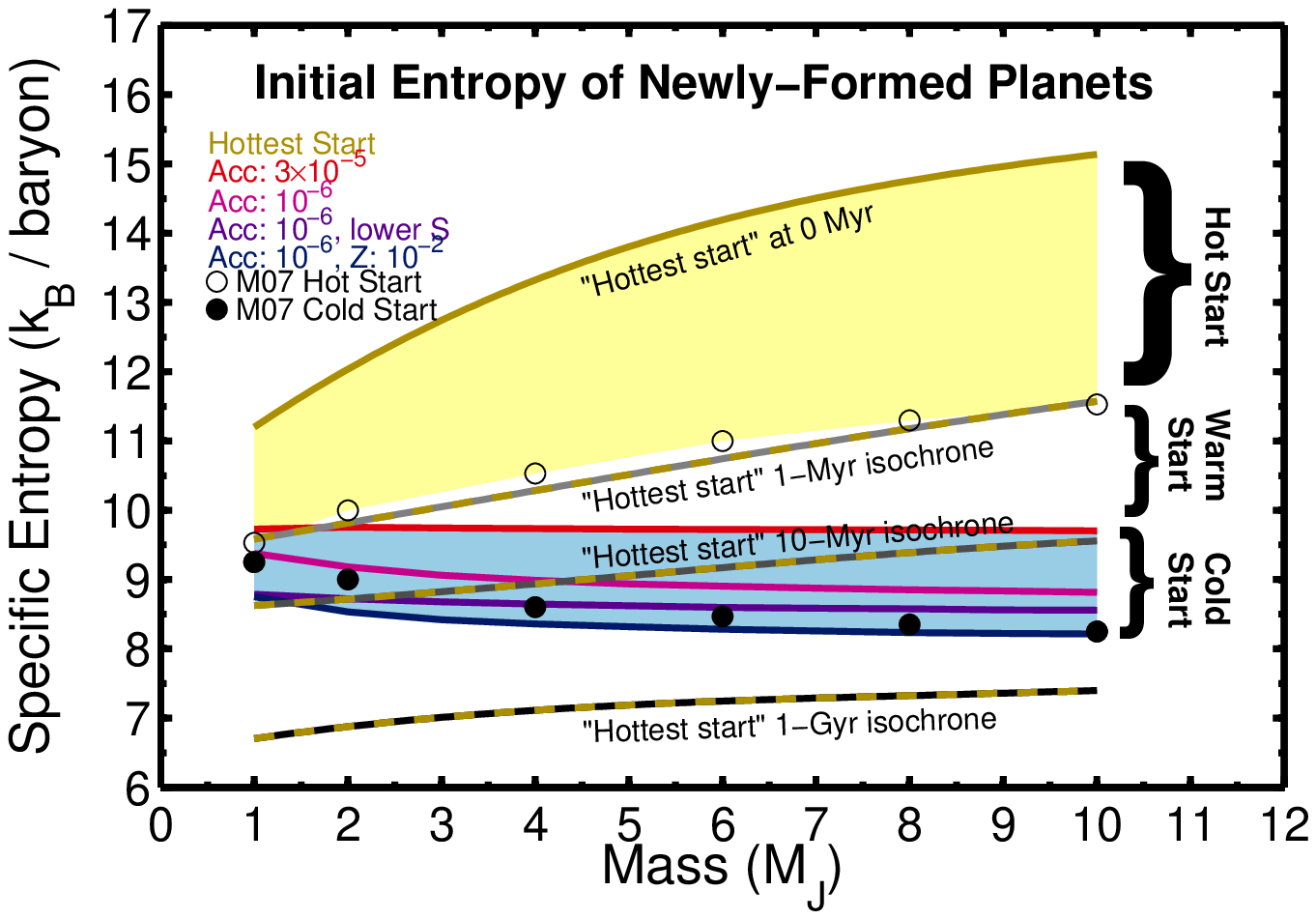}\\
\plotone
{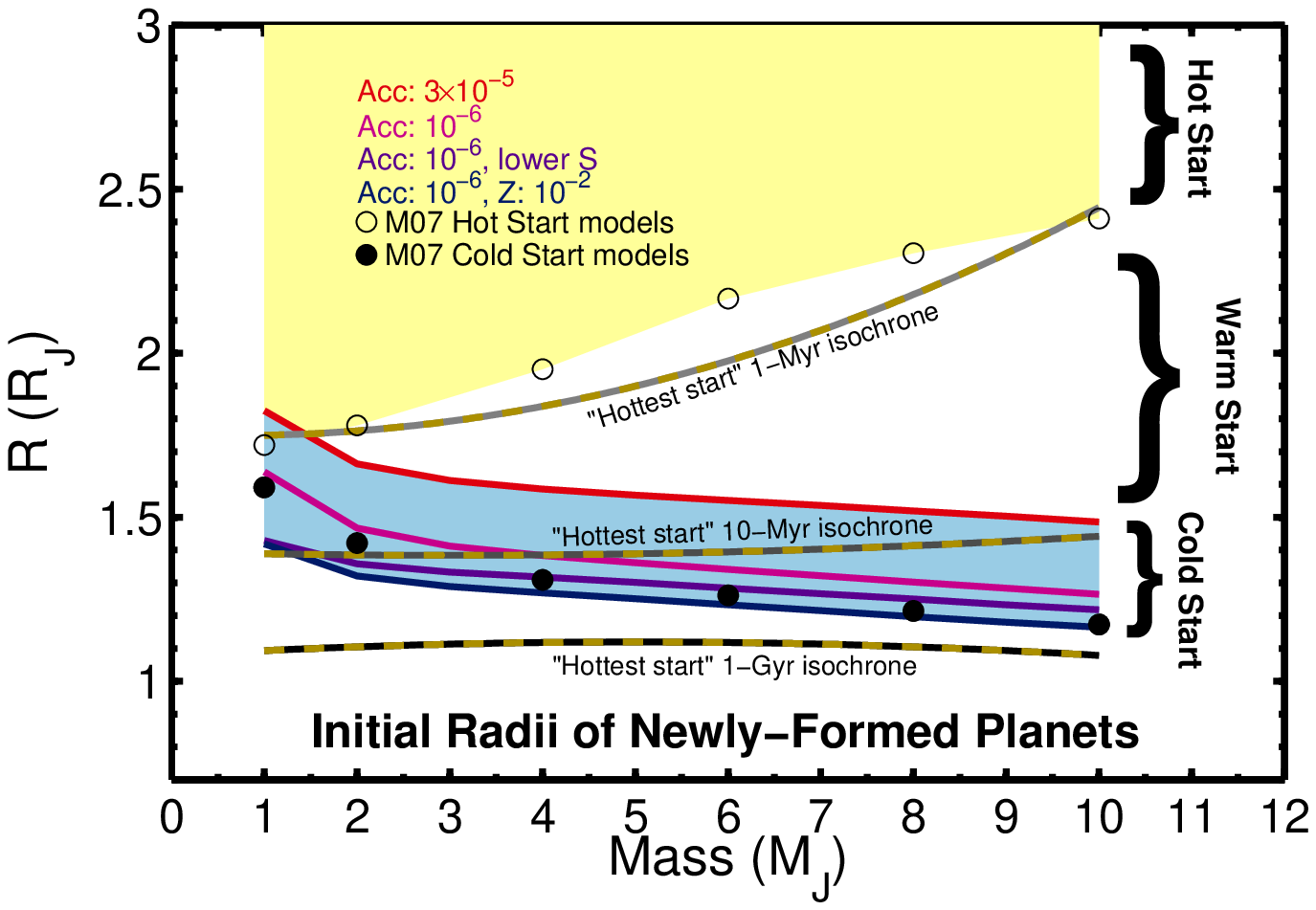}
\caption{Models of specific entropy ({\it top}) and radius ({\it
    bottom}) of newly-formed planets.  Values from M07 are shown as
    open (filled) circles for hot (cold) start models.  Four families
    of core accretion models are shown (accretion rates of $3\times
    10^{-5} M_J$/yr to $10^{-6} M_J$/yr, with various metallicities
    and initial entropies).  There is also a ``Hottest start'' family
    of models shown with extremely high initial entropies and radii.
    1-Myr, 10-Myr, and 1-Gyr isochrones of the ``Hottest start'' model
    are shown; the 1-Myr isochrone is fairly similar to the
    \citeauthor{marley_et_al2007} ``Hot Start'' models.
    Representative bands are shown for cold-start (blue), hot-start
    (yellow), and warm-start models.}
\label{fig:initSR1}
\end{figure}

We calculate a broad range of models, from extremely hot
(high-entropy, large radius) ones near the limit of
gravitationally-bound stability through several relatively cold models
that might represent possible outcomes of a core-accretion process.
These models are summarized in Figs.~\ref{fig:initSR1} and
\ref{fig:introduce}, with the top panel of each figure showing initial
entropy and the bottom panel showing initial radius as a function of
mass.  Figure~\ref{fig:initSR1} shows a total of seven families of
models.  Two are the ``hot'' and ``cold'' start models from M07, shown
with open and filled circles, respectively.  Another is the ``Hottest
Start'' family of models, which represents an approximate upper limit
to the initial entropy that might be expected of bound objects that
form via a disk-instability scenario.  The other four all represent
core-accretion scenarios, with various initial entropies, accretion
rates, and atmospheric properties.  All these models assume that
accretion flows onto an initial protoplanet of low mass, with a given
starting entropy.  Accretion proceeds with newly-accreted material
always at the specific entropy of the gas that has already accreted,
which is itself slowly cooling in accordance with the atmospheric
boundary condition.  Our accretion models have the following
properties:
\begin{itemize}
\item The ``Acc: $3\times 10^{-5}$'' models begin at an entropy of
  9.8, and accrete material at $3\times 10^{-5}M_J$/yr, similar to the
  assumed accretion rate in M07.  The \citet{burrows_et_al1997}
  atmospheric boundary condition is assumed.  Forming a 1-$M_J$ object
  takes $\sim$$3\times 10^4$~years.
\item The ``Acc: $10^{-6}$'' models are identical to the
  ``Acc. $3\times 10^{-5}$'' ones, except accretion proceeds at $1\times
  10^{-6}M_J$/yr, a factor of 30 slower.
\item The ``Acc: $10^{-6}$, lower S'' models are identical to the
  ``Acc. $10^{-6}$'' ones, except the initial entropy is 8.9 instead
  of 9.8.
\item The ``Acc: $10^{-6}$, Z: $10^{-2}$'' models are identical to the
  ``Acc. $10^{-6}$'' ones, except the atmosphere boundary condition is
  for $10^{-2}$-solar metallicity.  The reduced atmospheric opacity
  allows cooling to proceed more rapidly.
\end{itemize}

The entropy (top) and radius (bottom) panels of Fig.~\ref{fig:initSR1}
also show several isochrones, corresponding to the ``Hottest Start''
models at various stages of evolution, at 0, 1, and 10~Myr, and at
1~Gyr.  Interestingly, the ``Hottest Start'' 1-Myr isochrone is quite
similar to the \citeauthor{marley_et_al2007} hot start models (``M07
Hot Start'').

Note that the \citeauthor{marley_et_al2007} curves for hot and cold
models nearly converge at $1M_J$, and diverge progressively more with
increasing mass, leading to the tuning-fork shape seen in
Fig.~\ref{fig:initSR1}.  The lower track (``M07 Cold Start'')
decreases with time because 10-$M_J$ objects take longer to form than
1-$M_J$ objects, so the already-accreted gas has longer to cool.
Although the \citeauthor{marley_et_al2007} core-accretion models and
the ``Acc: $3\times 10^{-5}$'' models nominally have the same
accretion rate, the entropy of the \citeauthor{marley_et_al2007}
models clearly decreases with time, leading to the downward slope
towards higher-mass objects (which take longer in forming), whereas
the ``Acc: $3\times 10^{-5}$'' models' initial entropy curve is nearly
flat from 1-10~$M_J$; this difference might be attributable to the
opacities and boundary conditions used.

The upshot of Fig.~\ref{fig:initSR1} -- particularly of the radius
panel -- is that there is a very dramatic difference in the initial
properties of ``Hot Start'' and of ``Cold Start'' models.
Newly-formed cold-start models are predicted to have radii in the
$\sim$1.1-1.7-$R_J$ range, whereas newly-formed hot-start models might
have initial radii of more than 3~$R_J$.  Of course, since the
core-accretion process might take of order millions of years and the
disk-instability process is orders of magnitudes faster, the
newly-formed objects under one scenario might correspond to a stellar
age that is different by several million years from the newly-formed
objects under the other formation scenario \citep{baraffe_et_al2002,
fortney_et_al2005}.

It is also worthwhile to keep in mind that planet formation models
generally predict non-monotonic evolution of luminosity during the
formation process, with one or more large spikes in luminosity (e.g.,
\citealt{hubickyj_et_al2005, bromley+kenyon2011a}).  Our models begin
(i.e., have $t=0$) when the final mass assembly is complete.  Although
the very early evolution of still-forming planets might contain
interesting diagnostic observables, we do not explicitly calculate
this process, and our models, therefore, do not predict observables
for this (comparatively brief) portion of planetary evolution.

\subsection{Evolution of Radius, Temperature, and Entropy}
\label{ssec:evolve}

We track the evolution of a broad range of models, from ``Hot Start''
to ``Cold Start,'' through an age of 1~Gyr.
Figure~\ref{fig:introduce} portrays our set of initial conditions
(i.e., initial entropies and radii) for each mass.  The large red and
blue circles in Fig.~\ref{fig:introduce} indicate our ``Hot Start''
and ``Cold Start'' models, which roughly trace the hottest and coldest
starts shown in Fig.~\ref{fig:initSR1}.  The medium-sized yellow and
green dots represent our variants of the M07 ``Hot Start'' and ``Cold
Start'' models.

Figure~\ref{fig:M_evs} depicts the evolution of specific entropy (top
panel), radius (middle panel), and effective temperature (bottom
panel) for our Hot Start and Cold Start models.  Evolution curves are
shown for a range of masses (1, 2, 5, 10~$M_J$).  We also calculated
the evolution of objects modeled after the M07 hot and cold starts
(not shown).  The qualitative behavior of the evolutionary
trajectories of both our hot-/cold-start models and those of M07 is
the same.  At early times, objects that start hot are much larger and
have much higher effective temperatures than objects of equivalent
mass that start cold.  As ``Hot Start'' (red) and ``Cold Start''
(blue) objects of the same mass evolve, their entropies, radii, and
effective temperatures asymptotically converge.  The convergence seen
in Fig.~\ref{fig:M_evs} is more rapid at lower masses for two reasons:
(1) lower mass objects evolve more quickly, and (2) our initial
difference between hot and cold models is smaller for lower mass
objects (similar to M07, though our hot/cold models encompass a
somewhat larger range).  Within a few hundred million years, the
memory of the initial conditions has been lost even for 10-$M_J$
objects.  At early times, however, the difference is quite stark for
more massive objects (5, 10~$M_J$), where the initial radius and
effective temperature are each a factor of $\sim$2 larger for ``Hot
Start'' models than for ``Cold Start'' ones.  These large differences
in size (and, therefore, surface gravity) and effective temperature
lead to striking differences in photometry and spectra (as noted by
M07 and \citealt{fortney_et_al2008b}), and are discussed in
\S\ref{ssec:spec_ev}.

The difference between a ``Hot Start'' model and a ``Cold Start''
model may be thought of as a time difference ($\Delta t$) -- the
horizontal offset in Fig.~\ref{fig:M_evs} between the red and blue
curves of equal mass.  As objects reach ages comparable to or larger
than this $\Delta t$, it becomes progressively more difficult to
distinguish observationally between formation scenarios, on the
assumption that these are the only formation conditions possible, an
assumption that we revisit in \S\ref{sec:cont}.

\subsection{Spectral Evolution}
\label{ssec:spec_ev}

The differences in entropy, radius, and effective temperature between
``Hot Start'' and ``Cold Start'' models translate into differences in
spectra and broad-band magnitudes.  In order to compute spectra, we
assume various atmospheres.  In particular, we consider four
atmosphere types from \citet{burrows_et_al2011}: hybrid clouds at
solar metallicity (our ``fiducial'' atmospheres); hybrid clouds at
3$\times$ solar metallicity; cloud-free atmospheres at solar
metallicity; and cloud-free at 3$\times$ solar metallicity.  Our
planets are modeled as isolated objects, assumed to be in radiative
equilibrium, and their emergent spectra are calculated with the
line-by-line radiative transfer code {\tt COOLTLUSTY}
\citep{hubeny_et_al2003, burrows_et_al2006}.

Figure~\ref{fig:M_ev_spec} shows spectra as a function of mass and age
for our fiducial (i.e., hybrid clouds at solar metallicity) ``Hot
Start'' and ``Cold Start'' scenarios.  The red and blue (hot and cold)
curves in Fig.~\ref{fig:M_ev_spec} correspond to the large red and
blue circles, respectively, in Fig.~\ref{fig:introduce}.  Spectra are
shown (assuming the source is at 10~pc) for objects of mass 1, 2, 5,
and 10~$M_J$, and at ages of 1, 3, 10, 30, and 100 Myrs (spectra
become progressively dimmer at later ages).  At 1~$M_J$, hot- and
cold-start models are nearly overlapping even at the earliest times.
At higher masses, the strong difference between the fiducial ``Hot
Start'' and ``Cold Start'' scenarios becomes apparent.  At ages of a
few Myr, for instance, the 10-$M_J$ objects are several orders of
magnitudes brighter at short wavelengths and early times if they start
hot than if they start cold.  If the dichotomy between initial
conditions is as dramatic as in the fiducial models, then at early
times there should be prominent observational differences (cf. M07;
\citealt{fortney_et_al2008b}).

\begin{figure}[t!]
\plottwob
{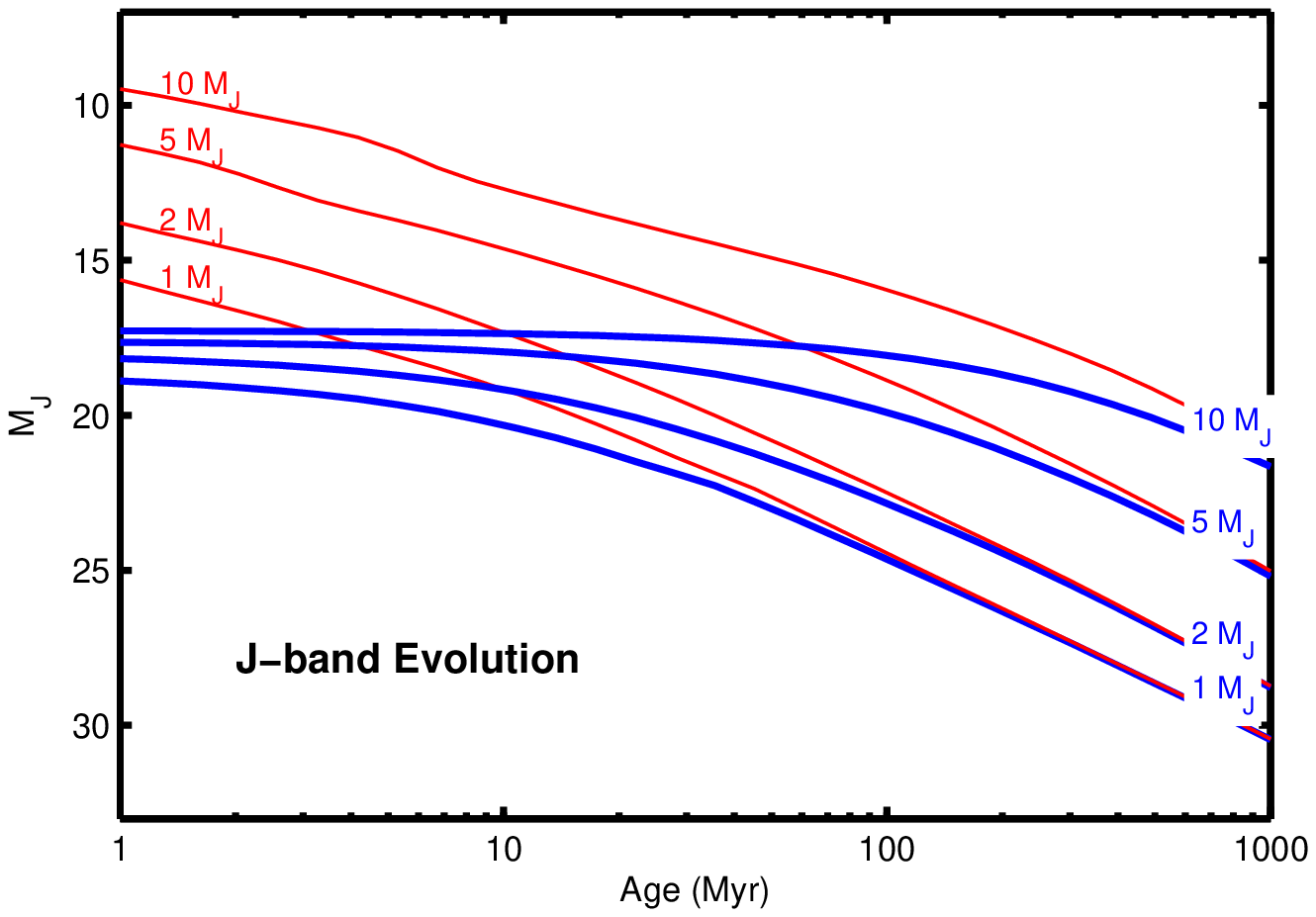}
{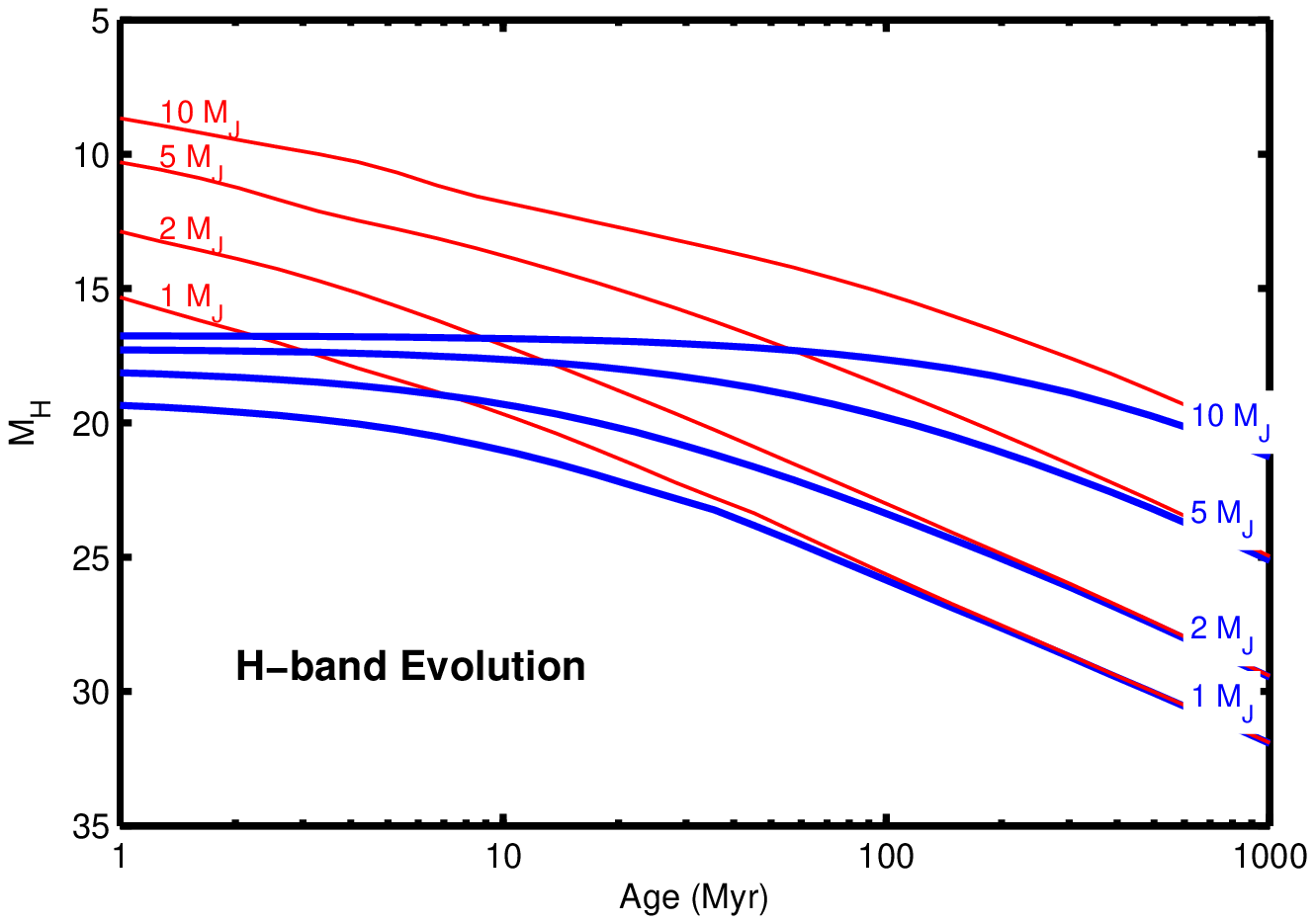}\\
\plottwob
{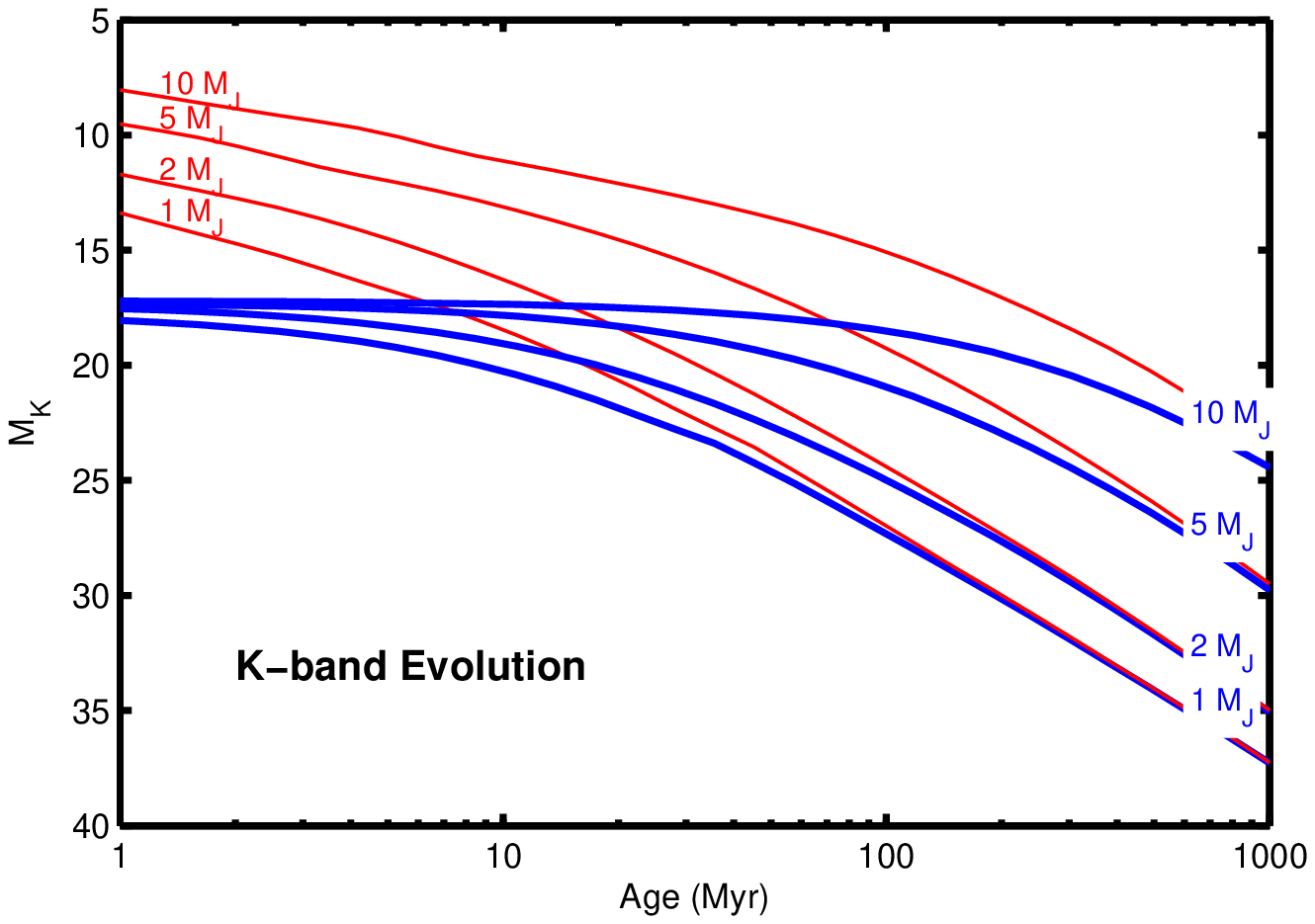}
{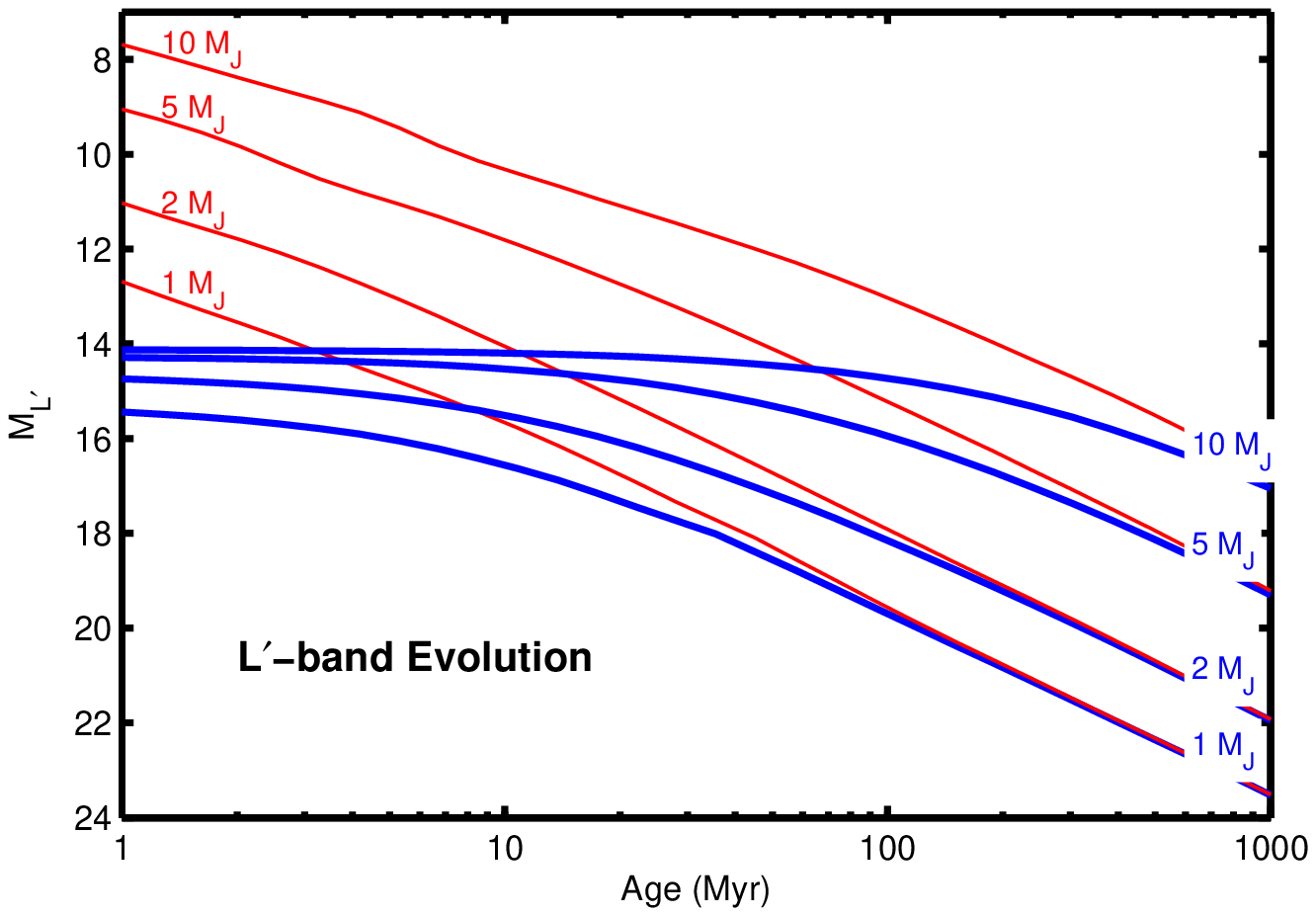}\\
\plottwob
{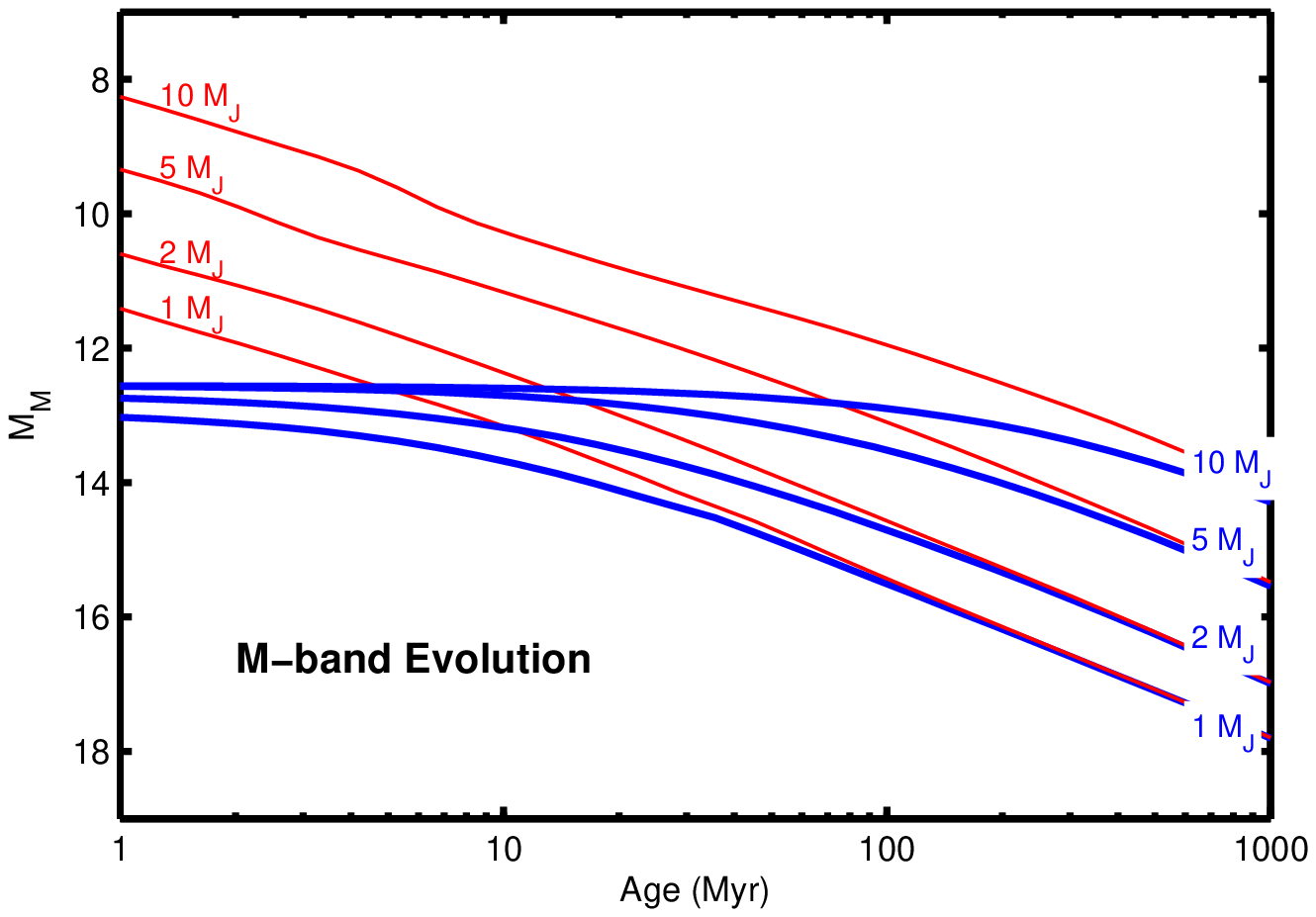}
{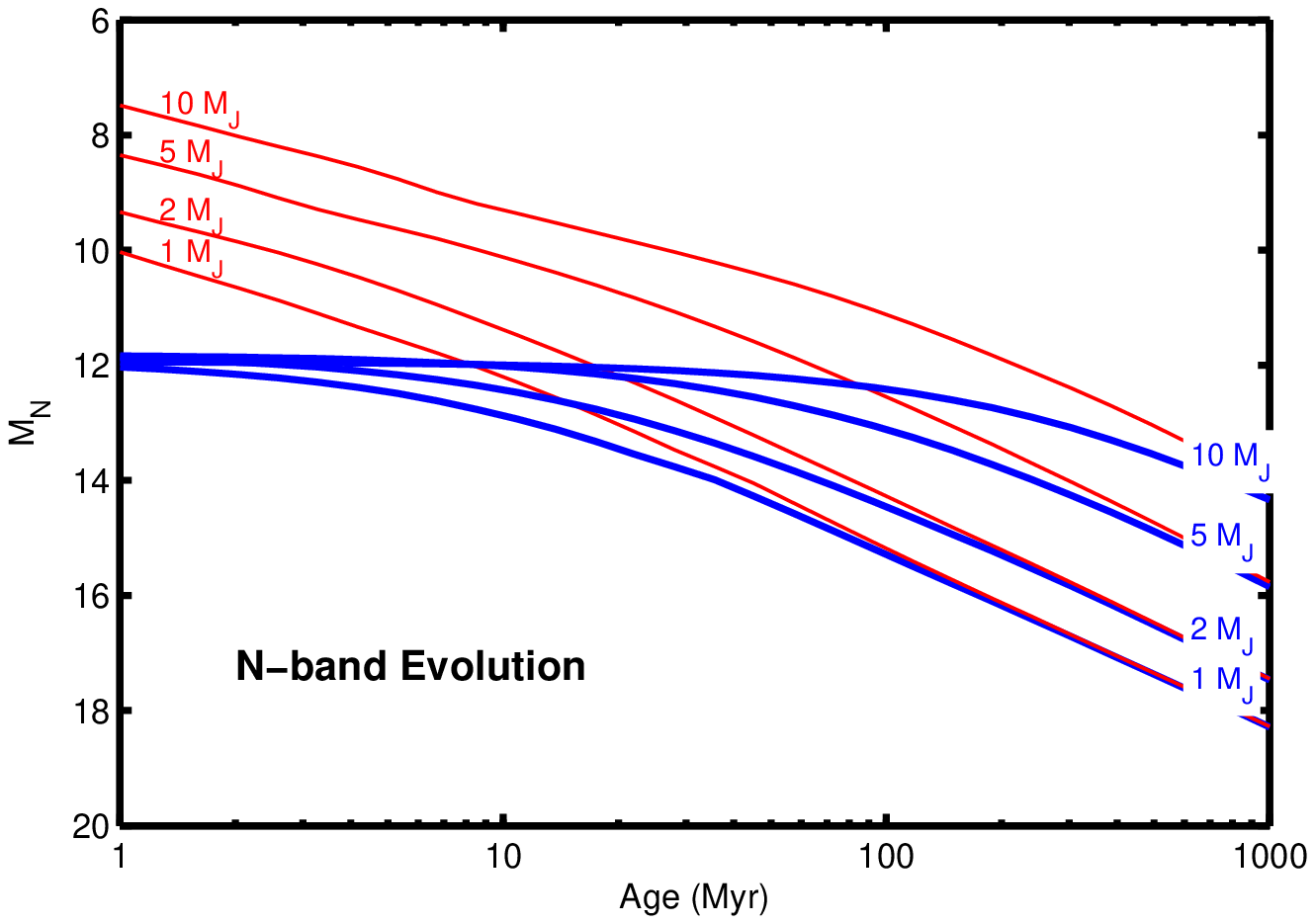}
\caption{Evolution of absolute magnitudes.  ``Hot Start'' (red) and
  ``Cold Start'' (blue) models are shown for 1, 2, 5, and $10M_J$.
  Evolutionary trajectories are shown for $J$ (\emph{top left}), $H$
  (\emph{top right}), $K$ (\emph{middle left}), $L'$ (\emph{middle
  right}), $M$ (\emph{bottom left}), and $N$ (\emph{bottom right})
  bands.  Hot and cold-start models begin at very different
  brightnesses, but as objects age the differences quickly approach
  zero.}
\label{fig:HC_ev_bands}
\end{figure}

These differences in spectra imply differences in broad-band
photometry.  The evolution of $J$, $H$, $K$, $L'$, $M$, and $N$-band
absolute magnitudes (for hybrid-clouds, solar-metallicity models) is
shown in Fig.~\ref{fig:HC_ev_bands}.  At young ages, the magnitude
differences range from small ($\sim$1) at $1M_J$ to more than 5 in
some bands (e.g., $J$, $H$, $K$) at $10M_J$.  As objects of a given
mass evolve, their ``hot-start'' and ``cold-start'' brightnesses
converge (in each band).  The data in Fig.~\ref{fig:HC_ev_bands} are
available in Table~\ref{ta:models_hy1s}, and similar tables for the
other three atmosphere types are available in electronic form
online.\footnote{The model spectra presented in this paper are
available in electronic form (at spectral resolution $\approx$204)
at\\
http://www.sns.ias.edu/$\sim$dave/\\
and at
http://www.astro.princeton.edu/$\sim$burrows/.}

\section{A Continuum of ``Warm-Start'' Planets}
\label{sec:cont}

Planet formation theory informs our understanding of the inital
conditions to be expected for core-accretion and for disk-instability
scenarios.  However, significant uncertainties remain.  Consider again
the range of hot-start and cold-start conditions represented in both
the entropy and radius panels in Fig.~\ref{fig:initSR1}.  A crucial
assumption leading to the negative slope that is characteristic both
of the \citeauthor{marley_et_al2007} core-accretion models and of our
own is that the accretion process adds no heat, meaning that each
parcel accretes only at the specific entropy of the gas that is
already present.  However, it is not clear that this assumption is
warranted.  Allowing infalling parcels to retain some of their nebular
entropy would allow higher-mass core-accretion models to have {\it
higher} initial entropy than lower-mass models, thereby potentially
filling in the ``Warm Start'' regions of the two panels of
Fig.~\ref{fig:initSR1}.

Since we do not know {\it a priori} the initial entropies (and radii
and temperatures) of objects that form via core accretion or via disk
instability, it is useful to consider a broad range of possibilities.
Indeed, if details of the accretion process that might differ from one
planet to another govern the fraction of nebular entropy that is
incorporated into the planet, then considering a broad range of
possible initial entropies might be essential to reflect not just our
current ignorance, but also the range of actual initial conditions
from which different newly-formed planets in our galaxy begin their
evolution.

Figure~\ref{fig:introduce} summarizes our framework for treating a
spectrum of initial conditions.  At each of 1, 2, 5, and 10~$M_J$, we
show a range of initial entropies (with different models spaced by
0.25 units): 8.75 to 10.5 at $1M_J$, 8.5 to 11.25 at $2M_J$, 8.25 to
12.25 at $5M_J$, and 8 to 13 at $10M_J$.  As discussed previously
(\S\ref{ssec:initial}), the large red/blue circles indicate our
hot/cold models and the medium-sized yellow/green circles indicate the
M07 hot and cold models.  These particular choices for hot and cold
models merely represent four values (at each mass) in a continuum of
possibilities.  The range of initial entropies in
Fig.~\ref{fig:introduce} corresponds to a range of initial radii:
$\sim$1.4-3.3~$R_J$ at $1M_J$, $\sim$1.3-3.0~$R_J$ at $2M_J$,
$\sim$1.2-3.6~$R_J$ at $5M_J$, and $\sim$1.1-4.2~$R_J$ at $10M_J$.

\subsection{A Dramatic Spread in Brightnesses and Spectra}
\label{ssec:spread}
\begin{figure}[t!]
\plotone
{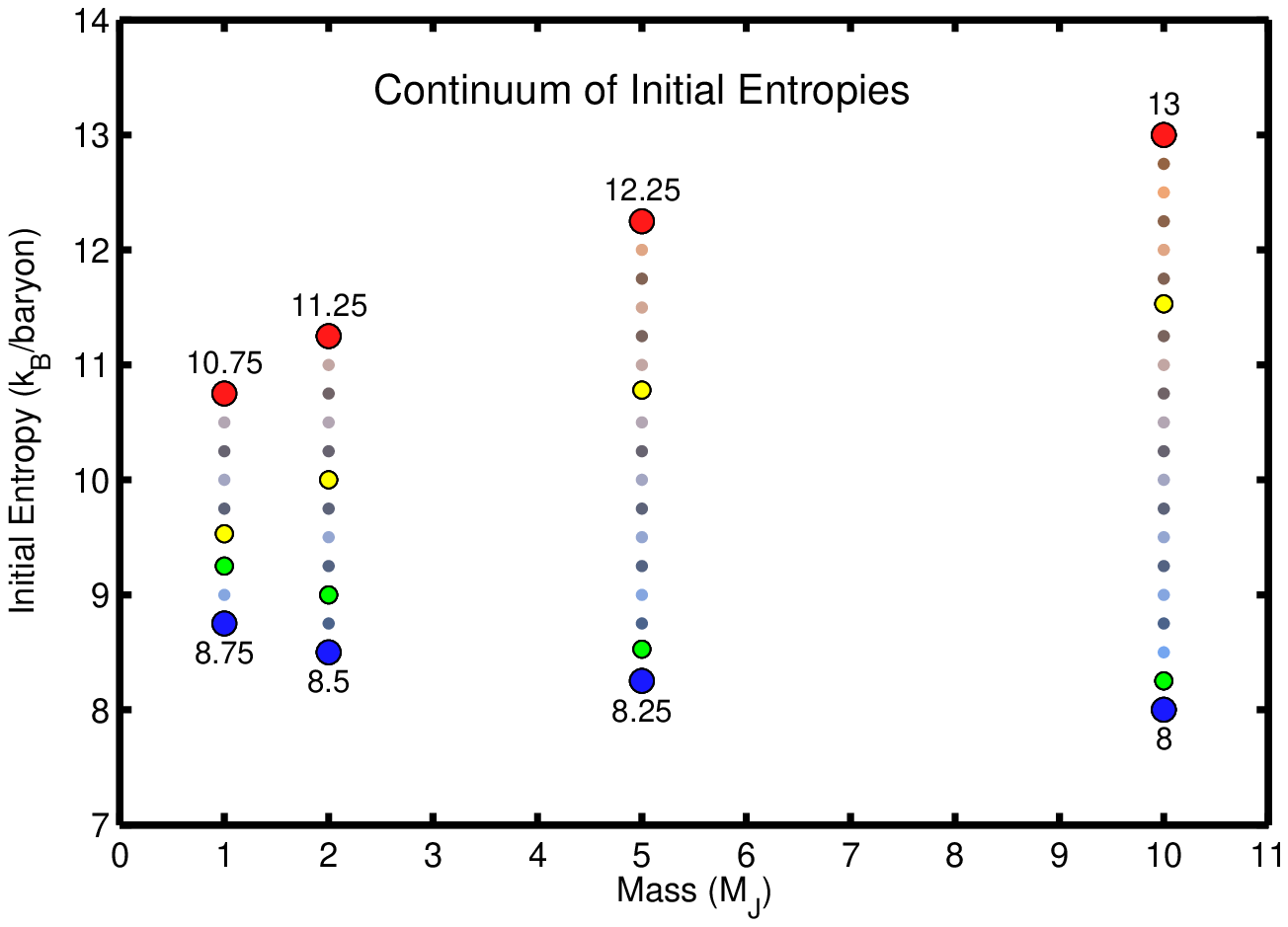}\\
\plotone
{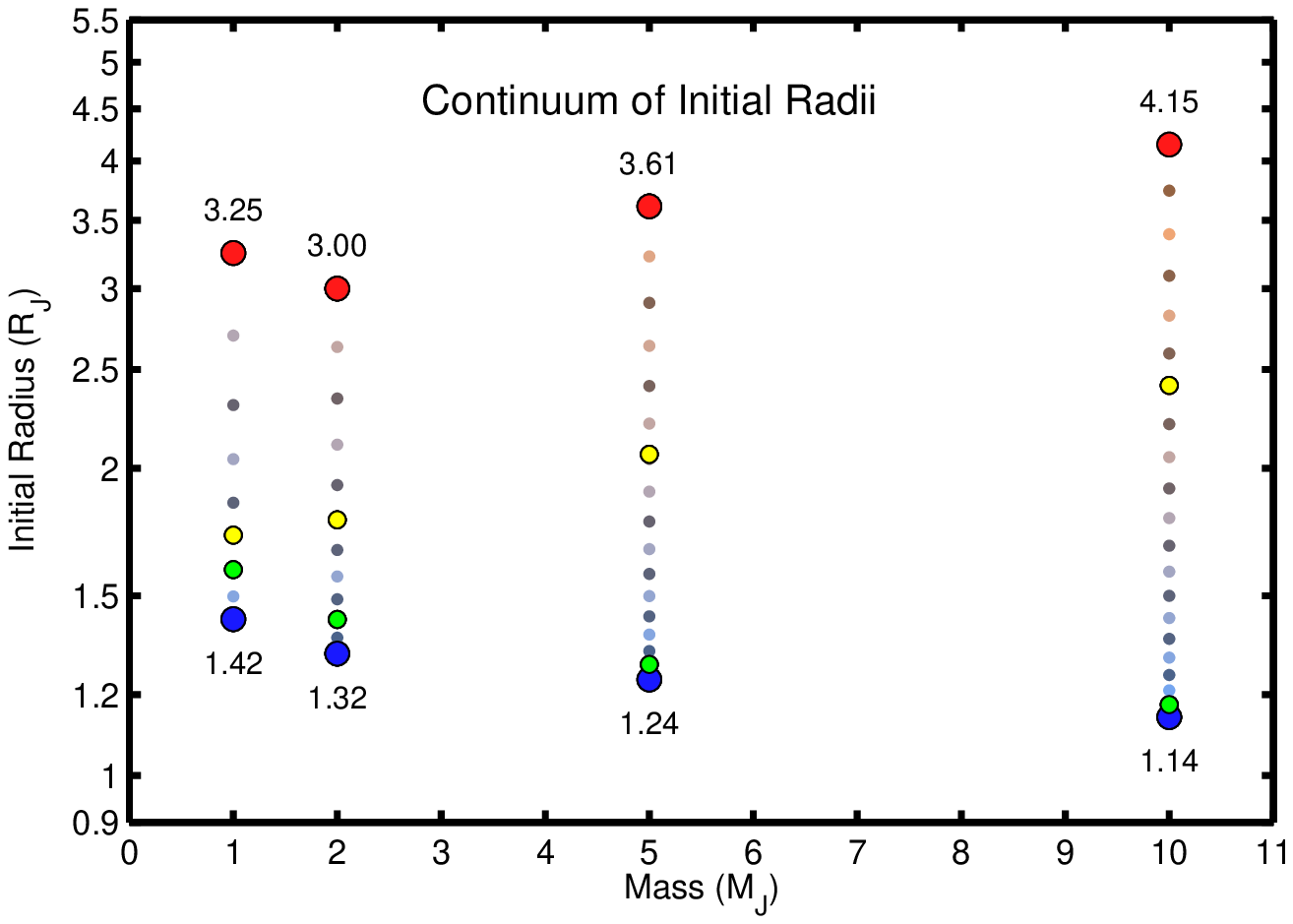}
\caption{Range of initial entropies (\emph{top}) and corresponding
  initial radii (\emph{bottom}) that we employ for evolution
  calculations are displayed .  For each of 1, 2, 5, and 10~$M_J$, a
  broad set of initial entropy values (in steps of 0.25) are listed.
  Very large red (blue) circles indicate our ``Hot Start'' (``Cold
  Start'') scenarios.  Medium-sized yellow (green) circles indicate
  our versions of the M07 ``hot-start'' (``cold-start'') models.  For
  more massive objects, the highest and the lowest plausible entropies
  both become more extreme.  The radii corresponding to the entropies
  of the top panel are shown in the bottom panel.}
\label{fig:introduce}
\end{figure}

\begin{figure}[t!]
\plottwo
{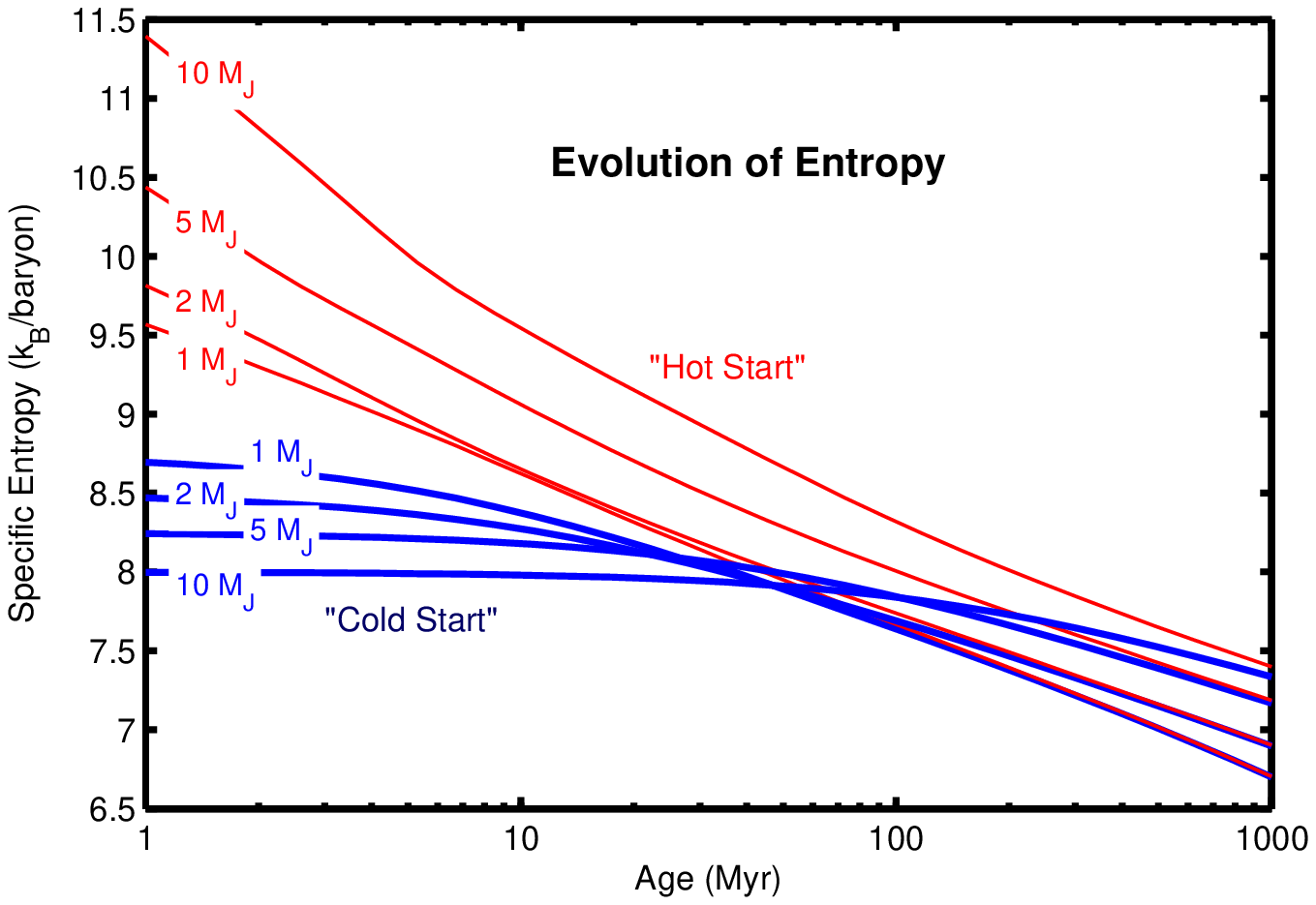}
{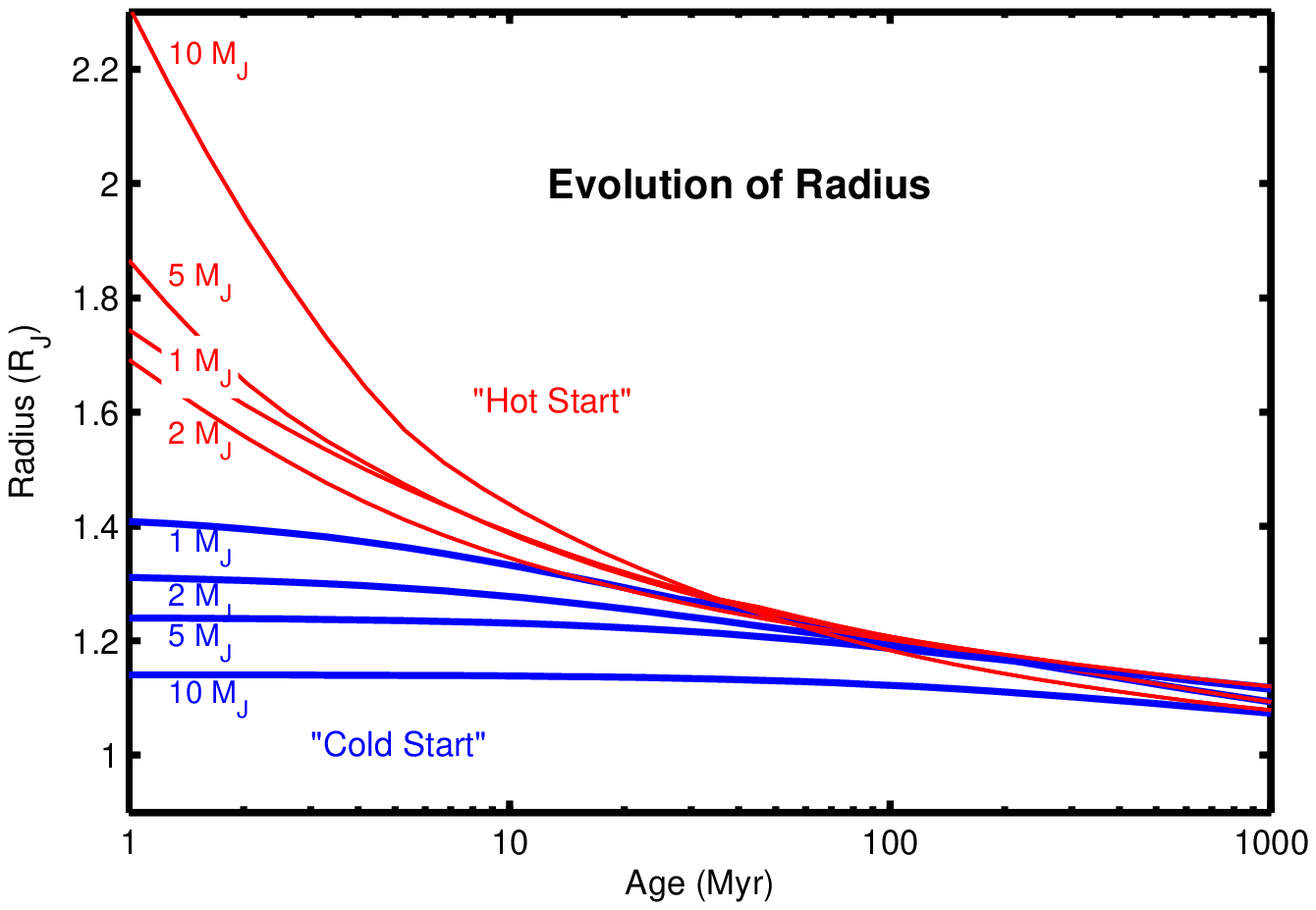} \\
\plotoneShrinkVSmall
{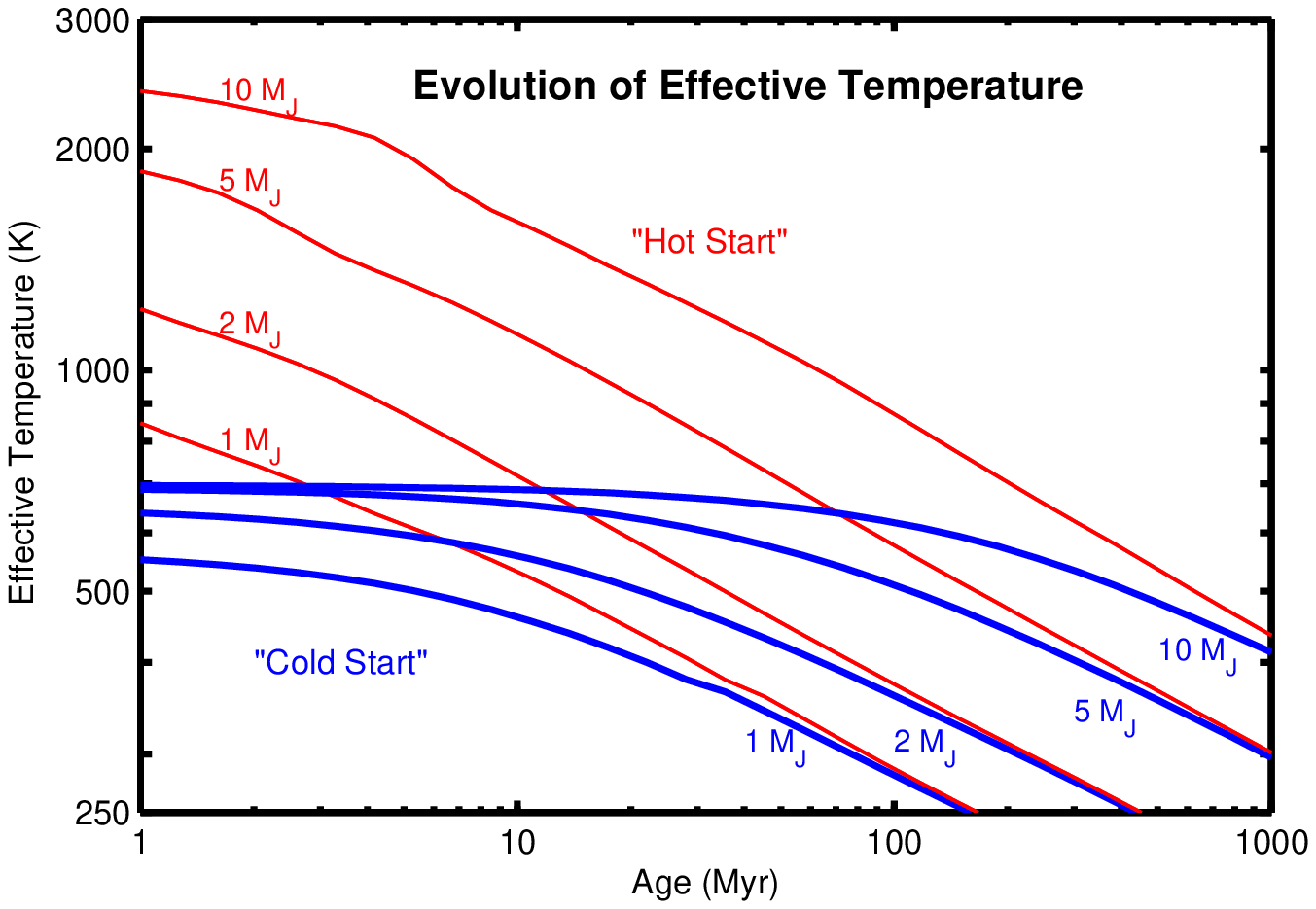}
\caption{Evolution of specific entropy ({\it top}), radius ({\it
    middle}), and effective temperature ({\it bottom}), for a range of
    masses (1, 2, 5, 10~$M_J$), for a particular set of ``Hot Start''
    (red) and ``Cold Start'' (blue) initial conditions, corresponding
    to the large red and blue dots in Fig.~\ref{fig:introduce},
    respectively.  The differences in radius and temperature between
    the ``Hot Start'' and the ``Cold Start'' objects is dramatic at
    early times, particularly for more massive objects.  As ``Hot
    Start'' and ``Cold Start'' objects of the same mass evolve, their
    entropies, radii, and effective temperatures asymptotically
    converge, such that within a few hundred million years the memory
    of the initial conditions has been lost even for 10-$M_J$ objects.
    Lower mass objects lose memory of their initial conditions within
    a few to a few tens of million years.}
\label{fig:M_evs}
\end{figure}

\begin{figure}[t!]
\plottwob
{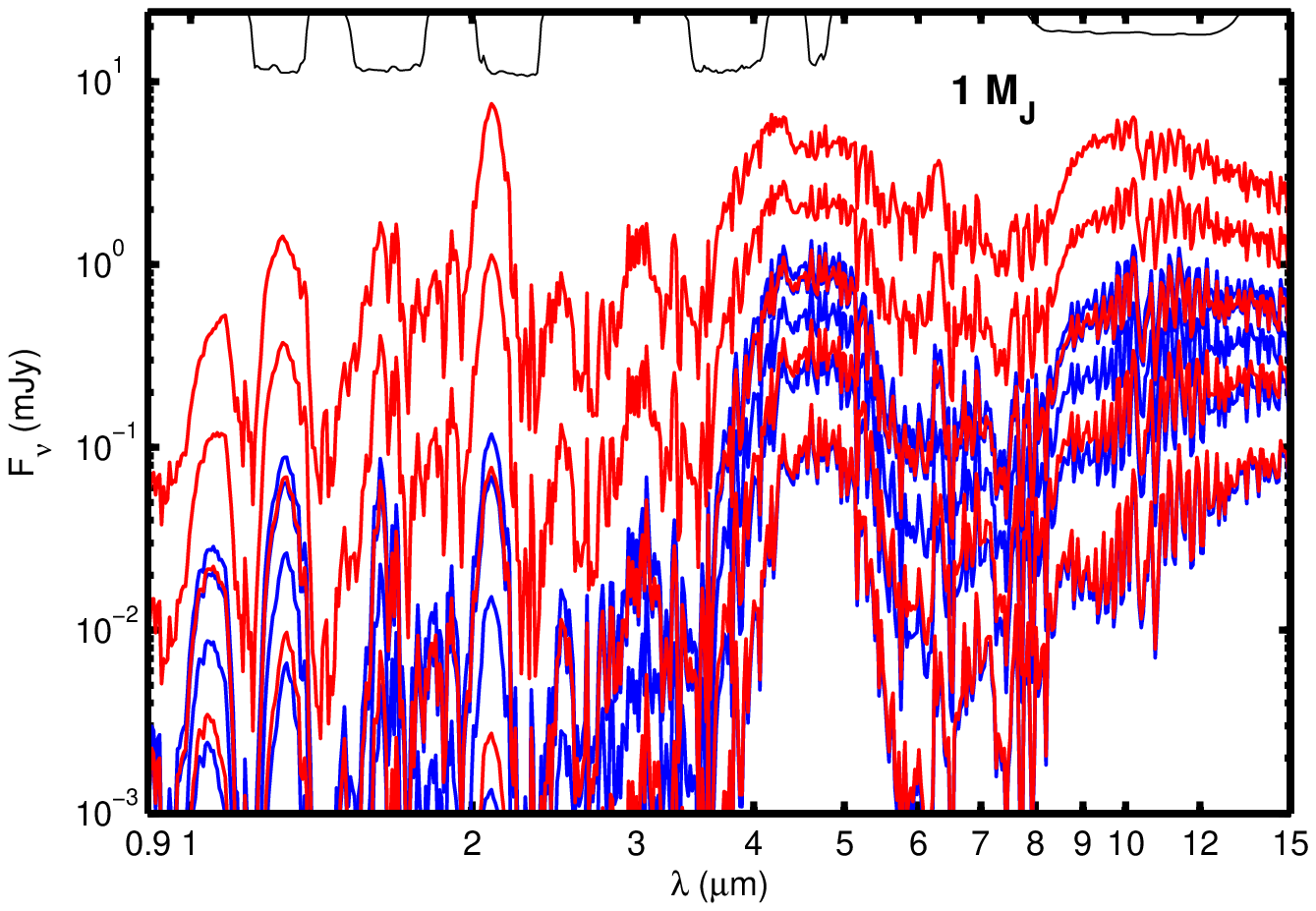}
{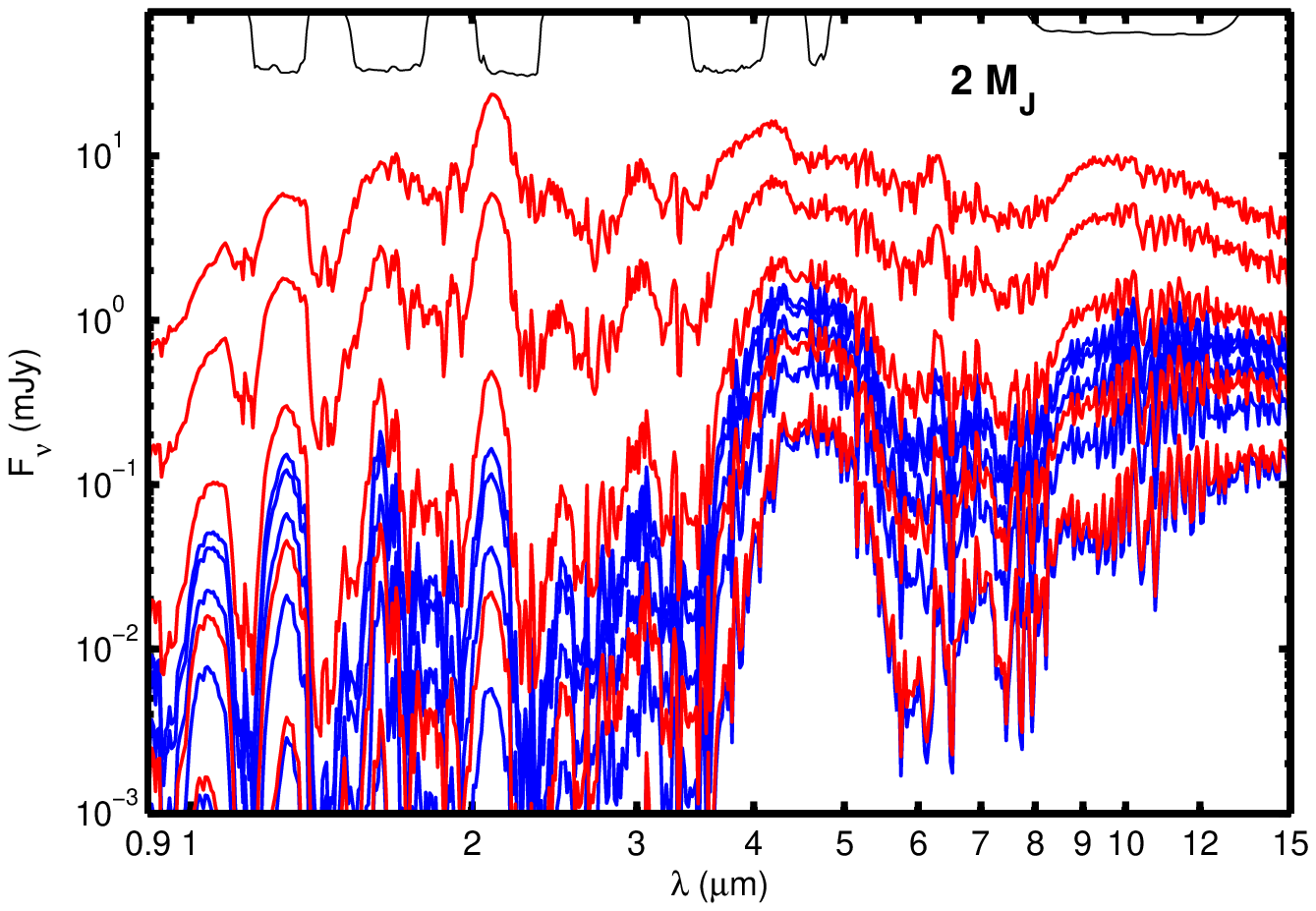}\\
\plottwob
{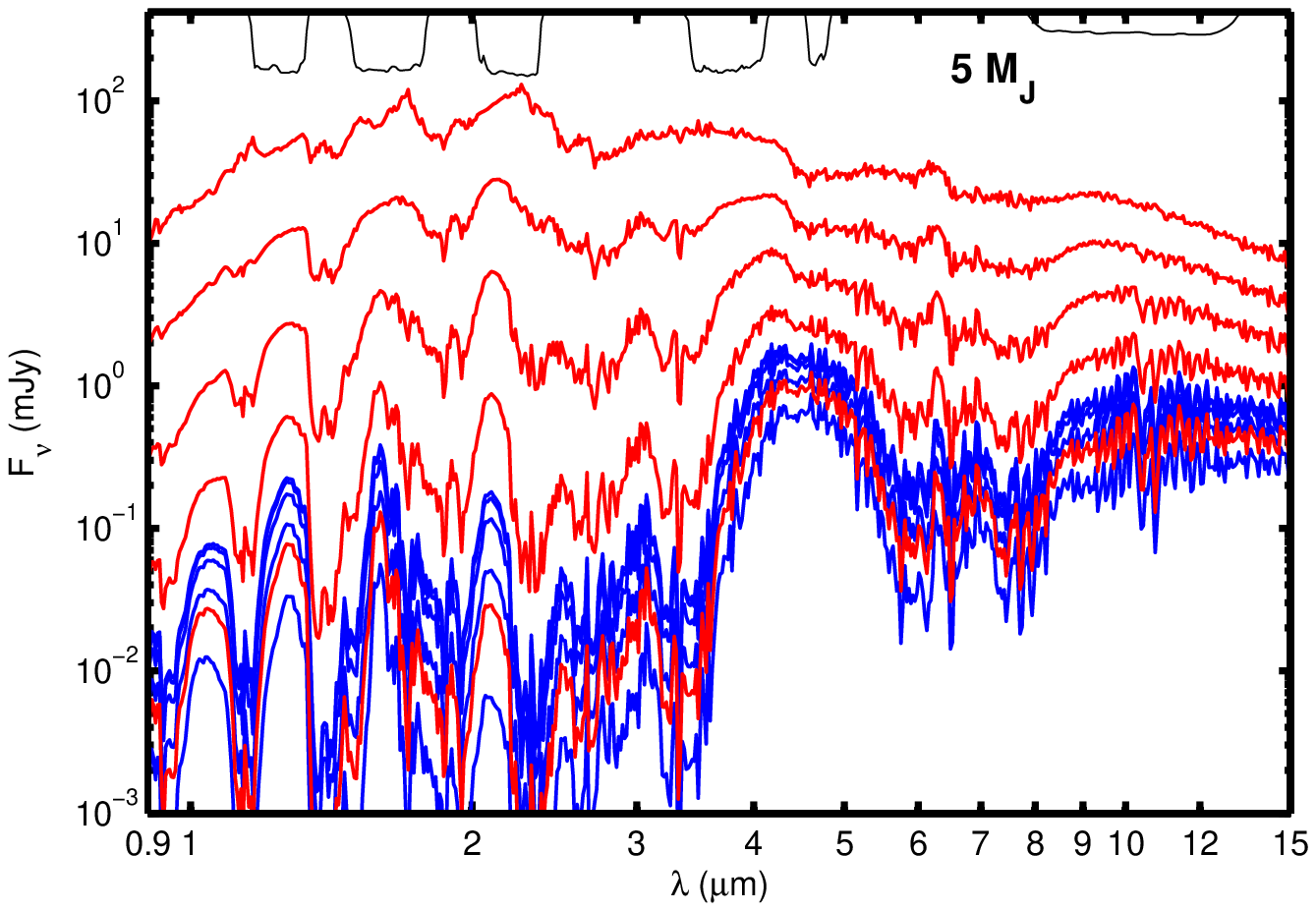}
{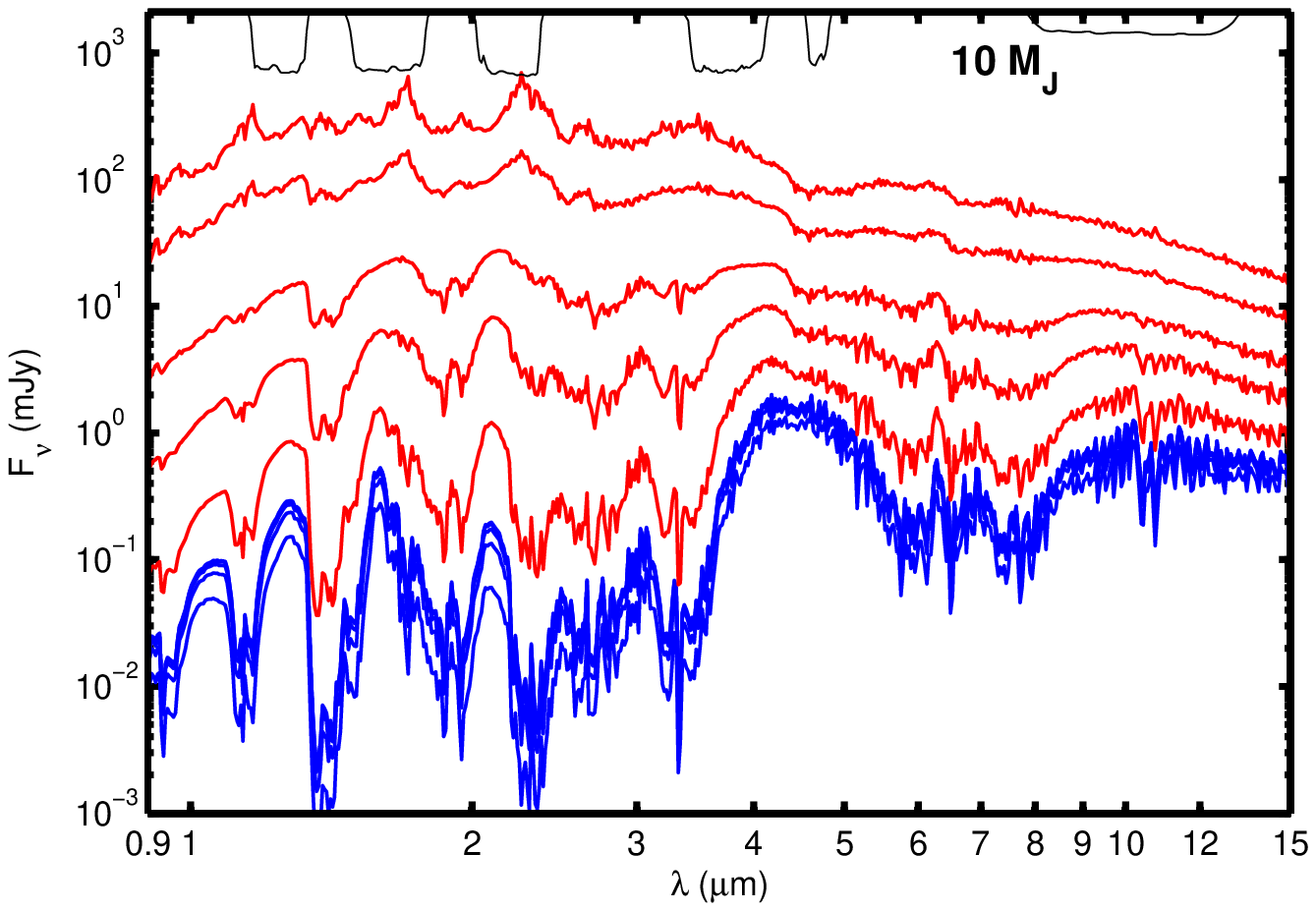}
\caption{Evolution of spectra for ``Hot Start'' (red) and ``Cold
  Start'' (blue) models for objects with hybrid-cloudy atmospheres at
  solar metallicity \citep{burrows_et_al2011}.  The initial conditions
  correspond to the large red and blue dots in
  Fig.~\ref{fig:introduce}.  Spectra are shown (assuming the source is
  at 10~pc) for objects of mass $1M_J$ ({\it top left}), $2M_J$ ({\it
  top right}), $5M_J$ ({\it bottom left}), and $10M_J$ ({\it bottom
  right}).  In each panel, 5 spectra are shown for each initial
  condition, indicating isochrones of 1, 3, 10, 30, and 100 Myrs.  At
  the top of each panel, thin black lines indicate the transmission
  functions for $J$, $H$, $K$, $L'$, $M$, and $N$ bands.}
\label{fig:M_ev_spec}
\end{figure}

\begin{figure}[t!]
\plottwob
{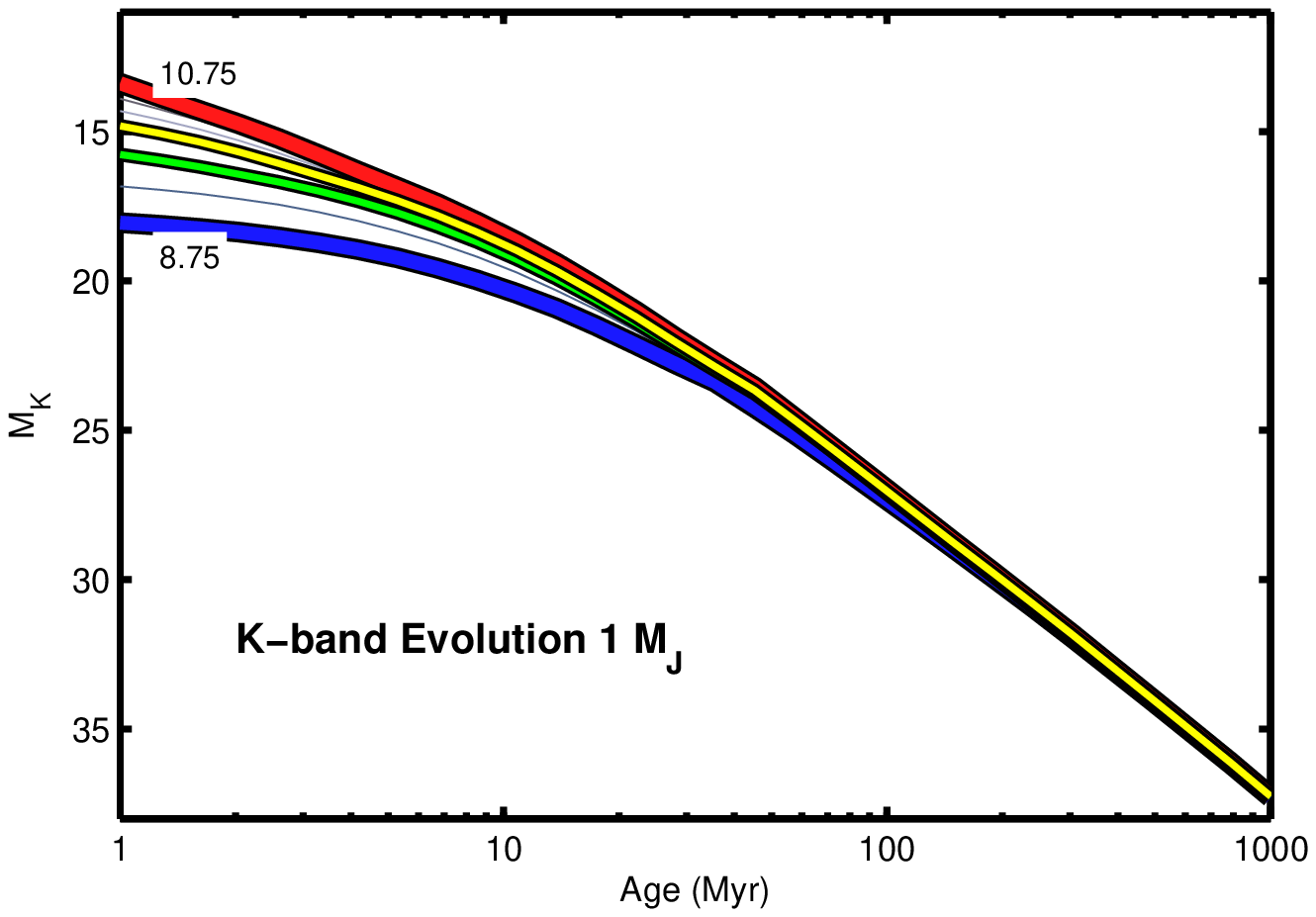}
{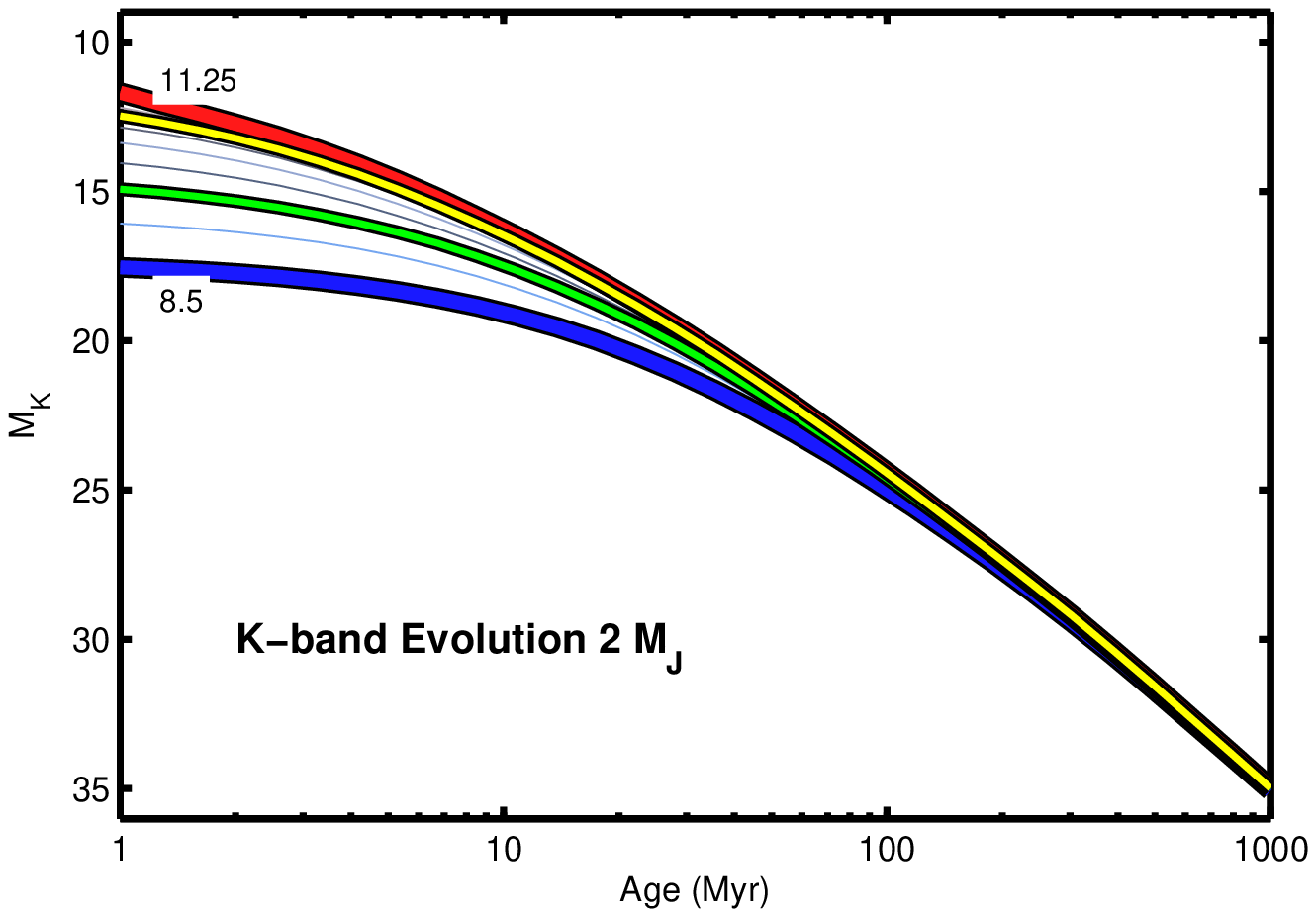}\\
\plottwob
{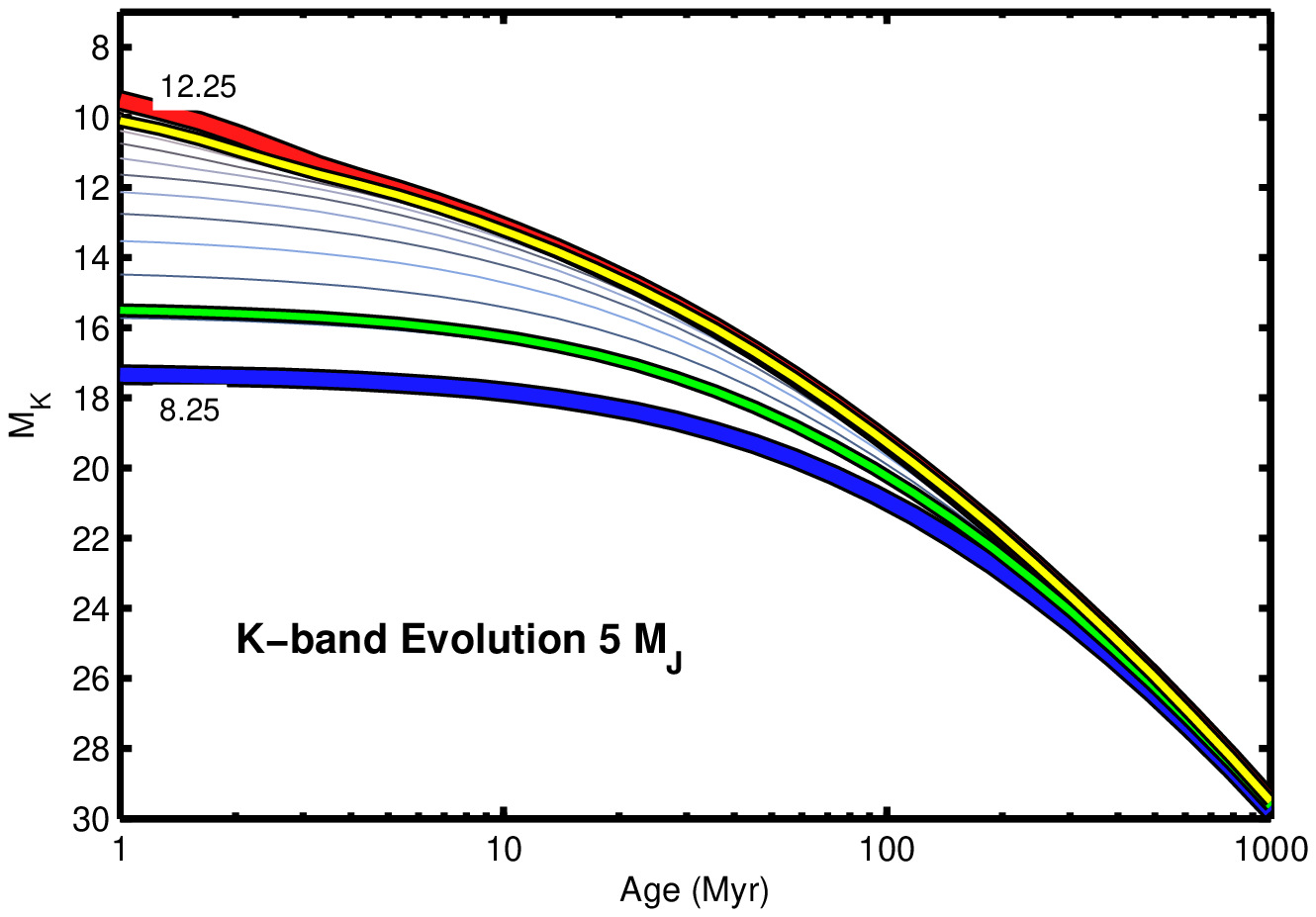}
{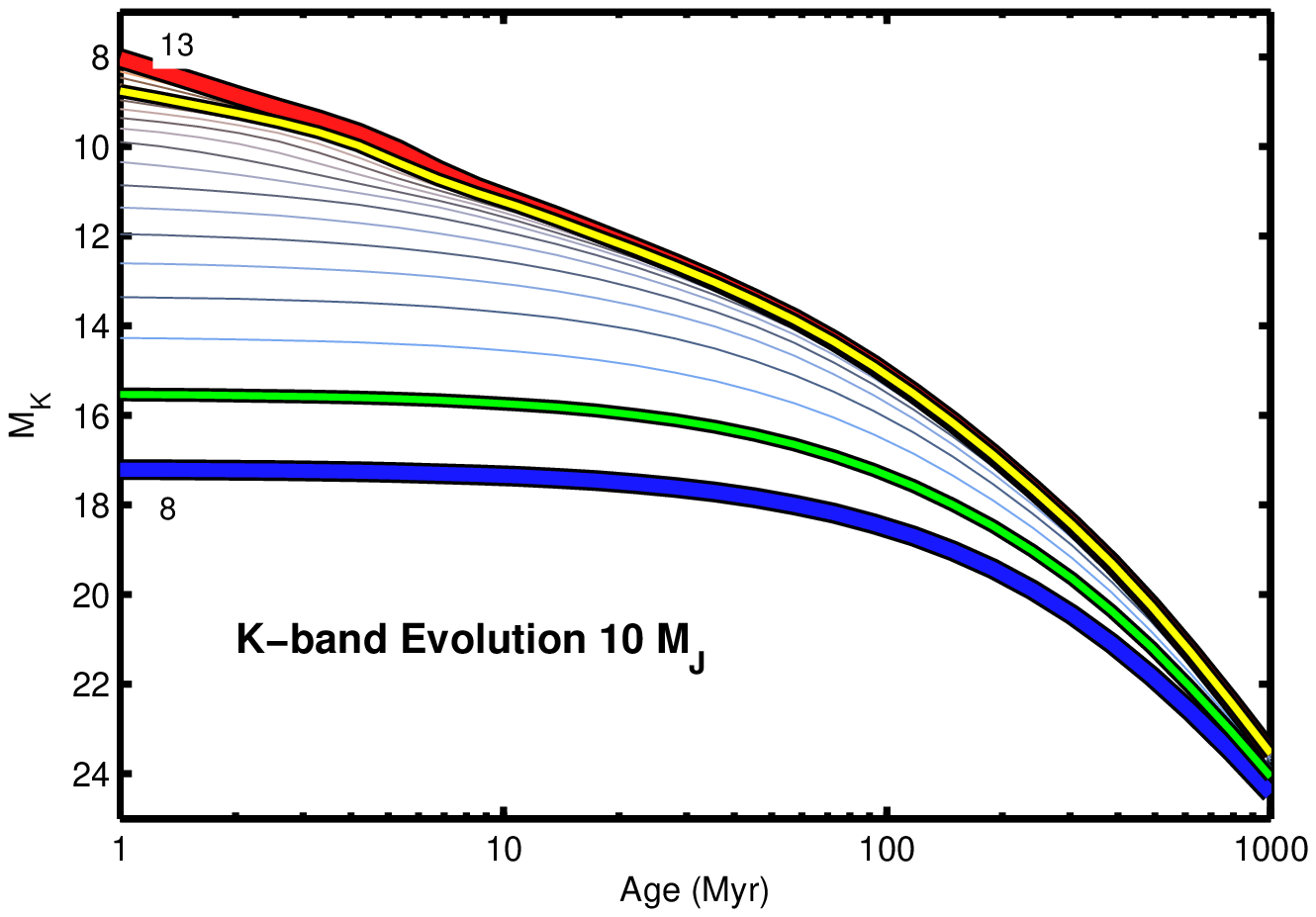}
\caption{Evolution of absolute $K$-band magnitude for different masses
  and different initial entropies.  Evolutionary tracks are shown for
  each of the models represented by dots in Fig.~\ref{fig:introduce}
  (with corresponding colors).  Thick red and blue curves represent
  our hottest- and coldest-start models, corresponding to the large
  dots of the same color in Fig.~\ref{fig:introduce}; medium-thick
  yellow and green curves similarly represent our version of the M07
  ``hot-start'' and ``cold-start'' models.  At each of 1~$M_J$
  (\emph{top left}), 2~$M_J$ (\emph{top right}), 5~$M_J$ (\emph{bottom
  left}), and 10~$M_J$ (\emph{bottom right}), planets begin bright in
  the $K$-band and cool with time.  Those that begin with higher
  initial entropy (and larger radius and higher effective temperature)
  are significantly brighter at early times than those that begin
  cooler.  As objects evolve, the memory of the initial conditions
  fades, but at a given mass and a given age there are a range of
  possible $K$-band magnitudes, depending on the initial entropy.}
\label{fig:K_ev_bands}
\end{figure}

At a given age, models that start with different initial entropies can
have very different brightnesses.  For example,
Fig.~\ref{fig:K_ev_bands} shows the evolution of the $K$-band absolute
magnitudes of the models denoted by dots in Fig.~\ref{fig:introduce}.
At each of four different masses (1, 2, 5, and 10~$M_J$), the
evolution of $K$-band brightnesses is shown for a range of initial
entropies.  As seen previously in Fig.~\ref{fig:HC_ev_bands},
the memory of the initial conditions fades as objects evolve.  At a
given mass and age there is a range of possible $K$-band absolute
magnitudes, depending on the initial entropy.  Cold-start and
warm-start planets of higher-mass ($\gtrsim$5~$M_J$) maintain
relatively constant $K$-band absolute magnitude during their early
evolution.  At 5~$M_J$, the near-constant brightness phase lasts for
$\lesssim$10~Myr, and at 10~$M_J$, this phase lasts for several tens
of Myr.  Model planets that begin with relatively high entropy are
significantly brighter in the early stages and evolve much faster, not
experiencing the early-evolution plateaus in brightness that
cooler-start models do.

\begin{figure}[t!]
\plottwob
{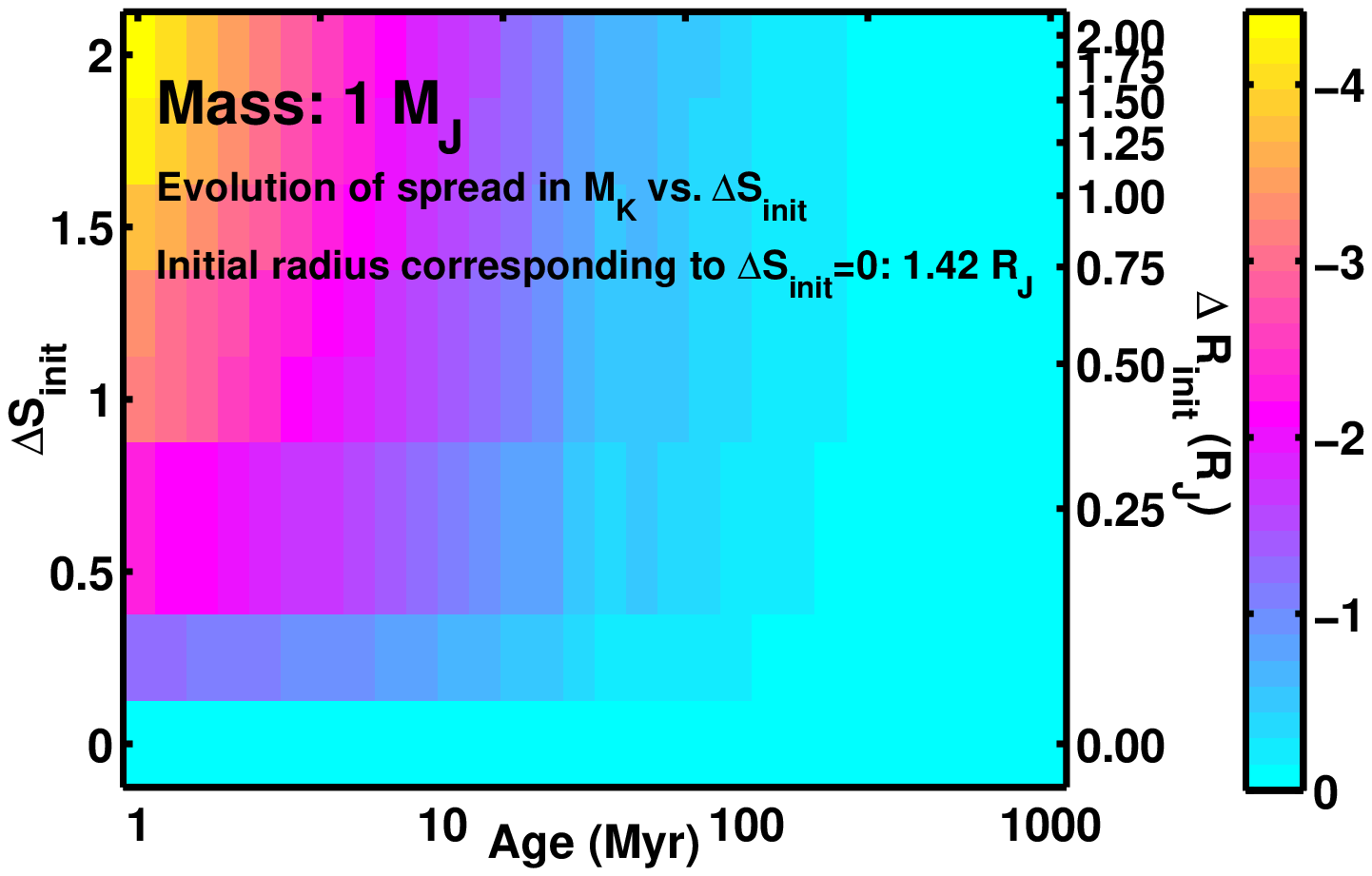}
{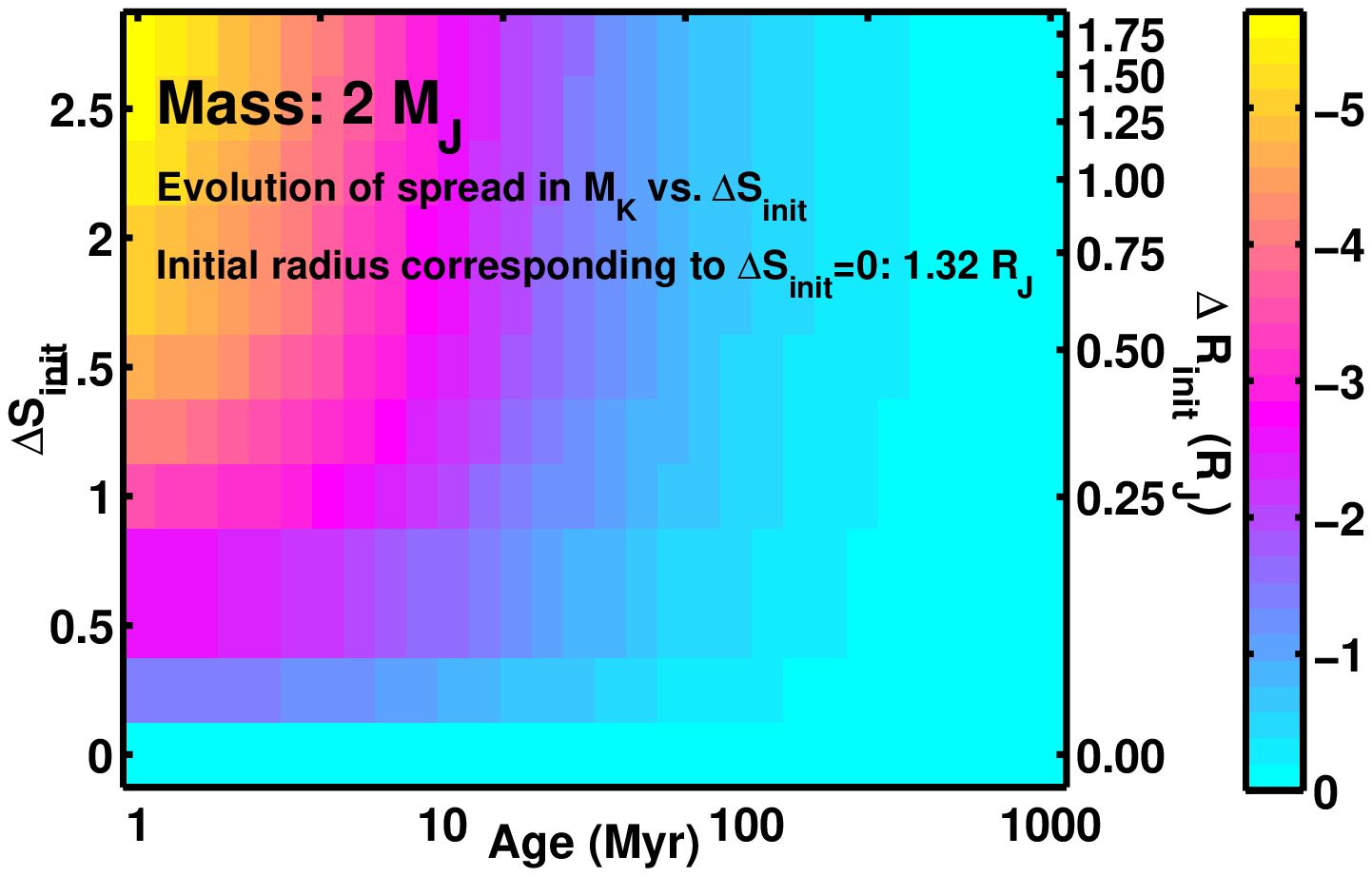}\\
\plottwob
{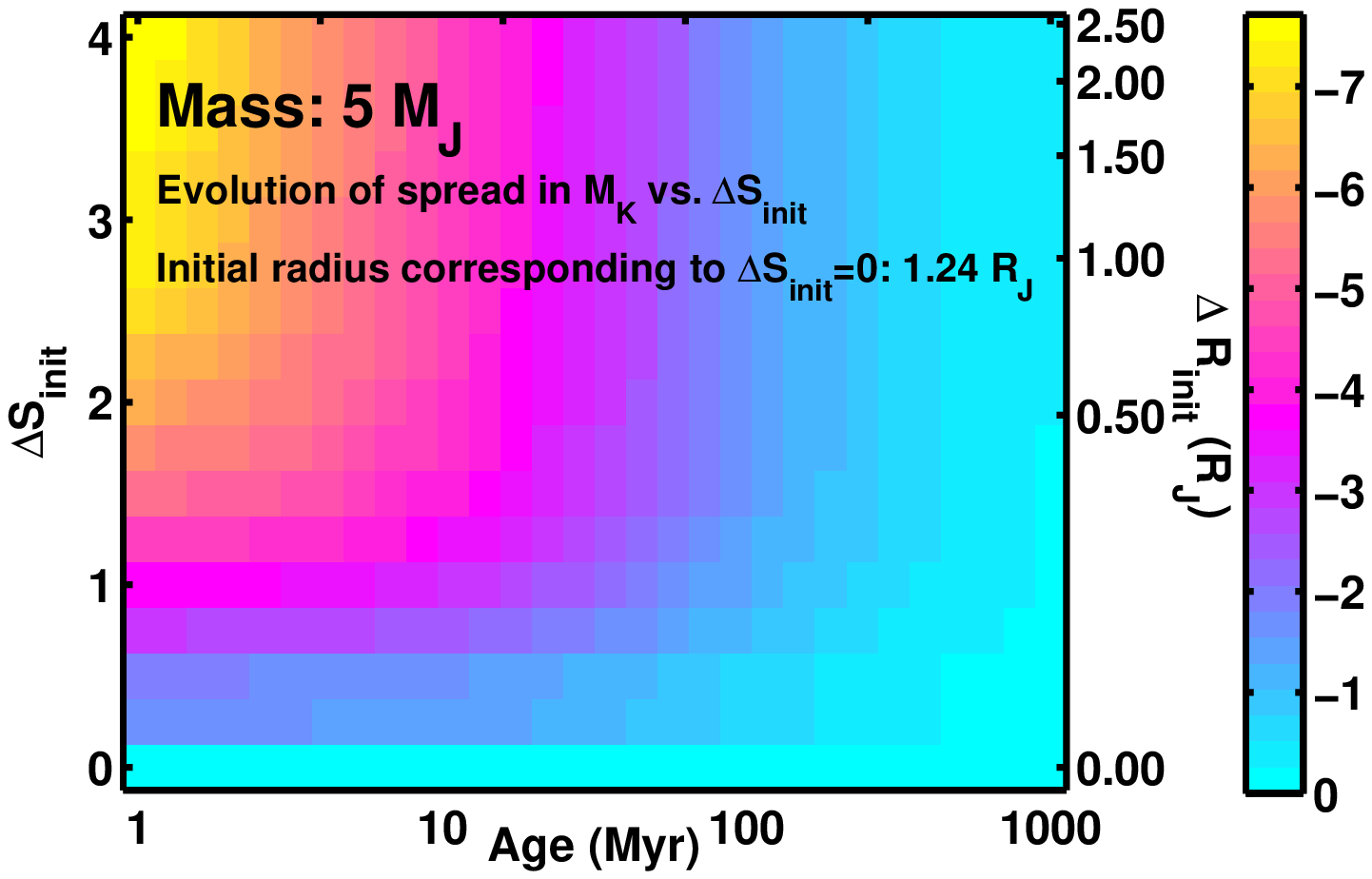}
{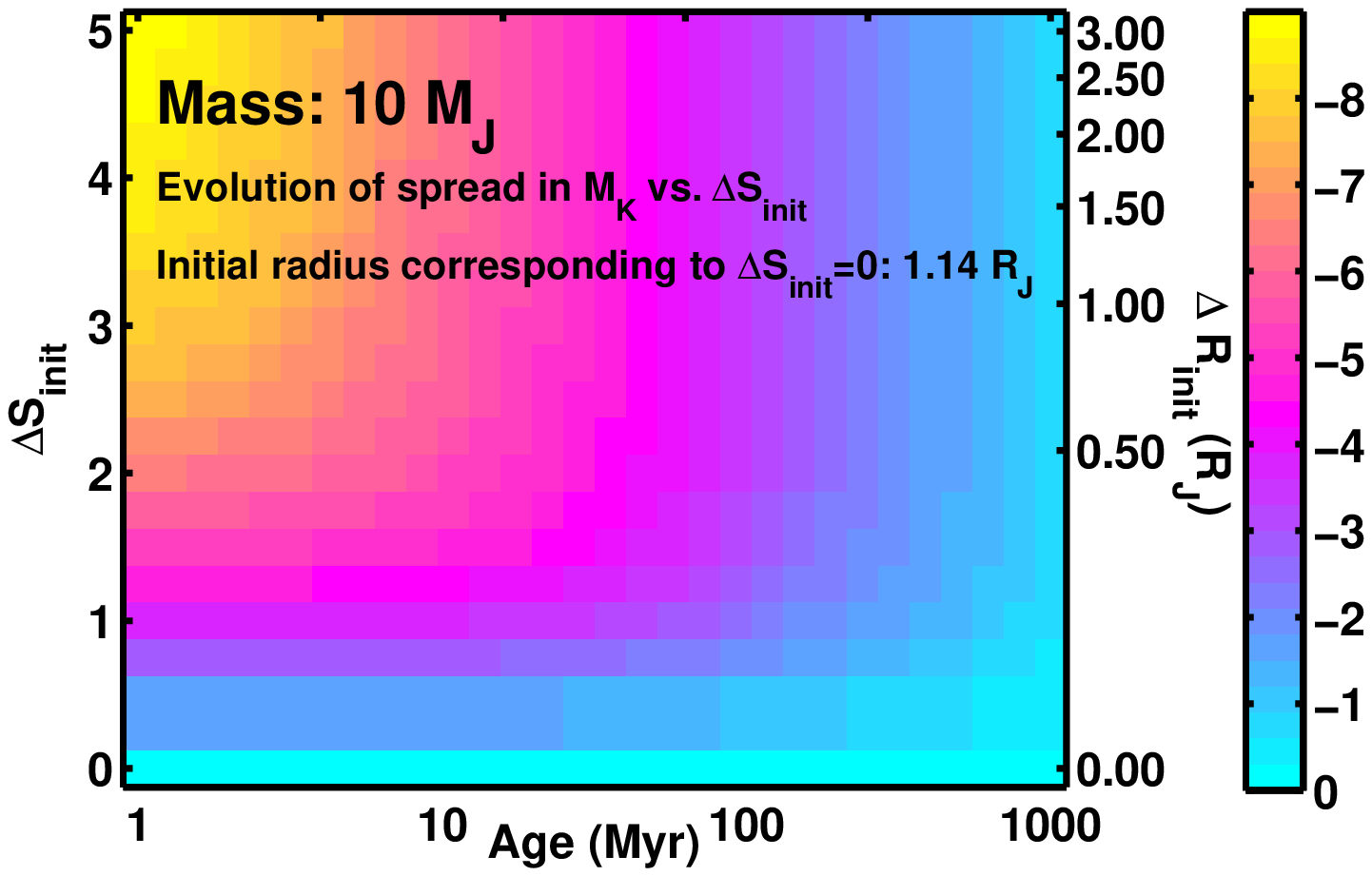}
\caption{Evolution of $K$-band absolute magnitude as a function of
  initial entropy and mass.  For each of four planet masses -- $1M_J$
  (\emph{top left}), $2M_J$ (\emph{top right}), $5M_J$ (\emph{bottom
    left}), $10M_J$ (\emph{bottom right}) -- the change in $K$-band
  absolute magnitude relative to the corresponding ``coldest start''
  case in Fig.~\ref{fig:introduce} is shown.  The coldest start cases
  have initial specific entropies of 8.75, 8.5, 8.25, and 8, and
  initial radii of 1.42, 1.32, 1.24, and 1.14~$R_J$, for masses of 1,
  2, 5, and 10~$M_J$, respectively.  The left $y$-axis of each panel
  shows the increase in specific entropy relative to the coldest
  start.  The right $y$-axis of each panel shows the change in radius
  relative to the coldest start.  At higher initial entropies (and
  larger initial radii), planets are brighter in the $K$ band.  As
  planets age, the change in $K$-band absolute magnitude relative to
  the coldest start case approaches zero.  Note that each panel has an
  independent color stretch.  More massive objects exhibit greater
  differences in brightness between the coldest and the hottest
  starts, and their differences persist for longer.}
\label{fig:spread_in_K}
\end{figure}

Figure~\ref{fig:spread_in_K} summarizes the spread in $K$-band
brightness as a function of initial entropy, initial radius, mass, and
age.  In each panel, the ordinate indicates the increase in entropy
relative to the coldest-start case (for the appropriate mass) depicted
in Fig.~\ref{fig:introduce}.  The corresponding initial radius scale
for each is shown on the right $y$-axis as $\Delta R_{\rm init}$, the
increase in initial radius relative to the coldest-start case.  The
uneven spacing of $\Delta R_{\rm init}$ results from the fact that
radius grows faster than linearly with entropy (see
Fig.~\ref{fig:SvsR}).  The color indicates the difference between the
$K$-band magnitude at the given entropy and the $K$-band magnitude for
the coldest-start case.  Coldest start cases have initial specific
entropies of 8.75, 8.5, 8.25, and 8, and initial radii of 1.42, 1.32,
1.24, and 1.14~$R_J$, for masses of 1, 2, 5, and 10~$M_J$,
respectively.  At higher initial entropies, planets are from $\sim$4.5
magnitudes (at Jupiter's mass) to $\sim$9 magnitudes (at 10~$M_J$)
brighter than the corresponding coldest-start cases.  As planets age,
the change in $K$-band absolute magnitude relative to the coldest
start case approaches zero.  More massive objects exhibit greater
differences in brightness between hottest- and coldest-start cases,
and maintain these differences for significantly longer.  At 1~$M_J$,
differences in brightness fade to less than a single magnitude by
$\sim$10~Myr, even for the hottest start.  At 10~$M_J$, it takes
several hundred Myr for the $K$-band magnitude difference between a
hottest-start and a coldest-start planet to reach 1~mag.  Of course,
at several hundred Myr, even 10-$M_J$ planets are extremely dim in the
$K$ band.  At longer wavelengths (e.g., $M$ and $N$ bands), cool
objects such as $\sim$1-Gyr old 10-$M_J$ planets are still brighter
than absolute magnitude 15 (not shown).

\begin{figure}[t!]
\plottwob
{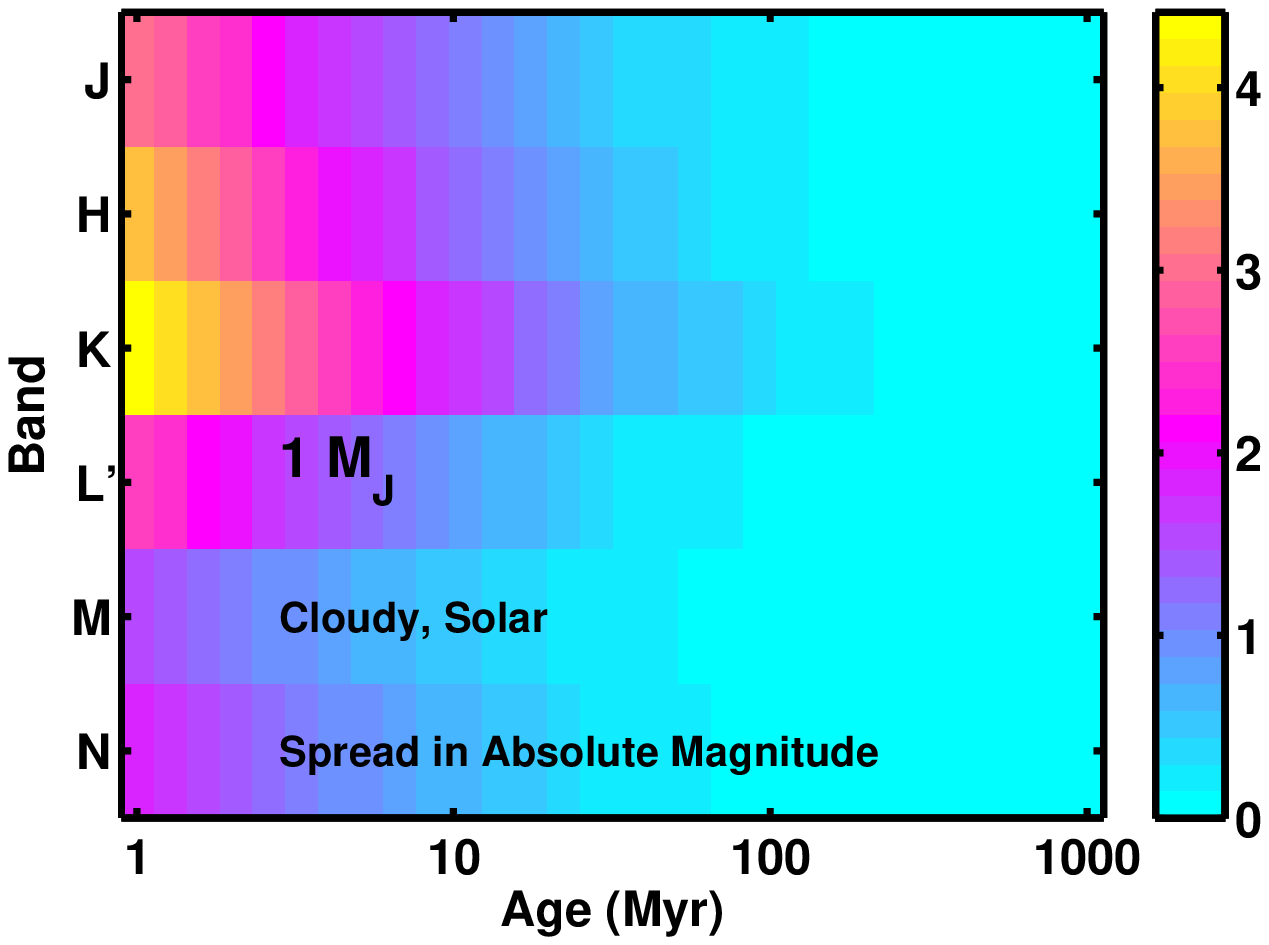}
{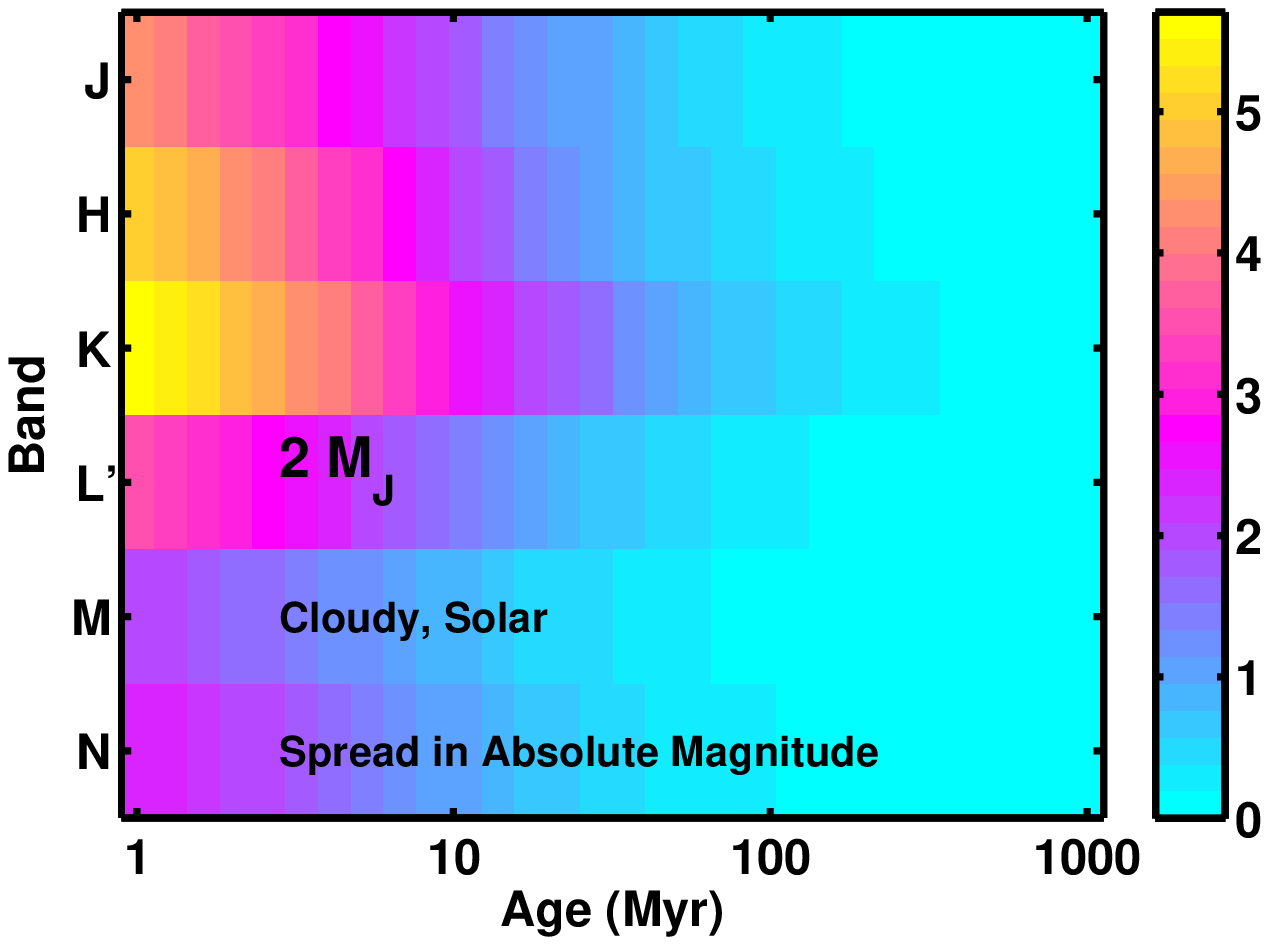}\\
\plottwob
{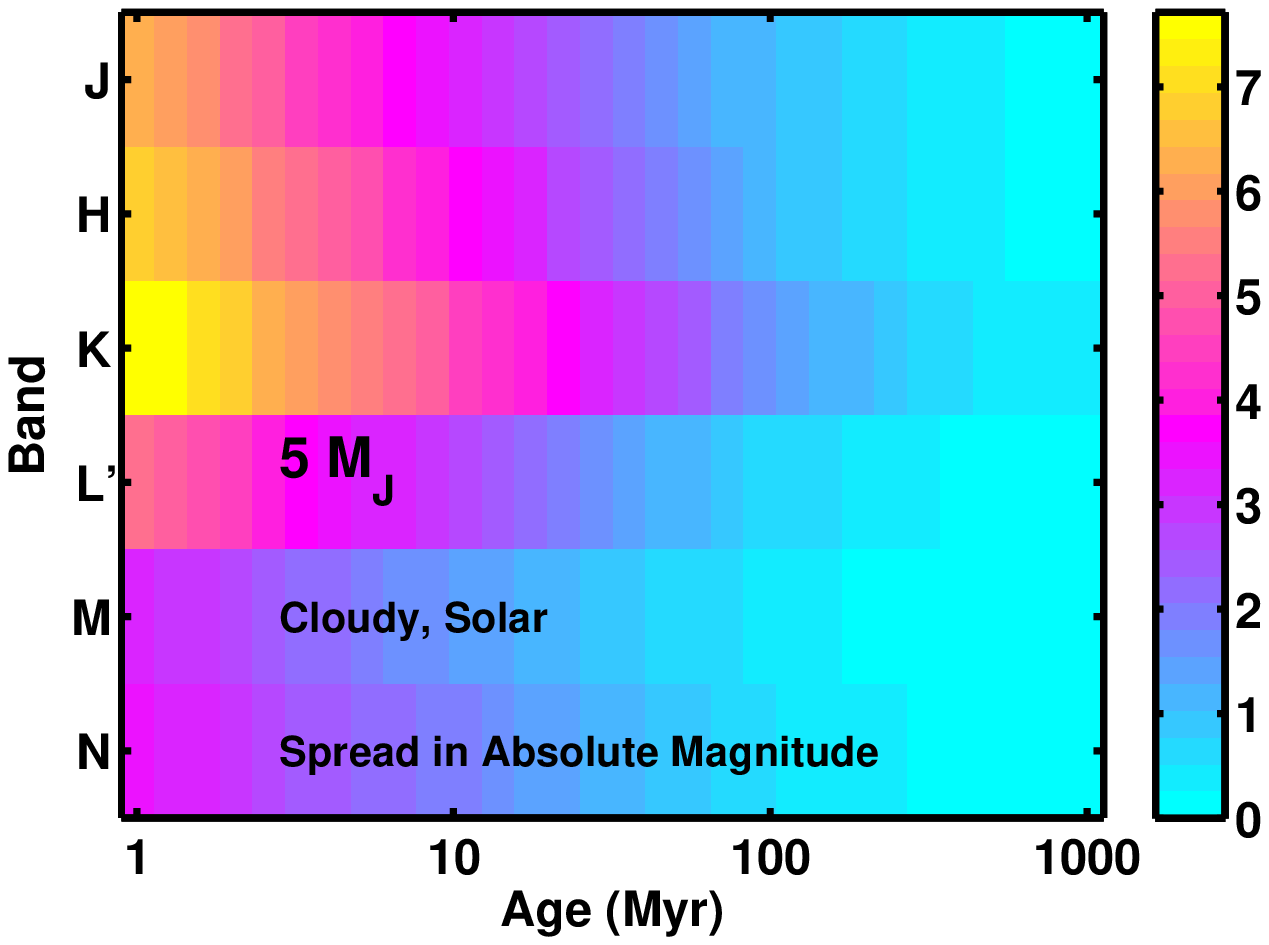}
{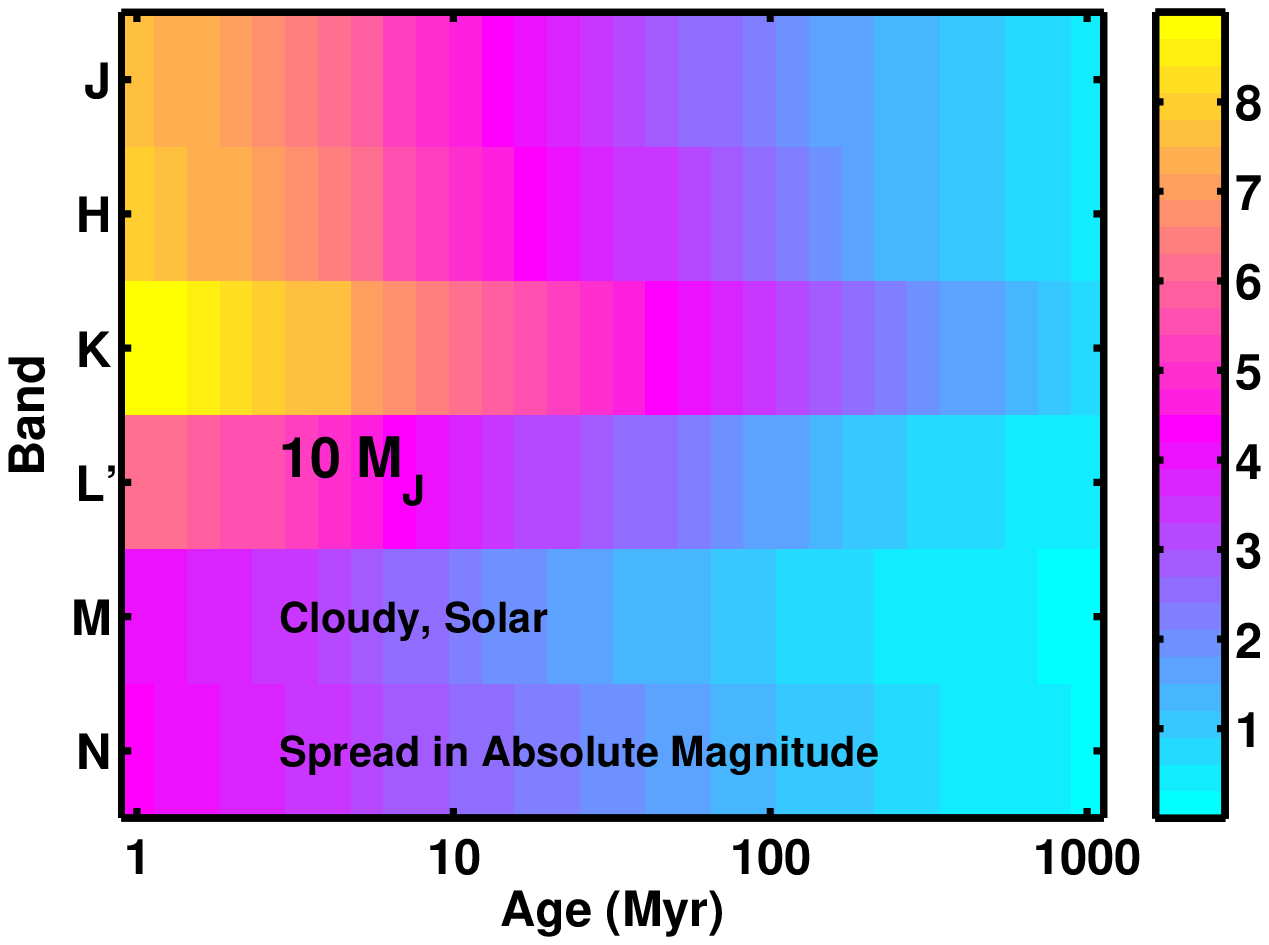}
\caption{Evolution of spread in absolute magnitude as a function of
  mass and spectral channel.  For each of four planet masses -- $1M_J$
  (\emph{top left}), $2M_J$ (\emph{top right}), $5M_J$ (\emph{bottom
    left}), $10M_J$ (\emph{bottom right}) -- the difference in
  absolute magnitude between the coldest start and the hottest start
  cases in Fig.~\ref{fig:introduce} is shown.  The atmospheres of all
  objects are the hybrid cloudy models at solar metallicity from
  \citet{burrows_et_al2011}.  Note that each panel has an independent
  color stretch.  In all six bands, more massive objects exhibit
  greater differences in brightness between the coldest and the
  hottest starts, and their differences persist for longer.  The
  spread in absolute magnitude is greatest at $K$ band throughout the
  range of masses.}
\label{fig:spread_of_mass}
\end{figure}

It is instructive to examine the evolution of the full spectral energy
distribution (SED) of model planets as a function of mass and initial
entropy.  Figure~\ref{fig:spread_of_mass} shows the full spread in
$K$-band absolute magnitude (from the hottest-start to the
coldest-start models in Fig.~\ref{fig:introduce}) for each of 6 bands,
from $J$ through $N$.  At all four planet masses shown (1, 2, 5, and
10~$M_J$), the difference in absolute magnitude is greatest in the $K$
band (which is why we presented $K$-band evolution plots in
Figs.~\ref{fig:K_ev_bands} and \ref{fig:spread_in_K}).  The spread in
magnitudes is smallest in $M$ band, where at 1~$M_J$ the difference
between the hottest-start and coldest-start models is only
$\approx$1.5~mags, even very early after ``formation'' (at 1~Myr).
The relatively small differences in brightness between hottest and
coldest models at $M$ and $N$ bands are in large part because these
bands are in the Rayleigh-Jeans tail of a planet's SED.  As a result,
the brightness in these bands is roughy proportional to effective
temperature, which varies by no more than a factor of a few at early
times, and significantly less at late times (see
Fig.~\ref{fig:M_evs}).

\subsection{The Effect of Atmosphere}
\label{ssec:atmos}

\begin{figure}[t!]
\plottwob
{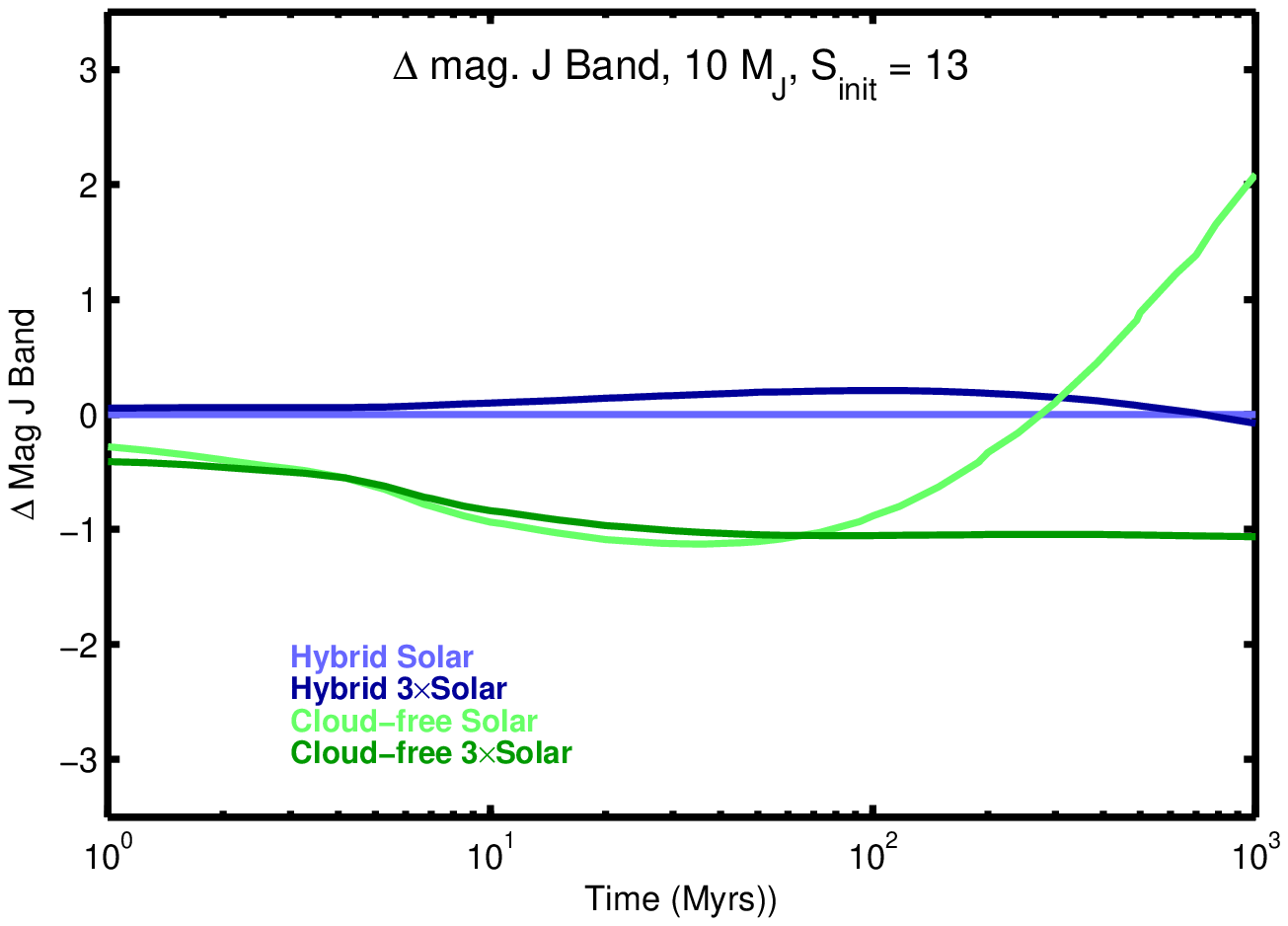}
{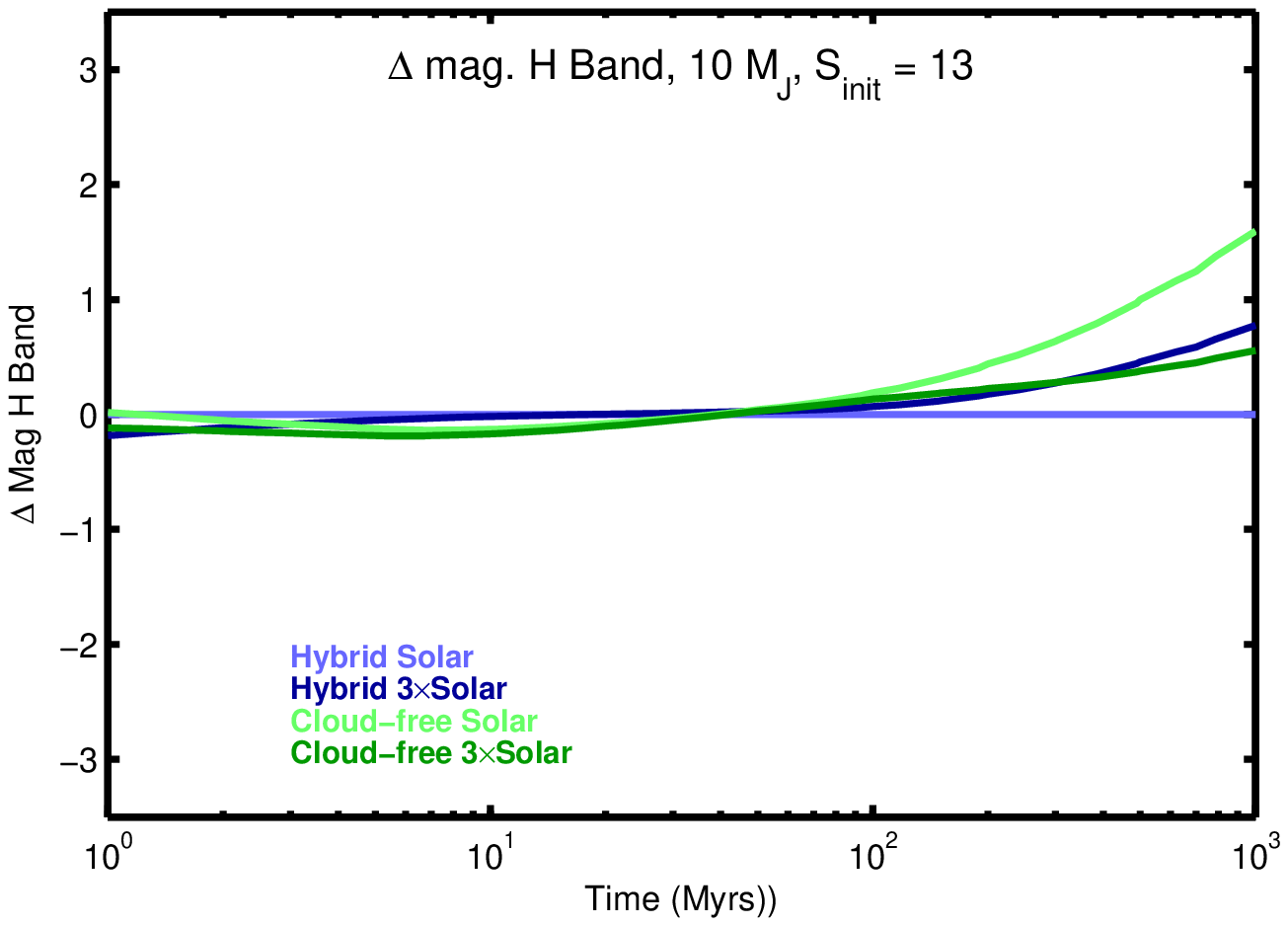} \\
\plottwob
{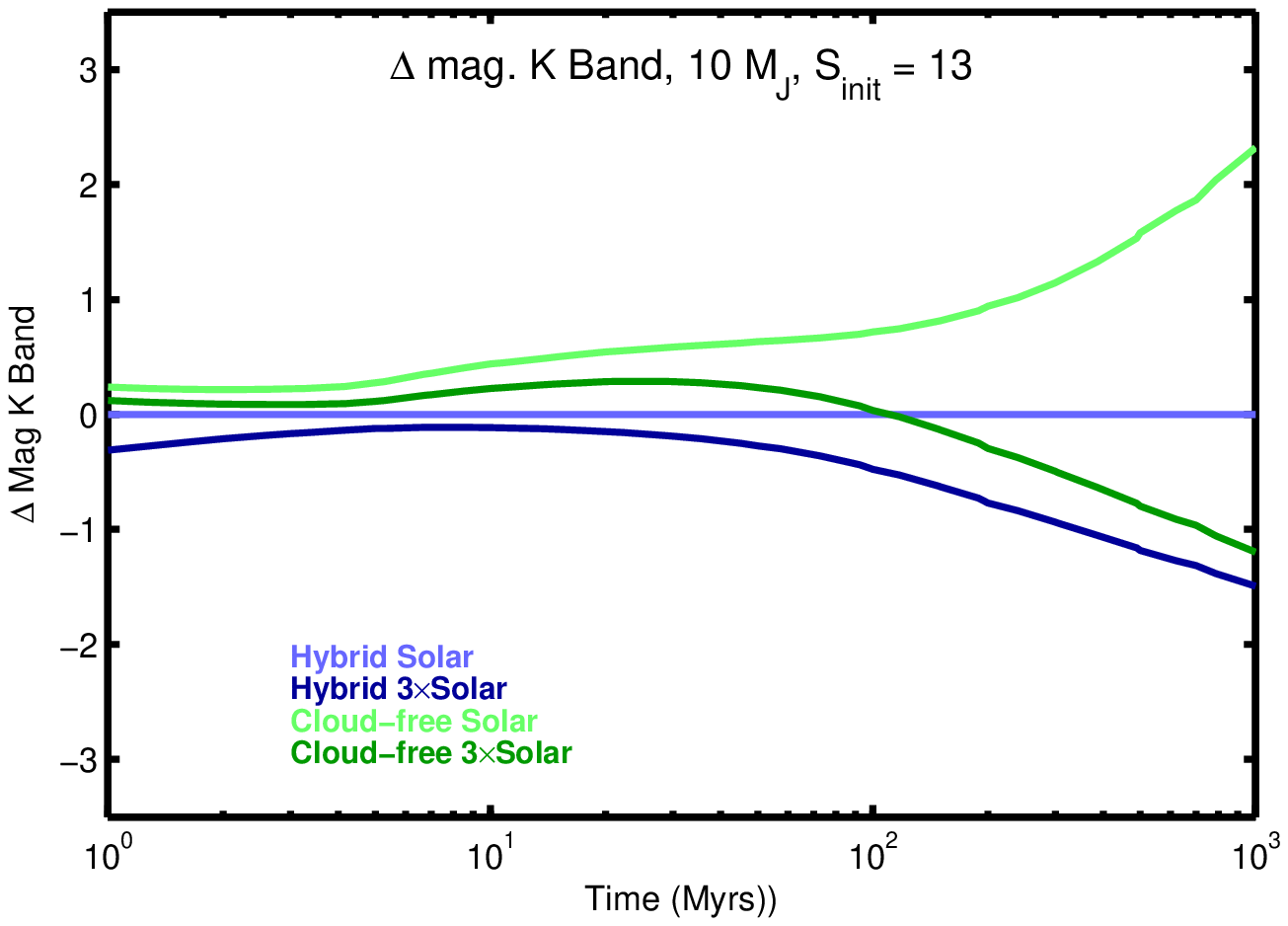}
{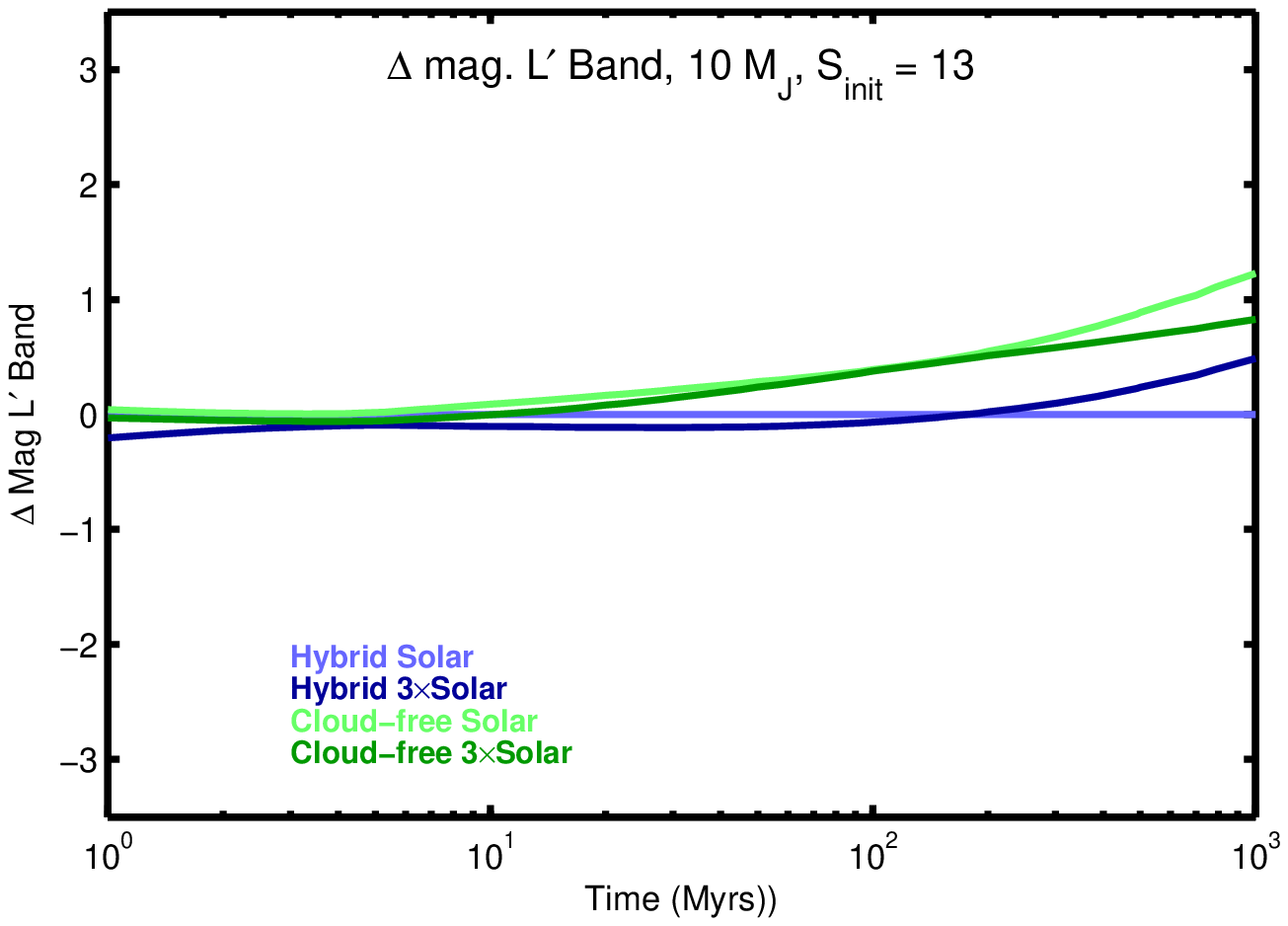} \\
\plottwob
{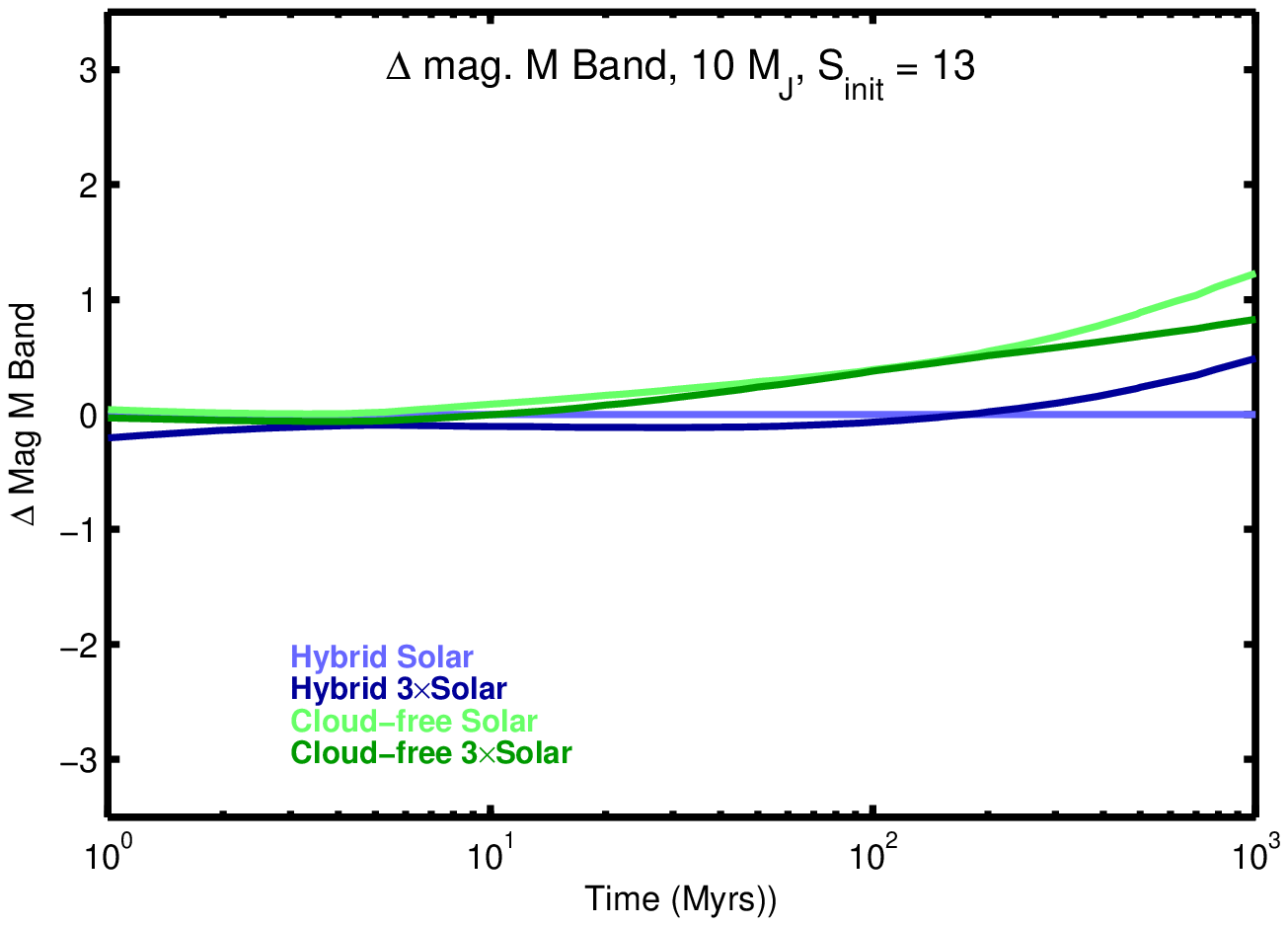}
{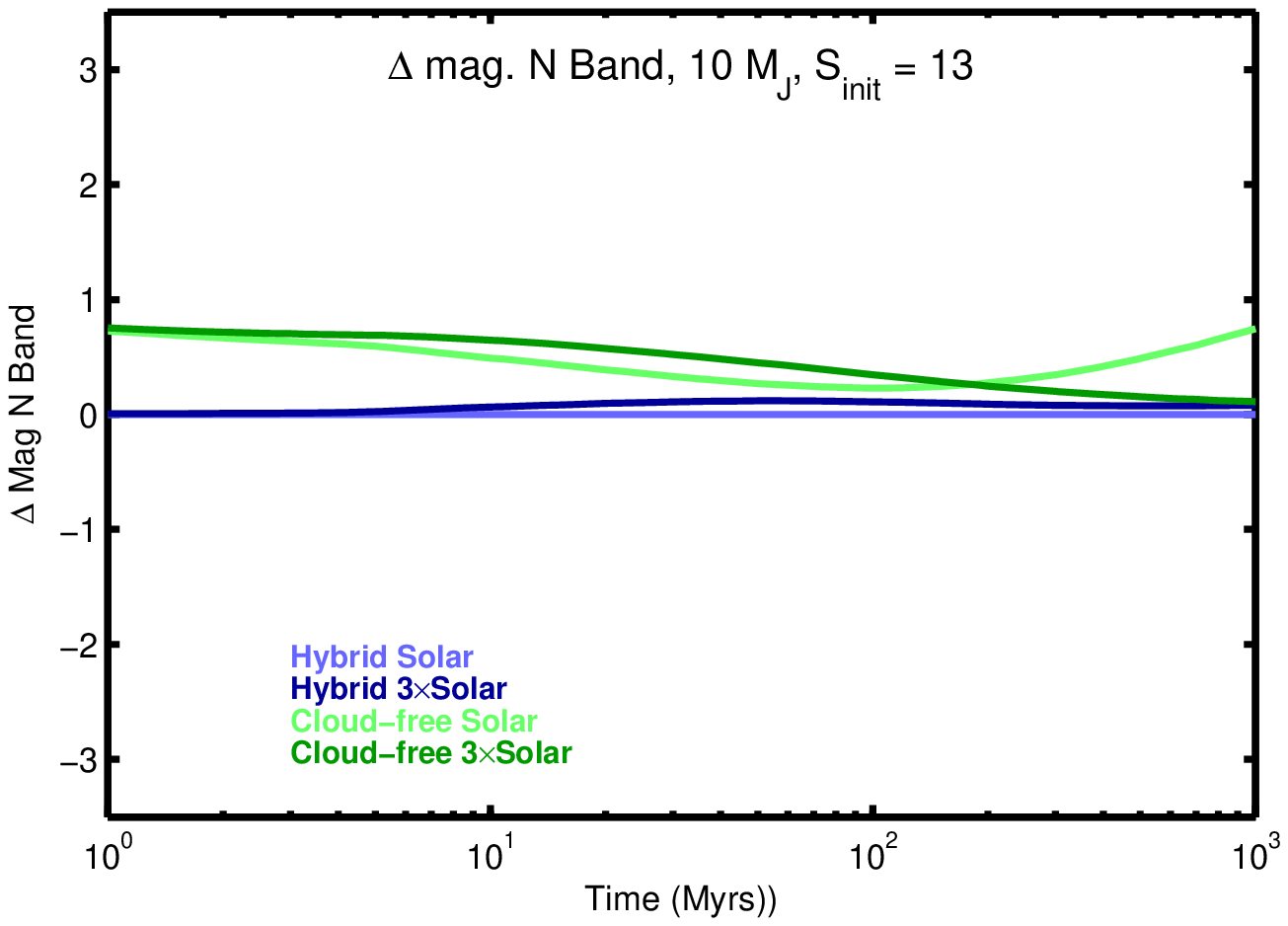}
\caption{Illustrative examples of evolution of the difference in
  magnitude relative to the fiducial atmosphere (i.e., relative to a
  hybrid clouds, solar metallcity atmosphere model).  This figure
  shows the differences for an object of mass $10M_J$ with initial
  entropy of 13.  In some bands, at some ages, the differences in
  brightness from one atmosphere model to another (at identical object
  mass and initial entropy) can be several magnitudes.}
\label{fig:Atm_ev_bands}
\end{figure}

It is worthwhile to consider the effect of different atmosphere types
(discussed in \S\ref{ssec:spec_ev}) on observables.  Equal-mass
objects with identical initial entropy can have very different colors
depending on their atmospheric properties.
Figure~\ref{fig:Atm_ev_bands} provides an illustrative example of
these differences, for a very hot start (initial entropy of 13)
10-$M_J$ model.  The ordinate represents the difference in magnitude,
as a function of age, between each of four atmosphere types (hybrid
clouds, solar metallicity; hybrid clouds, $3\times$~solar metallicity;
cloud-free, solar metallicity; cloud-free, $3\times$~solar
metallicity; these are the four atmospheres of
\citealt{burrows_et_al2011}) and our fiducial atmosphere (hybrid
clouds, solar).  In $J$ band, differences start small, but by late
times the cloud-free solar models are two magnitudes fainter than the
fiducial models, and 3 magnitudes fainter than the cloud-free
$3\times$~solar models.  In $K$ band, the differences between some
models and others grow to more than 3 magnitudes.  In $N$ band, by
contrast, the differences between cloudy and cloud-free models begin
as moderate ($\sim$1 magnitude) and slowly decrease with age.

\begin{figure}[t!]
\plottwob
{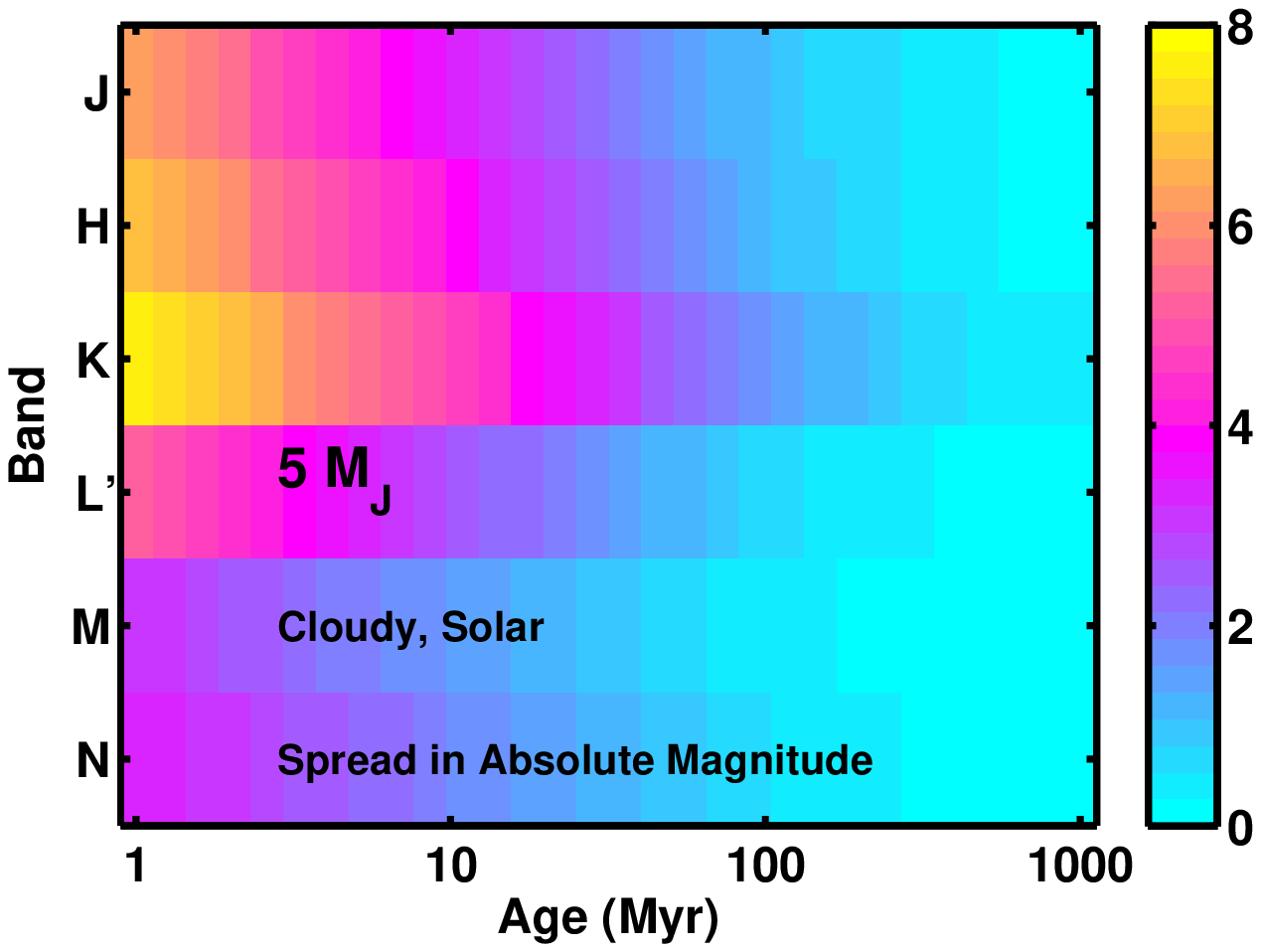}
{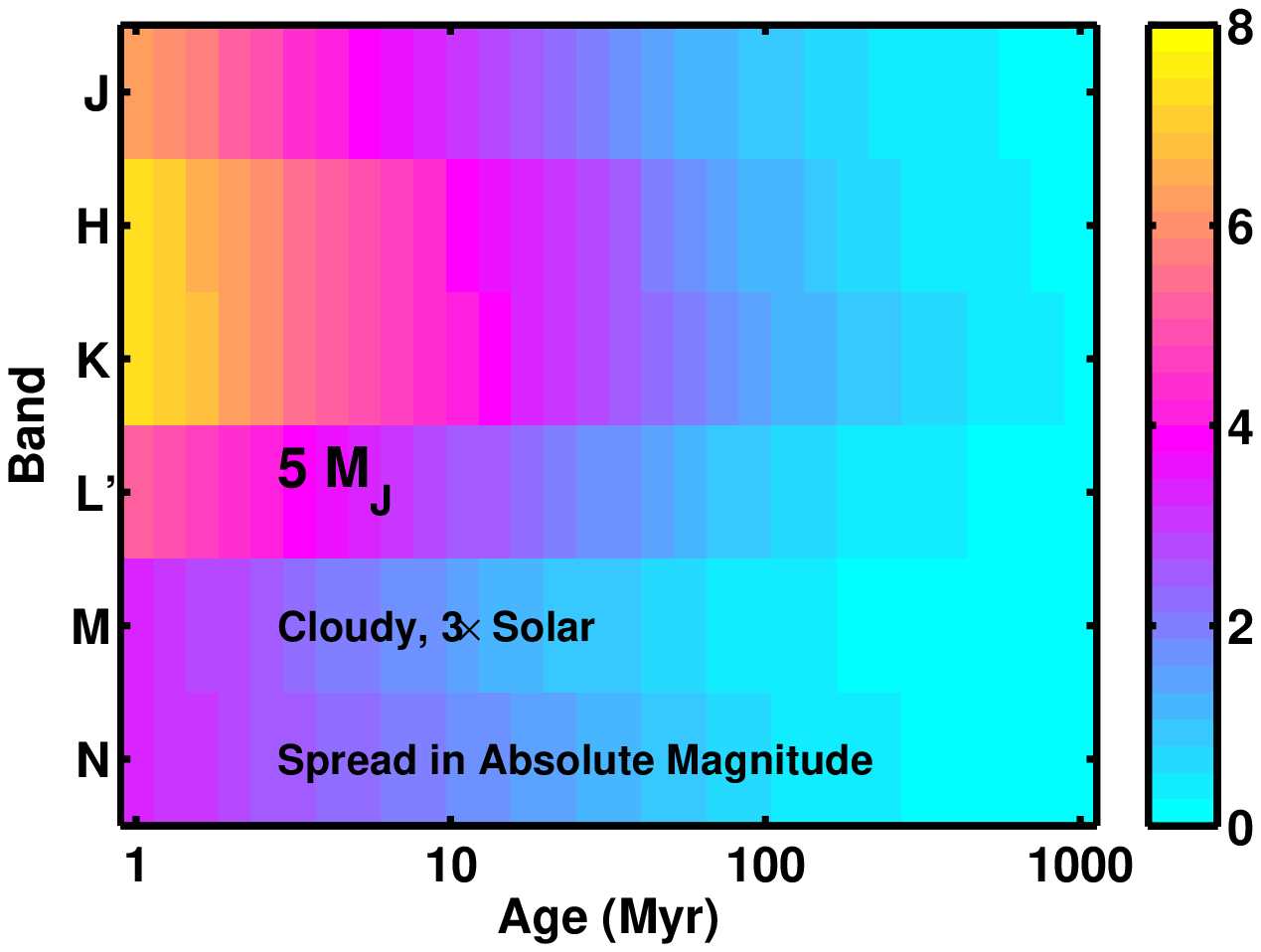}\\
\plottwob
{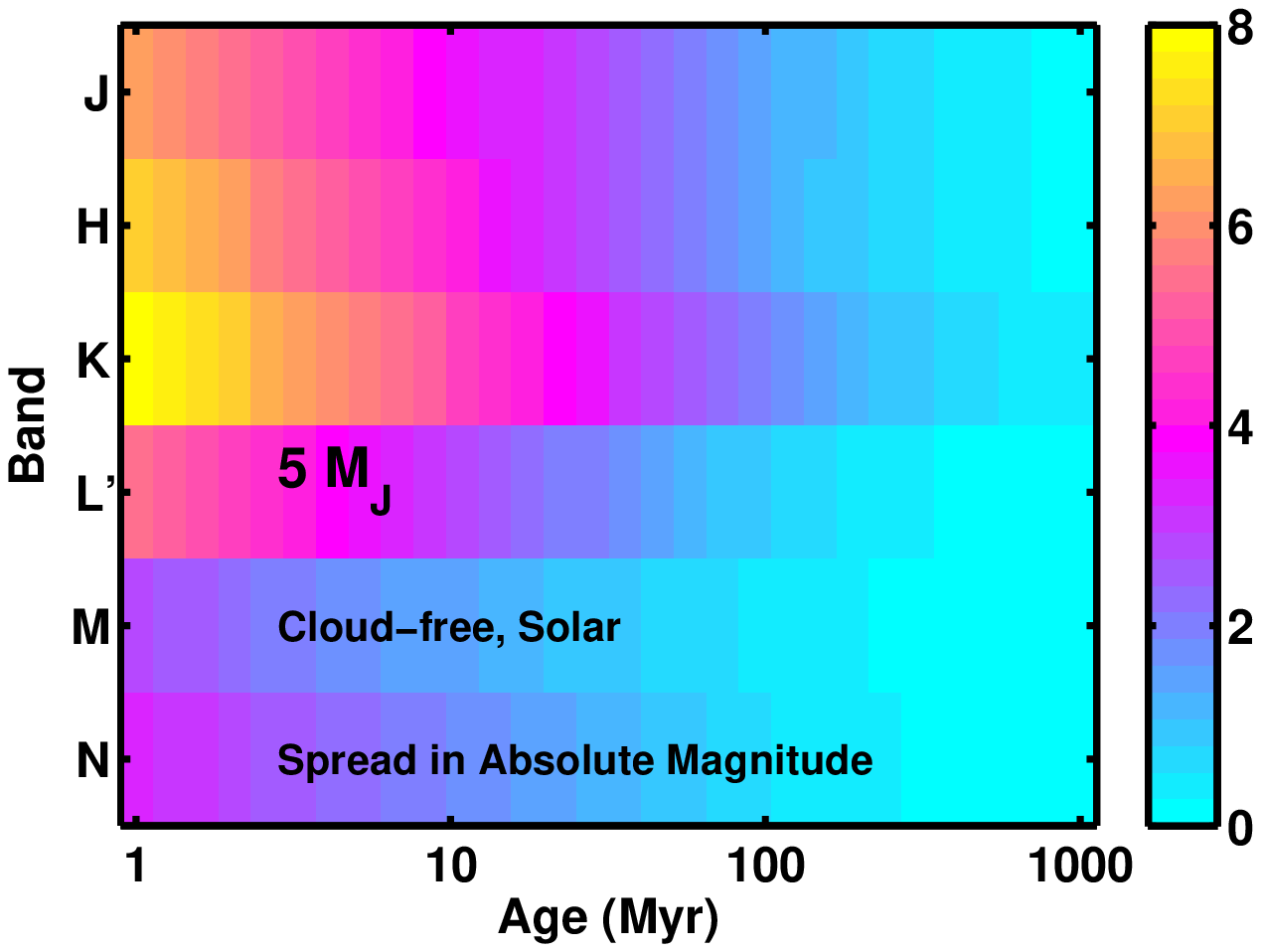}
{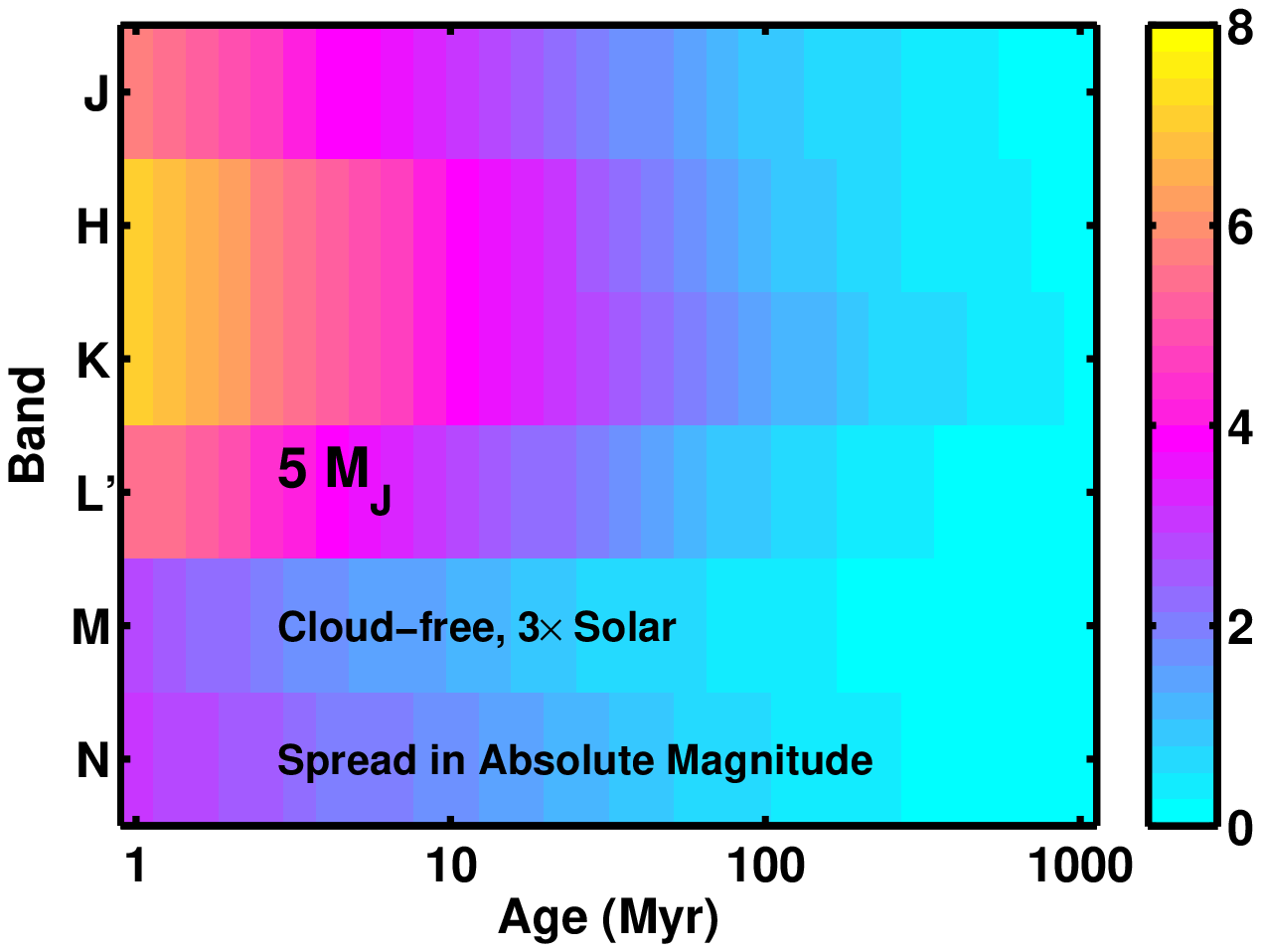}
\caption{Evolution of spread in absolute magnitude as a function of
  atmosphere type and spectral channel.  For each of four atmosphere
  types -- cloudy at solar metallicity (\emph{top left}), cloudy at
  three times solar metallicity (\emph{top right}), cloud-free at
  solar metallicity (\emph{bottom left}), cloud-free at three times
  solar metallicity (\emph{bottom right}) -- the difference in
  absolute magnitude between the coldest start and the hottest start
  cases in Fig.~\ref{fig:introduce} is shown.  Details regarding the
  atmosphere types are described in \citet{burrows_et_al2011}.  All
  model planets are 5~$M_J$.  The effect of atmosphere type on the
  spread in absolute magnitude is not dramatic, although significant
  differences exist from one spectrum to another which are not
  apparent in this figure.}
\label{fig:spread_of_atmosphere}
\end{figure}

Another perspective on the effect of atmosphere type is provided in
Fig.~\ref{fig:spread_of_atmosphere}, which presents the evolution of
the spread in absolute magnitude as a function of atmosphere type and
spectral band for 5~$M_J$ planets.  The four panels, representing the
four different atmosphere types, look broadly similar, with the
largest differences between hottest and coldest planets generally
occuring in $H$ and $K$ bands.  However, there are some quantitative
and qualitative differences among the four panels.  At solar
metallicity, both the cloudy and the cloud-free models exhibit the
largest hot-cold differences in $K$ band ($\sim$8 magnitudes at early
times), whereas at 3$\times$ solar metallicity $H$ and $K$ bands have
approximately equal hot-cold differences ($\sim$7 magnitudes at early
times).  For all four atmospheres, $M$ band shows the smallest
hot-cold difference throughout the 1~Gyr of evolution displayed.  A
key result shown in Fig.~\ref{fig:spread_of_atmosphere} is that the
effect of atmosphere type on the spread in absolute magnitude is not
nearly so strong as the effects of changes in mass, age, or initial
entropy, but spectra can nonetheless differ by several magnitudes
depending upon atmospheric properties.

One can view Figs.~\ref{fig:K_ev_bands}-\ref{fig:spread_of_atmosphere}
in a forward sense.  If we know the age, mass, initial entropy, and
atmosphere of a model, these figures and associated tabulated data
tell us the absolute magnitude in each of 6 photometric bands in the
infrared.  However, this information is probably more useful when
viewed in an inverse sense.  For comparing these models with data, the
magnitudes or spectra will be the observed quantities, and one will
try to make inferences about the properties of the object.  These
properties include the same parameters listed for the forward problem:
age (about which there might be some prior constraints from either the
star or its environment), mass, initial entropy, and atmosphere type.
By checking which models reasonably match observed data, we might hope
to find contours in a multidimensional parameter space that rule in or
rule out possible models.  In this way, we may use data to infer the
initial entropies of observed objects.  Note that an exploration of
our model space indicates that the thicker clouds employed in the
models of \citet{currie_et_al2011} and \citet{madhusudhan_et_al2011}
seem to be essential for finding self-consistent (coeval) models of
planets b, c, and d in the HR8799 system.

One contour that appears to be available for the HR8799 system is the
stellar age.  \citet{marois_et_al2008} argue that several lines of
evidence suggest that the system has an age in the range 30-160~Myr.
\citet{moya_et_al2010} find that asteroseismological data imply an age
closer to 1~Gyr.  However, if the planets were 1~Gyr old, their
luminosities would imply masses large enough to render the system
dynamically unstable \citep{fabrycky+murrayclay2010,
  moro-martin_et_al2010}.  Furthermore, \citet{zuckerman_et_al2011}
find that HR8799's probable association with a nearby group suggests
an age of $\sim$30~Myr.  We therefore assume, for now, that the
objects have ages in the range originally identified by
\citet{marois_et_al2008}.  The fourth planet found in the system,
planet e, was recently discovered in $K_s$ and $L'$ band observations
by \citet{marois_et_al2010}.  \citeauthor{marois_et_al2010} find that
the planet's luminosity suggests a mass of 7-10~$M_J$ if its age is in
the range 30-60~Myr.  We tried to fit the photometry (the absolute
magnitudes in these bands are $12.93\pm 0.22$ and $11.61 \pm 0.12$,
respectively) with both the AE cloud models of
\citet{madhusudhan_et_al2011} and with our own set of models.  Because
of the paucity of data for planet e, a range of masses, ages, initial
entropies, and atmospheres are consistent with the available data.
Both the \citeauthor{madhusudhan_et_al2011} AE cloud models and our
own hybrid cloud models (hy1s and hy3s) are consistent with the data
at the ages and masses suggested by \citet{marois_et_al2010}.  At
these ages and masses, consistency with the photometry requires
significantly higher initial entropies than our coldest start
scenarios.  Since our cloud-free models (cf1s and cf3s) are dimmer in
$K_s$ and $L'$ bands than the hybrid cloud models, fitting the data
with the cloud-free models requires higher masses within the age-range
(30-60~Myr) specified by \citet{marois_et_al2010}.  Additional
photometry at other bands could break some of the present degeneracies
and might rule in or rule out some atmosphere types.

\section{Conclusion}
\label{sec:conc}

Gas-giant planets are thought to form in one of two scenarios --
either runaway accretion onto a rocky or icy core, or direct
gravitational collapse via a Jeans-like instability triggered in the
protoplanetary disk.  The former process is generally thought to lead
to colder (i.e., lower entropy) objects than the latter, but
first-principles calculations cannot yet specify with certainty what
the initial (post-formation) entropies of objects should be in the
different formation scenarios, nor whether the range of entropies
resulting from the different mechanisms might overlap.

We seek observational determinants of the various models for giant
planet origins, including their ``initial conditions.''  Candidate
properties that might be discriminating include metallicity, rotation
rate, mass function, or orbits.  Core-accretion-formed objects might
generally be expected to have higher bulk metallicity than
disk-instability-formed objects; and the different modes of formation
might tend to lead to differences in angular momentum, typical object
mass, and initial orbit.  A wrinkle associated with the idea of using
rotation rate as a discriminant has to do with the fact that,
for 10~$M_J$ and slightly more massive objects, a deuterium-burning
phase occurs a few million years post-formation.  This phase delays
the evolutionary shrinkage, and when it has completed objects cool and
shrink rapidly.  It is conceivable that there is a signature of this
rapid shrinkage embedded in the rotation rates of these objects, where
massive objects burning deuterium might appear to be rotating
anomalously slowly, and conservation of angular momentum might cause a
spin-up in the post-deuterium stages.  But without knowing precisely
the initial total angular momentum budget of a massive planet, it is
unclear how such a signature could be interpreted.

However, more direct signatures involve broad-band colors and spectra.
We have presented a large suite of evolutionary models of
non-irradiated gas-giant planets, which are publicly available in
electronic form.  Our models span a range of ``initial conditions,''
with masses ranging from 1 to 15 times Jupiter's, entropies ranging
from 8 to 13 Boltzmann constants per baryon, four different atmosphere
types, and a dense range of ages up to a gigayear.  Motivated by
(\emph{i}) the strong preference for the core-accretion mode of
gas-giant planet formation within the inner tens of AU, (\emph{ii})
the recent observations of several young exoplanets that seem to have
been born hotter than traditional ``cold start'' models, and
(\emph{iii}) the uncertainties inherent in the current generation of
both core-accretion and disk-instability models of planet formation,
we introduce a set of warm-start models that are at intermediate
entropies between traditional ``hot starts'' and ``cold starts.''  We
find the following robust conclusions:
\begin{itemize}
\item Objects that begin at larger radius and higher entropy can be
  significantly brighter in the first few million years (by 2-9
  magnitudes, depending on mass and spectral band).
\item These photometric differences are most prominent in $H$ and $K$
  bands, and least in $M$ and $N$ bands.
\item For lower-mass (1 and 2~$M_J$) gas giant planets, these
  differences fade to barely observable levels within $\sim$10-30~Myr,
  whereas for more massive (5-10~$M_J$) planets, a hottest-start model
  can remain several magnitudes brighter than a coldest-start model in
  the $K$ band for more than 100~Myr.
\item Changes in atmosphere type can result in changes in brightness
  of a few magnitudes for some masses, ages, initial entropies, and
  spectral bands, but atmosphere type has less influence on a planet's
  emergent flux than mass, age, and initial entropy.
\end{itemize}

Some caveats remain.  For one, our exploration of atmosphere types is
far from complete, comprising only a single kind of cloudy atmosphere
and a single cloud-free atmosphere, each at both solar and three-times
solar metallicity.  Other types of clouds, non-solar mixtures of
elements, and nonequlibrium chemical processes could all have
important affects on spectra that are not captured by our models.
Perhaps more significant, we do not model in detail the formation
processes, including the large spike in luminosity that might occur at
very early ``pre-formation'' times, due to an accretion shock during
either direct collapse or core accretion.  Furthermore, since disk
instability is thought to form planets orders of magnitude faster than
the $\sim$several million year timescale for core accretion, there is
a hidden age ambiguity that must be accounted for when our model
planets are compared with observed systems.  That is to say, perhaps a
model that is supposed to represent disk instability should be taken
to be several million years older than a model that is taken to
represent core accretion.  Rather than try to build this into our
public models, we prefer to simply start the models at time zero with
a particular initial entropy and allow others to adjust the clocks as
they consider appropriate.

We hope that these models will prove to be useful in diagnosing
observed young planets.  By comparing data to model atmospheres, one
can infer the initial, post-formation conditions of gas-giant planets.
Eventually, this type of analysis might constrain planet formation
mechanisms by, for instance, quantifying how much entropy nebular gas
retains as it accretes onto a core.  In this way, models such as those
in this paper contribute toward developing a coherent picture of the
evolution of planetary systems, from their formation through the
epochs when we observe them.

\vspace{0.5in}

\acknowledgments

We thank Scott Kenyon for helpful conversations and for sharing models
of giant planet formation.  We thank Kevin Heng, Jack Lissauer, Peter
Bodenheimer, Morris Podolak, Ivan Hubeny, Sean Raymond, Madhu Nikku,
and Jason Nordhaus for useful discussions.  We also thank an anonymous
referee for helpful comments and suggestions.  The authors would like
to acknowledge support in part under NASA ATP grant NNX07AG80G, HST
grants HST-GO-12181.04-A and HST-GO-12314.03-A, and JPL/Spitzer
Agreements 1417122, 1348668, 1371432, and 1377197.  DSS acknowledges
support from NSF grant AST-0807444.  Models can be found at
http://www.astro.princeton.edu/$\sim$burrows/ and
http://www.sns.ias.edu/$\sim$dave/.

\newpage

\vspace{0.5in}

\bibliography{biblio.bib}
\clearpage

\begin{landscape}
\begin{table}[t!] 
\small 
\begin{center} 
\caption{Evolution of Multi-band Magnitudes (hybrid clouds, solar abundance)}
\label{ta:models_hy1s}
\begin{tabular}{rr|rrrrrrr|rrrrrrr} 
\hline 
\hline 
             &             &             &             &             &      Hot    &             &             &             &             &             &             &    Cold     &             &             &             \\ 
   Age (Myr) & $M$ ($M_J$) & $R$ ($R_J$) &       $J$   &       $H$   &       $K$   &     $L'$    &     $M$     &     $N$     & $R$ ($R_J$) &      $J$    &      $H$    &      $K$    &    $L'$     &    $M$      &      $N$      \\ 
\hline 
        1.00 &        1.00 &        1.74 &       15.64 &       15.33 &       13.38 &       12.69 &       11.41 &       10.03 &        1.41 &       18.89 &       19.35 &       18.05 &       15.44 &       13.03 &       12.03 \\ [0.05cm] 
        1.00 &        2.00 &        1.69 &       13.80 &       12.87 &       11.70 &       11.03 &       10.60 &        9.34 &        1.31 &       18.18 &       18.14 &       17.54 &       14.74 &       12.75 &       11.86 \\ [0.05cm] 
        1.00 &        5.00 &        1.87 &       11.28 &       10.29 &        9.52 &        9.05 &        9.34 &        8.35 &        1.24 &       17.64 &       17.28 &       17.34 &       14.29 &       12.57 &       11.84 \\ [0.05cm] 
        1.00 &       10.00 &        2.31 &        9.48 &        8.66 &        8.04 &        7.69 &        8.26 &        7.48 &        1.14 &       17.28 &       16.76 &       17.21 &       14.13 &       12.56 &       11.94 \\ [0.05cm] 
\hline 
        2.00 &        1.00 &        1.61 &       16.61 &       16.56 &       14.71 &       13.54 &       11.92 &       10.64 &        1.40 &       19.09 &       19.59 &       18.36 &       15.60 &       13.12 &       12.15 \\ [0.05cm] 
        2.00 &        2.00 &        1.56 &       14.66 &       13.89 &       12.74 &       11.79 &       11.06 &        9.84 &        1.31 &       18.31 &       18.30 &       17.75 &       14.85 &       12.81 &       11.94 \\ [0.05cm] 
        2.00 &        5.00 &        1.66 &       12.19 &       11.23 &       10.47 &        9.82 &        9.88 &        8.86 &        1.24 &       17.68 &       17.32 &       17.40 &       14.32 &       12.58 &       11.86 \\ [0.05cm] 
        2.00 &       10.00 &        1.94 &       10.19 &        9.45 &        8.86 &        8.38 &        8.78 &        8.01 &        1.14 &       17.29 &       16.77 &       17.23 &       14.14 &       12.56 &       11.95 \\ [0.05cm] 
\hline 
        3.00 &        1.00 &        1.55 &       17.20 &       17.30 &       15.54 &       14.05 &       12.22 &       11.01 &        1.38 &       19.28 &       19.81 &       18.64 &       15.75 &       13.21 &       12.26 \\ [0.05cm] 
        3.00 &        2.00 &        1.49 &       15.20 &       14.53 &       13.42 &       12.26 &       11.35 &       10.16 &        1.30 &       18.44 &       18.45 &       17.94 &       14.94 &       12.86 &       12.01 \\ [0.05cm] 
        3.00 &        5.00 &        1.57 &       12.91 &       11.95 &       11.19 &       10.39 &       10.27 &        9.22 &        1.24 &       17.71 &       17.36 &       17.45 &       14.34 &       12.60 &       11.88 \\ [0.05cm] 
        3.00 &       10.00 &        1.77 &       10.63 &        9.89 &        9.30 &        8.77 &        9.08 &        8.30 &        1.14 &       17.30 &       16.78 &       17.24 &       14.14 &       12.57 &       11.96 \\ [0.05cm] 
\hline 
        5.00 &        1.00 &        1.48 &       18.03 &       18.32 &       16.76 &       14.75 &       12.62 &       11.51 &        1.37 &       19.61 &       20.21 &       19.17 &       16.02 &       13.36 &       12.46 \\ [0.05cm] 
        5.00 &        2.00 &        1.42 &       16.05 &       15.56 &       14.52 &       12.99 &       11.77 &       10.65 &        1.29 &       18.67 &       18.73 &       18.30 &       15.13 &       12.97 &       12.15 \\ [0.05cm] 
        5.00 &        5.00 &        1.48 &       13.65 &       12.72 &       11.99 &       11.00 &       10.66 &        9.59 &        1.24 &       17.79 &       17.45 &       17.57 &       14.40 &       12.63 &       11.92 \\ [0.05cm] 
        5.00 &       10.00 &        1.59 &       11.37 &       10.59 &        9.98 &        9.36 &        9.55 &        8.71 &        1.14 &       17.32 &       16.81 &       17.27 &       14.16 &       12.58 &       11.97 \\ [0.05cm] 
\hline 
        7.00 &        1.00 &        1.43 &       18.54 &       18.92 &       17.50 &       15.16 &       12.87 &       11.82 &        1.35 &       19.91 &       20.56 &       19.63 &       16.25 &       13.50 &       12.64 \\ [0.05cm] 
        7.00 &        2.00 &        1.38 &       16.64 &       16.27 &       15.31 &       13.48 &       12.05 &       10.99 &        1.29 &       18.88 &       18.97 &       18.62 &       15.29 &       13.06 &       12.27 \\ [0.05cm] 
        7.00 &        5.00 &        1.43 &       14.09 &       13.19 &       12.48 &       11.36 &       10.90 &        9.83 &        1.23 &       17.85 &       17.53 &       17.67 &       14.45 &       12.66 &       11.96 \\ [0.05cm] 
        7.00 &       10.00 &        1.50 &       12.08 &       11.23 &       10.58 &        9.87 &        9.94 &        9.03 &        1.14 &       17.33 &       16.83 &       17.31 &       14.17 &       12.59 &       11.98 \\ [0.05cm] 
\hline 
       10.00 &        1.00 &        1.39 &       19.19 &       19.71 &       18.51 &       15.68 &       13.17 &       12.21 &        1.33 &       20.33 &       21.05 &       20.29 &       16.57 &       13.68 &       12.88 \\ [0.05cm] 
       10.00 &        2.00 &        1.34 &       17.35 &       17.15 &       16.33 &       14.08 &       12.38 &       11.39 &        1.28 &       19.17 &       19.31 &       19.07 &       15.51 &       13.18 &       12.43 \\ [0.05cm] 
       10.00 &        5.00 &        1.39 &       14.65 &       13.80 &       13.14 &       11.82 &       11.17 &       10.13 &        1.23 &       17.95 &       17.64 &       17.83 &       14.53 &       12.70 &       12.02 \\ [0.05cm] 
       10.00 &       10.00 &        1.44 &       12.72 &       11.80 &       11.14 &       10.33 &       10.28 &        9.31 &        1.14 &       17.36 &       16.86 &       17.35 &       14.20 &       12.60 &       12.00 \\ [0.05cm] 
\hline 
       15.00 &        1.00 &        1.35 &       19.97 &       20.62 &       19.70 &       16.29 &       13.53 &       12.67 &        1.31 &       20.86 &       21.66 &       21.13 &       16.98 &       13.92 &       13.19 \\ [0.05cm] 
       15.00 &        2.00 &        1.31 &       18.14 &       18.09 &       17.47 &       14.71 &       12.73 &       11.85 &        1.27 &       19.58 &       19.78 &       19.70 &       15.82 &       13.36 &       12.66 \\ [0.05cm] 
       15.00 &        5.00 &        1.34 &       15.27 &       14.50 &       13.92 &       12.35 &       11.48 &       10.47 &        1.23 &       18.11 &       17.82 &       18.08 &       14.65 &       12.77 &       12.11 \\ [0.05cm] 
       15.00 &       10.00 &        1.38 &       13.29 &       12.33 &       11.69 &       10.77 &       10.59 &        9.59 &        1.14 &       17.41 &       16.91 &       17.42 &       14.23 &       12.62 &       12.02 \\ [0.05cm] 
\hline 
       20.00 &        1.00 &        1.32 &       20.61 &       21.37 &       20.74 &       16.79 &       13.81 &       13.04 &        1.29 &       21.34 &       22.21 &       21.91 &       17.33 &       14.12 &       13.46 \\ [0.05cm] 
       20.00 &        2.00 &        1.29 &       18.77 &       18.85 &       18.45 &       15.21 &       13.01 &       12.20 &        1.26 &       19.95 &       20.21 &       20.30 &       16.10 &       13.52 &       12.87 \\ [0.05cm] 
       20.00 &        5.00 &        1.32 &       15.76 &       15.07 &       14.58 &       12.76 &       11.71 &       10.75 &        1.22 &       18.27 &       17.99 &       18.32 &       14.77 &       12.84 &       12.20 \\ [0.05cm] 
       20.00 &       10.00 &        1.34 &       13.71 &       12.74 &       12.13 &       11.10 &       10.81 &        9.80 &        1.14 &       17.45 &       16.96 &       17.49 &       14.26 &       12.64 &       12.05 \\ [0.05cm] 
\hline 
       30.00 &        1.00 &        1.29 &       21.51 &       22.40 &       22.16 &       17.47 &       14.21 &       13.56 &        1.27 &       21.97 &       22.92 &       22.92 &       17.80 &       14.40 &       13.82 \\ [0.05cm] 
       30.00 &        2.00 &        1.26 &       19.63 &       19.83 &       19.76 &       15.86 &       13.38 &       12.69 &        1.24 &       20.54 &       20.87 &       21.23 &       16.53 &       13.76 &       13.20 \\ [0.05cm] 
       30.00 &        5.00 &        1.28 &       16.44 &       15.87 &       15.54 &       13.33 &       12.04 &       11.15 &        1.22 &       18.54 &       18.30 &       18.74 &       14.97 &       12.95 &       12.35 \\ [0.05cm] 
       30.00 &       10.00 &        1.29 &       14.24 &       13.28 &       12.73 &       11.54 &       11.09 &       10.08 &        1.13 &       17.54 &       17.06 &       17.64 &       14.33 &       12.67 &       12.10 \\ [0.05cm] 
\hline 
       50.00 &        1.00 &        1.25 &       22.69 &       23.73 &       24.11 &       18.32 &       14.70 &       14.22 &        1.24 &       23.05 &       24.13 &       24.71 &       18.58 &       14.85 &       14.42 \\ [0.05cm] 
       50.00 &        2.00 &        1.23 &       20.86 &       21.22 &       21.74 &       16.75 &       13.89 &       13.37 &        1.22 &       21.45 &       21.88 &       22.72 &       17.18 &       14.14 &       13.70 \\ [0.05cm] 
       50.00 &        5.00 &        1.25 &       17.44 &       17.04 &       17.03 &       14.13 &       12.48 &       11.72 &        1.21 &       19.01 &       18.83 &       19.51 &       15.32 &       13.15 &       12.62 \\ [0.05cm] 
       50.00 &       10.00 &        1.24 &       14.94 &       14.03 &       13.61 &       12.13 &       11.45 &       10.49 &        1.13 &       17.70 &       17.24 &       17.91 &       14.46 &       12.74 &       12.20 \\ [0.05cm] 
\hline 
      100.00 &        1.00 &        1.21 &       24.50 &       25.71 &       27.13 &       19.60 &       15.44 &       15.21 &        1.20 &       24.69 &       25.91 &       27.44 &       19.72 &       15.52 &       15.31 \\ [0.05cm] 
      100.00 &        2.00 &        1.20 &       22.55 &       23.07 &       24.53 &       17.94 &       14.58 &       14.29 &        1.20 &       22.88 &       23.43 &       25.09 &       18.17 &       14.72 &       14.47 \\ [0.05cm] 
      100.00 &        5.00 &        1.21 &       18.92 &       18.73 &       19.36 &       15.25 &       13.11 &       12.57 &        1.19 &       19.92 &       19.82 &       20.98 &       15.96 &       13.52 &       13.12 \\ [0.05cm] 
      100.00 &       10.00 &        1.18 &       15.98 &       15.24 &       15.13 &       13.05 &       11.96 &       11.13 &        1.12 &       18.07 &       17.65 &       18.52 &       14.73 &       12.90 &       12.42 \\ [0.05cm] 
\end{tabular} 
\vspace{0.1in} 
\end{center} 
\end{table}
\clearpage
\end{landscape}
\clearpage

\clearpage

\clearpage

\clearpage

\vspace{1.2in}

\begin{appendix}
\label{sec:app}
\section*{An Analytic Fit for Radius}
\label{ssec:analytic}
Here, we present a fit to the radius-entropy-mass relation that works
quite well for masses between $0.3~M_J$ and $15~M_J$ (and for
reasonable radii).  Because of a distinct change in the behavior of
the $R[S,M]$ surface near $2~M_J$, we present two separate fits in the
two domain regions.  Each is an exponential of a polynomial (3rd order
in $S$ and 2nd order in $M$).

For $M \le 2~M_J$,
\begin{eqnarray}
\nonumber        p_{00} & = &       -1.27 \\
\nonumber        p_{10} & = &      0.5404 \\
\nonumber        p_{01} & = &    -0.09388 \\
\nonumber        p_{20} & = &    -0.08935 \\
\nonumber        p_{11} & = &      0.1549 \\
\nonumber        p_{02} & = &     -0.3515 \\
\nonumber        p_{30} & = &    0.005578 \\
\nonumber        p_{21} & = &    -0.02021 \\
                 p_{12} & = &     0.05055 \, .
\end{eqnarray}
Likewise, for $M > 2~M_J$,
\begin{eqnarray}
\nonumber        p_{00} & = &      0.2273 \\
\nonumber        p_{10} & = &    -0.03987 \\
\nonumber        p_{01} & = &    -0.02136 \\
\nonumber        p_{20} & = &   -0.002942 \\
\nonumber        p_{11} & = &     0.00655 \\
\nonumber        p_{02} & = &   -0.001173 \\
\nonumber        p_{30} & = &   0.0007376 \\
\nonumber        p_{21} & = &  -0.0005432 \\
                 p_{12} & = &   0.0001446 \, .
\end{eqnarray}
Then, for $M$ in $M_J$ and $S$ in $k_B$ per baryon, the following is
an excellent fit to the $R[S,M]$ function when $R \le 2~R_J$:
\begin{eqnarray}
           {\rm LogR} & =       & p_{00} + p_{10}S + p_{01}M + p_{20}S^2 + p_{11}S M + p_{02} M^2 + p_{30}S^3 + p_{21}S^2M + p_{12}S M^2 \\
\label{eq:RMS} R[S,M] & \approx & 10^{\rm LogR}~R_J \, .
\end{eqnarray}

\end{appendix}

\end{document}